\documentclass[11pt,a4paper]{article}
\pdfoutput=1
\usepackage{jheppub}
\usepackage{amsfonts}
\usepackage{bbm}
\usepackage{graphicx}
\usepackage{caption}
\usepackage{subcaption}
\usepackage{bigints}
\usepackage[normalem]{ulem}
\usepackage{bm}
\usepackage{mathtools}
\usepackage{booktabs}
\usepackage{enumitem}

\def\a{\alpha}
\def\b{\beta}
 \def\g{\gamma}

\def\e{\epsilon}

\def\l{\lambda}    
\def\m{\mu}
\def\n{\nu}

\def\t{\tau}

\def\z{\zeta}
\def\vth{\vartheta}

\newcommand{\dd}{{\mathrm d}}

\newcommand{\CE}{\mathcal{E}}

\newcommand{\CK}{\mathcal{K}}

\newcommand{\CM}{\mathcal{M}}
\newcommand{\CN}{\mathcal{N}}

\newcommand{\CH}{\mathcal{H}}

\newcommand{\SL}{\mathrm{SL}}

\renewcommand{\Im}{{\rm Im}}
\renewcommand{\Re}{{\rm Re}}

\newcommand{\sgn}{\mbox{sgn}}

\def\CP{\mathbb{CP}}

\newcommand{\IR}{\mathbb{R}}
\newcommand{\IC}{\mathbb{C}}
\newcommand{\IZ}{\mathbb{Z}}
\newcommand{\IH}{\mathbb{H}}

\newcommand{\IP}{\mathbb{P}}

\newcommand{\half}{\frac{1}{2}}
\newcommand{\ndt}{\noindent}

\newcommand{\nn}{\nonumber}

\def\e{\epsilon}

\def\p{\partial}

\def\bea{\begin{eqnarray}}
\def\eea{\end{eqnarray}}
\def\be{\begin{equation}}
\def\ee{\end{equation}}
\def\ba{\begin{align}}
\def\ea{\end{align}}

\newcommand{\bem}{\begin{pmatrix}}
\newcommand{\eem}{\end{pmatrix}}

\def\qb {\bar{q}}

\def\={\;  = \;}
\def\+{\, + \,}

\def\wt{\widetilde}
\def\wh{\widehat}
\def\bar{\overline}

\def\rt2{\sqrt{2}}

\renewcommand{\Im}{\mathrm{Im}}
\renewcommand{\Re}{\mathrm{Re}}

\newcommand{\bPhi}{\overline{\Phi}}
\newcommand{\bP}{\overline{P}}

\newcommand{\bSigma}{\overline{\Sigma}}

\newcommand{\taubar}{\overline{\tau}}

\newcommand{\smat}[4]{\bigl(\smallmatrix #1&#2\\ #3&#4\endsmallmatrix\bigr)}

\newcommand{\hol}{\text{hol}}

\setlist[itemize]{noitemsep, topsep=0pt}

\newcommand{\andd}{\quad \mbox{ and } \quad}
\newcommand{\where}{\quad \mbox{ where }}

\newcommand{\del}{\partial}
\newcommand{\ddd}{\mathrm{d}}

\newcommand{\pmat}[1]{\left( \smallmatrix #1 \endsmallmatrix \right)}
\newcommand{\mat}[1]{\left( \begin{matrix} #1 \end{matrix} \right)}

\def\chl{\chi_{{\mathrm{ell}}}}
\newcommand{\tb}{{\bar \tau}}

\newcommand{\Ait}{{\wt{\mathrm{A}}_1}}
\newcommand{\CZtwo}{{\IC /\IZ_2}}
\newcommand{\CZtwot}{{\wt{\IC /\IZ_2}}}

\newcommand{\rr}{r}
\newcommand{\ww}{w}
\newcommand{\mm}{m}
\newcommand{\nnn}{n}
\newcommand{\uu}{u}
\newcommand{\uub}{\bar{u}}
\newcommand{\vv}{v}

\newcommand{\mmm}{\mu}
\newcommand{\zzz}{\mathfrak{z}}
\newcommand{\zzb}{\bar{\mathfrak{z}}}
\newcommand{\zs}{s}
\newcommand{\zt}{t}
\newcommand{\tim}{\tau_2}
\newcommand{\tre}{\tau_1}
\newcommand{\thone}{\vartheta_{1}}
\newcommand{\yy}{\zeta}

\newcommand{\intt}{\int\displaylimits}
\newcommand{\erf}{\mathrm{erf}}
\newcommand{\qr}{Q_N}

\newcommand{\sign}[1]{\mathrm{sgn}\left( #1 \right)}
\newcommand{\CMM}{\mathcal{M}}
\newcommand{\im}{\mathrm{Im}\,}
\newcommand{\kk}{k}
\newcommand{\kkk}{\mathbf{k}}

\def\lp{\left(}
\def\rp{\right)}
\def\lb{\left[}
\def\rb{\right]}

\title{Squashed Toric Manifolds and Higher Depth Mock Modular Forms}

\author{Rajesh Kumar Gupta$^1$, Sameer Murthy$^1$, and Caner Nazaroglu$^2$}
\affiliation{$^1$Department of Mathematics, King's College London,The Strand, London WC2R 2LS, UK}
\affiliation{$^2$Mathematical Institute, University of Cologne,  Weyertal 86-90, 50931 Cologne, Germany}

\emailAdd{rajesh.gupta, sameer.murthy at kcl.ac.uk, cnazarog  at math.uni-koeln.de}

\abstract{
Squashed toric sigma models are a class of sigma models
whose target space is a toric manifold in which the torus fibration is squashed away from the 
fixed points so as to produce a neck-like region. The elliptic genera of squashed toric-Calabi-Yau manifolds 
are known to 
obey the modular transformation property of holomorphic 
Jacobi forms, but have an explicit non-holomorphic dependence on the modular parameter. 
The elliptic genus of the simplest one-dimensional example is known to be a mixed mock Jacobi form, 
but the precise automorphic nature for the general case remained to be understood. 
We show that these elliptic genera fall precisely into a class of functions called \emph{higher-depth} 
mock modular forms that have been formulated recently in terms of indefinite theta series.
We also compute a generalization of the elliptic genera of these models corresponding to an additional set of 
charges corresponding to the toric symmetries. Finally we speculate on some relations of the elliptic genera of 
squashed toric models with the Vafa-Witten partition functions of~$\CN=4$ SYM theory on~$\CP^2$. 
}

\begin{document}

%

\maketitle

\section{Introduction and Summary}

The context underlying this paper is a three-way relation that was established in the 1980s between 
compact Calabi-Yau (CY) manifolds, two-dimensional~$\CN=(2,2)$ superconformal field 
theories (SCFTs), and modular and Jacobi 
forms~\cite{Schellekens:1986yi,Schellekens:1986xh,Pilch:1986en, Witten:1986bf, Witten:1987cg, Alvarez:1987wg,Alvarez:1987de}. 
A central object in this story is the \emph{elliptic genus}, which is a generating function of a 
sequence of Dirac indices associated to the manifold in question.
The relation between geometry and automorphic forms arises because the elliptic genus of 
a compact CY manifold of complex dimension~$d$ is a Jacobi form of weight~0 and index~$d/2$. 
The relation with physics arises by thinking of this CY manifold as the target space of 
two-dimensional~$\CN=(2,2)$ SCFTs of central charge~$c=3d$. The elliptic genus 
encodes the information about BPS states of this SCFT, and it can be computed as a 
functional integral using various physics techniques.

The paper~\cite{Gupta:2017bcp} began a systematic study of similar ideas for a class of 
theories which generalized the above story. On the geometric side, one now considered a class 
of SCFTs corresponding to~\emph{squashed toric-Calabi-Yau manifolds}, a certain deformation 
of toric CY manifolds, introduced in~\cite{Hori:2001ax}, which we explain below.
The resulting elliptic genera, which were computed in~\cite{Gupta:2017bcp} as a torus functional integral using the gauged 
linear sigma model (GLSM) description of the theory, obey the modular transformation properties 
of a Jacobi form as expected. The interesting feature is that they explicitly depend on~$\overline{\tau}$, the 
complex conjugate of the modular parameter~$\tau$ of the torus. The dependence on~$\taubar$ 
is intuitively understood by the fact that these target spaces are necessarily non-compact and 
that the density of bosonic and fermionic states in the corresponding continuum are not necessarily equal.

This behavior is characteristic of a class of automorphic functions called mock modular forms~\cite{Zwegers:2008zna, BruinierFunke}, 
which, along with the closely related mock Jacobi forms, have been discussed with great interest in recent 
years in diverse contexts~\cite{Zagier2009, Dabholkar:2012nd, Bringmann:2017book}.
However, the details of the modular transformation properties of the elliptic genera of squashed toric models do 
not quite match the basic definitions of mock modular and mock Jacobi forms and suggest some 
sort of a generalization. 
The precise nature of their modular transformation properties, beyond the general properties 
obeyed by the partition function of a SCFT, was left open in~\cite{Gupta:2017bcp}. This is the question 
that we address and answer in the current paper. The answer is related to a class of functions, called 
\emph{higher-depth mock modular forms}, that have been formulated recently in terms of 
indefinite theta functions~\cite{ZagierZwegers, Alexandrov:2016enp}, and further developed 
in~\cite{kudla2016theta, westerholt2016indefinite, Nazaroglu:2016lmr, funke2017theta}.
We show here that the elliptic genera of squashed toric sigma models belong precisely to this class.  
In the rest of this introduction we explain some of the details of this correspondence.

\vspace{0.4cm}

Squashed toric manifolds are an interesting subclass of toric manifolds.
Their construction begins by considering a (real) $2d$-dimensional toric manifold~$M_\text{tor}$, which 
has the structure of a~$d$-dimensional torus fibered on a~$d$-dimensional base. The size of this torus fiber varies along the base and at
distinguished fixed points it shrinks 
to zero size. The corresponding squashed toric manifold~$\wt M_\text{tor}$ looks the same 
as the unsquashed~$M_\text{tor}$ near the fixed 
points, but the torus now has a constant size in the deep interior parts of the manifold.  
In this paper we study situations in which the initial toric manifold~$M_\text{tor}$ obeys the CY condition. 
Such manifolds are necessarily non-compact and have an asymptotic cone-like structure.
The squashing deformation reaches the asymptotic region and deforms it into a cylindrical shape. 
Although the SCFT corresponding to the squashed model is anomaly-free, the target space geometry is 
not Ricci flat, and the sigma model is expected to be supported by a non-trivial dilaton profile.
The simplest example corresponds to a two-dimensional manifold where we begin with a toric space that is 
asymptotically of the form~$\IC/\IZ_{2}$, and the corresponding squashed deformation has a cigar-like shape.

Such deformations were studied in~\cite{Hori:2001ax} using an~$\CN=(2,2)$ GLSM description~\cite{Witten:1993yc}. 
The GLSM for the undeformed situation has~$n$ chiral superfields and~$n-d$ gauge superfields, and its $2d$-dimensional 
vacuum manifold is the toric manifold~$M_\text{tor}$. The squashing deformation adds~$d$ compensator-chiral superfields 
and gauges the $d$-dimensional flavor symmetry of the original model which now also acts on the compensator fields
as a shift symmetry.
The resulting~$2d$-dimensional vacuum manifold is a squashed toric manifold. The data that define the 
squashed toric model are the original data of the toric model plus the strengths of the couplings~$k_\ell$,
$\ell = 1,\cdots, d$, of the compensators. 
The squashed models are not Ricci-flat even if we begin with a toric CY. However, 
the original CY condition in the GLSM description ensures the presence of non-anomalous chiral~$U(1)$ 
R-symmetries even in the squashed models, so that they can flow to~$\CN=(2,2)$ SCFTs. 
In the simplest two-dimensional example mentioned above, the squashed model~$\wt{\IC/\IZ_{2}}$ is conjectured to flow 
to the~$SL(2,\IR)_{k}/U(1)$ cigar coset, where the level~$k$ is the effective coupling of the compensator 
superfield. 

The elliptic genus of these models was computed in~\cite{Gupta:2017bcp} using the technique of 
supersymmetric localization applied to the GLSM description. The resulting expression~$\chi_\text{ell}(\wt M_\text{tor}; \tau,z)$
transforms like a holomorphic Jacobi form, but it explicitly depends on~$\taubar$.
Its~$\taubar$-derivative equals an integral of a function that has both holomorphic and anti-holomorphic 
dependence on~$\tau$ over a~$(d-1)$-dimensional torus. 
For the squashed~$\wt{\IC/\IZ_{2}}$ model discussed above, the holomorphic anomaly equation is precisely 
the one obeyed by mixed mock Jacobi forms. 
Moreover, the~$\taubar$-derivative can be identified as coming from momentum and winding modes of the compensator 
superfield about the asymptotic cylinder~\cite{Murthy:2013mya}. 
The elliptic genus of this model is governed by~$A_{1,k}(\t,z)$, the so-called Appell-Lerch sum in two 
variables~\cite{Zwegers:2008zna, Dabholkar:2012nd}, and is equal to the elliptic genus of the cigar theory in accordance 
with the conjecture.
In the current paper we generalize the above calculation to that of the \emph{flavored elliptic genus}, 
which is the torus partition function with an additional set of chemical potentials~$\{\beta_{\ell}\}$,
$\ell = 1,\cdots, d$ coupled to the toric symmetries. In the~$\wt{\IC/\IZ_{2}}$ model, we find that 
these are governed by the three-variable functions~$A_{1,k}(\tau,u,v)$ introduced by Zwegers in \cite{Zwegers:2011}, 
thus confirming the results of~\cite{Ashok:2014nua} on the flavored elliptic genus of the cigar theory
using current algebra techniques in the cigar SCFT.

\vspace{0.2cm}

The main result of this paper is that the elliptic genera of squashed toric manifolds of dimension~$n$ 
are precisely modular completions of mock modular forms of depth~$n$. These are functions 
that generalize the usual notion of a mock modular form in that they are modular but not quite holomorphic.
We recall that the holomorphic completion of mock modular forms obey a differential equation which says that 
their~$\taubar$-derivative is an anti-holomorphic modular form. The generalized mock modular forms at depth~$n$ 
are defined recursively as follows. 
At depth~one they are simply linear combinations of products of holomorphic modular forms and mock modular
forms so that the~$\taubar$-derivative of their holomorphic completion is a linear combination of holomorphic 
and anti-holomorphic modular forms, and are called mixed mock modular forms~\cite{Dabholkar:2012nd}.
At depth~$n>1$ the~$\taubar$-derivative of the holomorphic completion is a linear combination of mock 
modular forms of depth~$(n-1)$ and an anti-holomorphic modular form.

In fact one can be more specific. In the well-understood one-dimensional case, the elliptic genus 
can be written in terms of the Appell-Lerch sum~$A_{1,k}$ as mentioned above. The 
functions~$A_{1,k}$ are examples of another wider class of functions called indefinite theta series.
These are functions which are like the usual theta series, but the  
associated lattice has indefinite signature which means that the full lattice sum is divergent. 
When the lattice has signature~$(r,1)$, it was shown in~\cite{Zwegers:2008zna} that one 
can define theta series associated to such lattices by summing over only a cone inside the indefinite lattice. 
The above Appell-Lerch sum falls in the class of~$(1,1)$ lattices and this leads precisely to mixed mock 
modular forms in the sense discussed above. 
The generalization of this construction to lattices of arbitrary signature was discussed more 
recently~\cite{ZagierZwegers, Alexandrov:2016enp, kudla2016theta, westerholt2016indefinite, Nazaroglu:2016lmr, funke2017theta} leading to examples of mock modular forms of higher depth. 
We show in this paper that the (flavored) elliptic genera of squashed toric models are precisely 
of this form, i.e.~the elliptic genus of an~$n$-dimensional squashed toric manifold is built out of 
indefinite theta functions associated with a lattice of signature~$(n,n)$. In the asymptotic region, this lattice can 
be thought of as the lattice of left and right-moving momenta around the~$n$-dimensional torus of the 
squashed toric model.

The plan of the paper is as follows. 
In Section~\ref{sec:STM} we review the computation of the elliptic genus of squashed toric GLSMs 
and generalize it to include chemical potentials for flavor symmetries. We give an expression 
for the elliptic genus as an integral of a meromorphic function on a~$d$-dimensional torus, 
and discuss its modular properties.
In Section~\ref{sec:MMF} we review the notion of mock modular forms of higher depth, and their 
construction in terms of indefinite theta functions. 
In Section~\ref{sec:EllGenMMF} we evaluate the integral expressions for the elliptic genera and show 
that they can be written in terms of completions of indefinite theta functions. 
In Section~\ref{sec:VWrel} we comment on possible relations between Vafa-Witten partition functions 
on~$\IC\IP^2$ and the elliptic genera of squashed toric models.

\section{Squashed Toric Manifolds and Their Elliptic Genera \label{sec:STM}}

In this section we briefly review the notion of squashed toric manifolds and 
the computation of their elliptic genera using the GLSM construction~\cite{Gupta:2017bcp}. 
We then present some new results, generalizing the calculations of~\cite{Gupta:2017bcp}, 
on the flavored elliptic genera of these manifolds.
These functions are the result of supersymmetric functional integrals with fields twisted by the 
R-symmetry (as in the elliptic genus), as well as by the global toric symmetries of the manifold. 
We show that these functions transform like holomorphic multi-variable Jacobi forms 
of weight~zero.
As the first two subsections are purely a review of~\cite{Gupta:2017bcp}, we will be very brief 
and refer the reader to the original paper for a more detailed explanation of the relevant concepts.

\subsection{A Brief Review of Squashed Toric Models}

Our starting point is the~$\CN=(2,2)$ GLSM description of $2d$-dimensional (unsquashed) toric manifolds. 
The theory contains $n-d$ abelian vector superfields~$V_{a}$, $a=1,\ldots,n-d$ with corresponding  twisted 
chiral superfields~$\Sigma_{a}$ and $n$ chiral superfields $\Phi_{i}$, $i=1,\ldots,n$ with charges~$Q^{i}_{a}$ 
with respect to the gauge group. Its action is:
\begin{align} 
S_{0} \= &\frac{1}{2\pi} \int\dd^2x\,  \int \dd^4\theta \, \biggl( \sum_{i=1}^{n} \bPhi_i \, 
  \exp \Bigl(\, \sum_{a=1}^{n-d}Q_{i}^{a} V_{a} \Bigr) \Phi_i
 \, - \, \sum_{a=1}^{n-d}\,{1\over 2e_{a}^2} \bSigma_{a} \, \Sigma_{a} \biggr)  
 \notag \\ & \quad
 \+ \frac{1}{4\pi} \int\dd^2x\, \int\dd^2\wt\theta \, \sum_{a=1}^{n-d} t_{a} \Sigma_{a}  \+ \text{c.c.} \, ,
 \label{action0}
\end{align}
where~$t_{a}=r_{a}-i \vth_{a}$ is the complexified Fayet-Iliopoulos parameter for the vector superfield~$V_{a}$.
The D-terms can be integrated out by setting:
\be\label{Dterm1} 
D_{a} \= - e_{a}^{2} \, \mu_{a} \, , \where \  
\mu_{a} \= \sum_{i=1}^{n} Q_{i}^{a} \, |\phi_{i}|^{2} - r_{a}  \, , \quad a = 1, \dots, n-d \, . 
\ee
The classical vacuum manifold of the theory is given by~$\mu_a^{-1}(0)$ modulo gauge transformations, which is 
immediately recognized to be a~$2d$-dimensional toric manifold, according to the symplectic quotient construction 
of toric manifolds~\cite{daSilva:2001}. In the quantum theory this account is modified, and the situation depends on 
the existence of conserved chiral R-symmetries. The existence of such symmetries is ensured by the 
\emph{anomaly-cancellation condition}, namely the sum of charges for all the gauge fields vanishes, i.e.
\begin{equation}
\sum_{i=1}^n Q_i^a = 0, \quad a = 1, \ldots, n-d \, . 
\end{equation}
In this situation the GLSM flows to a SCFT whose target space is a toric Calabi-Yau manifold which is in the same
K\"ahler class as the vacuum manifold~\eqref{Dterm1}. This will be the situation for all the models that we consider in this paper. 

The toric GLSM defined by the action~\eqref{action0} also has a global $U(1)^{d}$ flavor symmetry under which 
chiral superfields $\Phi_{i}$ have charges $F_{i}^{\ell}$, 
$\ell = 1, \ldots, d$. We perform a squashing deformation~\cite{Hori:2001ax} by gauging this $U(1)^{d}$ symmetry 
and adding $d$ compensator chiral superfields $P_\ell$, $\ell = 1,\ldots,d$, on which the $U(1)^{d}$ flavor symmetry 
acts as shift symmetries. Denoting the new vector superfields by $V'_{\ell}$, $\ell = 1, \ldots, d$, (and the corresponding 
twisted chiral superfields by $\Sigma'_{\ell}$) the action for the deformed theory is obtained by adding the appropriate canonical kinetic terms:
\begin{align} 
&S_{\mathrm{squashed}} \= \frac{1}{2\pi} \int\dd^2x\,  \int \dd^4\theta \, \biggl[ \sum_{i=1}^{n} \bPhi_i \, 
  \exp \Bigl(\, \sum_{a=1}^{n-d}Q_{i}^{a} V_{a} + \sum_{\ell=1}^{d}F_{i}^{\ell} V'_{\ell} \Bigr) \Phi_i
 \, - \, \sum_{a=1}^{n-d}{1\over 2e_{a}^2} \bSigma_{a} \, \Sigma_{a}
 \notag \\ &  \ \ 
 \, - \, \sum_{\ell=1}^{d} \,{1\over 2e_{\ell}'^2}\,\bSigma'_{\ell} \, \Sigma'_{\ell}
 \+
  \sum_{\ell=1}^{d} \, {k_{\ell}\over 4}(P_{\ell}+\bP_{\ell}+V'_{\ell})^2
  \biggr]  
 \+ \frac{1}{4\pi} \int\dd^2x\, \int\dd^2\wt\theta \, \sum_{a=1}^{n-d} t_{a} \Sigma_{a}  \+ \text{c.c.} \, .
 \label{Ssquashed}
\end{align}
The D-terms corresponding to the new gauge superfields can be integrated out by setting
\begin{equation}\label{DtermP}
D'_\ell\=-e_{\ell}'^{\,2}\mu'_\ell \, , \where \ 
 \mu'_\ell\=\sum_{i=1}^nF^\ell_i \, |\phi_i|^2+k_{\ell} \, \Re P_{\ell} \, ,
 \qquad \ell=1,...,d\,.
\end{equation}
The vacuum manifold of the deformed theory is then found by setting both D-terms in \eqref{Dterm1} 
and \eqref{DtermP} to zero and modding gauge symmetries out:
\be \label{sqVac}
\mu^{-1}(0)/(U(1)^{n-d}\times U(1)^d)\,, \qquad \mu\; \coloneqq\;(\mu_a,\mu'_\ell) \,.
\ee
This has the same symplectic quotient structure as the original toric manifold, and therefore yields 
toric manifolds in their own right. 
The vacuum manifolds of the squashed models are called the squashed toric manifolds~\cite{Hori:2001ax}.

For squashed models, the base of the vacuum manifold can be parametrized by~$\Re P_{\ell}$ with 
$\Im P_{\ell}$ parametrizing circle fibers over this base (with an appropriate gauge choice). These fibers 
have fixed sizes of order~$\sqrt{k_{\ell}}$ in the interior part of the base but can degenerate to zero size 
at the boundaries. Importantly for our purposes, the squashed theory has a $U(1)^d$ toric symmetry 
which acts as translations along these circle fibers.

\subsection{Elliptic Genera of Squashed Toric Models}

From the physics point of view, the elliptic genus is the partition function of the theory on a 
two-dimensional torus with periodic boundary conditions, coupled to a constant
background R-symmetry gauge field~$A^R_\mu$. As this is a quantity protected by supersymmetry, 
one can compute the elliptic genus using the GLSM description of the previous subsection.
This was done in~\cite{Gupta:2017bcp} for the squashed toric model~\eqref{Ssquashed} using the technique of 
supersymmetric localization applied to GLSMs~\cite{Benini:2013nda, Benini:2013xpa}.

Localization reduces the infinite dimensional path integral required to compute the partition function to a 
finite dimensional integral over the \emph{localization manifold}, i.e.~the set of solutions
to the off-shell BPS equations for the right moving supercharge.
For our model~\eqref{Ssquashed}, the localization manifold is parametrized by the holonomies 
of the gauge fields~$V^a$, $a=1,\cdots, n-d$, $V'^\ell$, $\ell = 1,\cdots, d$, 
along the two cycles of the torus, i.e.,
\be \label{defuupr}
u^{a}\; \coloneqq \; \oint_A V^{a}-\tau\oint_B V^{a} \, , \qquad 
u'^{\ell} \; \coloneqq \; \oint_A V'^{\ell}-\tau\oint_B V'^{\ell} \, ,
\ee 
with all other modes set to zero. 
The answer also depends on the holonomy of the background R-symmetry gauge field 
\be \label{defz}
z \; \coloneqq \; \oint_A A^R-\tau\oint_B A^R \, .
\ee 
Due to the large gauge transformation symmetries, the holonomies~$u^{a},u'^{\ell}$, 
and~$z$ take values in~$E_\tau \coloneqq \mathbb C/(\mathbb Z\tau+\mathbb Z)$.

The final expression for the elliptic genus of the squashed model is~\cite{Gupta:2017bcp} 
\bea\label{SquashElliptic2}
\chi_\text{ell}(\wt M_\text{tor}; \tau,z)  \= 
\int_{E_\tau^{d}} \, \prod_{\ell=1}^d   \frac{\dd^2 u'_\ell}{\t_{2}} \,  
\wt H_\ell (\tau, z, u'_\ell)  \; \chi_\text{ell}(M_\text{tor};\tau,z, u') \, ,
\eea
where~$\chi_\text{ell}(M_\text{tor};\tau,z, u')$ is elliptic genus of the unsquashed toric sigma model, and 
\be\label{def.Hell} 
\wt H_\ell (\tau, z, u) \= k_{\ell} \sum_{m,w \, \in \, \mathbb{Z}} \,
e^{2\pi i b_\ell w z - \frac{\pi k_\ell}{\tau_2} \bigl( w\tau+m+u+\frac{b_\ell z}{k_\ell} \bigr)
\bigl(w \bar{\tau}+m+\overline{u}+\frac{b_\ell z}{k_\ell } \bigr)} \,,
\ee 
with~$b_\ell = \sum_{i=1}^n F_i^\ell$. 
The elliptic genus~$\chi_\text{ell}(M_\text{tor};\tau,z, u')$ of the unsquashed toric sigma model is 
obtained by picking up the residues of the one loop determinant $Z_{\text{1-loop}}(\tau,z,u,u')$, 
calculated for all the fields, at the set of poles~$\mathfrak{M^*}_{\text{sing}}$
\be\label{JKresidue1}
\chi_\text{ell}(M_{\text{tor}};\tau,z,u')\=
-\sum_{u_*\in\mathfrak{M^*}_{\text{sing}}}\underset{u=u_*}{\text{JK-Res}} (Q(u_*),\eta)\,
Z_{\text{1-loop}}(\tau,z,u,u') \, , 
\ee
where $\text{JK-Res}(Q(u_*),\eta)$ is a residue operation called the Jeffrey-Kirwan residue. 
Here the one-loop determinant in question is
\be \label{Z1loop}
Z_{\text{1-loop}}(\tau,z,u,u')\=\biggl(\frac{i \, \eta(\tau)^{3}}{\vth_1(\tau,z)}\biggr)^{n-d} \, 
\prod_{i=1}^n\frac{\vth_1(\tau,-z+Q_i\cdot u
+F_i\cdot u')}{\vth_1(\tau,Q_i\cdot u+F_i\cdot u')} \, ,
\ee
where $Q_i\cdot u=\sum_{a=1}^{n-d}Q^{a}_{i}\,u^{a}$ and $F_i\cdot u'=\sum_{\ell=1}^{d}F^{\ell}_{i}u^{\ell'}$. 
The first factor in~\eqref{Z1loop} comes from~$(n-d)$ vector multiplets and the second factor comes from~$n$ chiral multiplets. 

We note here that the elliptic genus of the unsquashed model $\chi_\text{ell}(M_{\text{tor}};\tau,z,u')$ 
is a meromorphic function of $u'$. The poles in~$u'$ are related to the non-compactness of the underlying toric manifold.
This non-compactness leads to a divergence in the the naive definition of the elliptic genus, and in order to regulate 
this divergence, we turn on a non-zero holonomy~$u'$ of the background flavor symmetry gauge field.
Now, the integral over $u'$ in the squashed model~\eqref{SquashElliptic2} smoothens the pole of the unsquashed model, 
and as as result, the elliptic genus $\chi_\text{ell}(\wt M_\text{tor}; \tau,z)$ is a well-defined holomorphic function of $z$. 
We refer the reader to~\cite{Gupta:2017bcp} for more details.
As we will see below, we can further introduce the chemical potentials~$\{\beta_{\ell}\}$ corresponding to the global 
symmetries of the squashed toric manifold and this will introduce non-holomorphicity in the chemical potential.

\vspace{0.4cm}

\ndt {\bf Modular and elliptic properties:} \hfill \break
 The modular and elliptic properties of the elliptic genus~$\chi_\text{ell}(\wt M_\text{tor}; \tau,z)$ 
were discussed in~\cite{Gupta:2017bcp}.
In order to compute its elliptic transformation properties, it is useful to unfold the integrals over $E_\tau$ 
for each $\ell$  in \eqref{SquashElliptic2} to the entire complex plane:
\be\label{unfoldedEq2}
\chi_\text{ell}(\wt M_\text{tor}; \tau,z)  \= (\prod_{\ell=1}^d k_{\ell})
\bigintsss_{\IC^{d}} \, \prod_{\ell=1}^d   \frac{\dd^2 u'^{\ell}}{\t_{2}} \, 
e^{ - \frac{\pi k_\ell}{\tau_2} \bigl( u'^{\ell}+\frac{b_\ell z}{k_\ell} \bigr)
\bigl(\overline{u'^{\ell}}+\frac{b_\ell z}{k_\ell } \bigr)} \, 
\chi_\text{ell}(M_\text{tor};\tau,z, u') \,. 
\ee
Assuming for convenience that $\frac{b_{\ell}}{k_{\ell}}$ is an integer for each $\ell$,\footnote{If~$b_\ell/k_\ell$ is not an integer, 
one can change the elliptic variable $z$ to $(\prod_{\ell=1}^{d}k_{\ell})\,z'$ and consider the elliptic transformations $z'\rightarrow z'+\l\t+\m$. 
In this case the index of $\chi_\text{ell}(\wt M_\text{tor}; \tau,z')$ is $(\prod_{\ell=1}^{d}k^{2}_{\ell})(\frac{d}{2}+\sum_{n=1}^{d}\frac{b^{2}_{n}}{k_{n}})$.} 
it is easy to see that, for~$\l, \mu \in \IZ$,
\be \label{elltranschi}
\chi_\text{ell}(\wt M_\text{tor}; \tau,z+\l\t+\m)
\=e^{-2\pi i(\frac{d}{2}+\sum_{\ell=1}^{d}\frac{b^{2}_{\ell}}{k_{\ell}})(\l^{2}\t+2\l z)}\chi_\text{ell}(\wt M_\text{tor}; \tau,z) \,.
\ee
Under the modular transformations the elliptic genus transforms as
\be \label{modtranschi}
\begin{split}
&\chi_\text{ell}  \Bigl( \wt M_\text{tor}; \tau+1,z\Bigr) \=\chi_\text{ell}  \Bigl( \wt M_\text{tor}; \tau,z\Bigr)\,,\\
&\chi_\text{ell} \Bigl( \wt M_\text{tor}; -\frac{1}{\tau},\frac{z}{\tau}\Bigr) \= e^{\frac{2\pi i}{\t}z^{2}(\frac{d}{2}
+\sum_{\ell=1}^{d}\frac{1}{\wt k_{\ell}})}\chi_\text{ell}(\wt M_\text{tor}; \tau,z) \,,
\end{split}
\ee
with $\frac{1}{\wt k_{\ell}}=\frac{b^{2}_{\ell}}{k_{\ell}}$. The first equality follows trivially from \eqref{SquashElliptic2}. 
To see the second equality we start from \eqref{unfoldedEq2}, change variables from $u'^{\ell}$ to $\frac{u'^{\ell}}{\t}$, 
and then use the modular properties of $\chi_\text{ell}(M_\text{tor};\tau,z, u')$, which transforms like a Jacobi 
form\footnote{We will review the basic notions of Jacobi forms in the following section.} 
of weight zero and (matrix) index
\be
M_{00}=\frac{d}{2} \,, \qquad M_{\ell0}=M_{0\ell}=-\frac{b_{\ell}}{2} \,,\qquad M_{\ell\ell'}=0 \,, \qquad \ell,\ell' \= 1, \cdots, d \,.
\ee 

The content of Equations~\eqref{elltranschi},~\eqref{modtranschi} can be summarized by the statement that  
the elliptic genus of the squashed model transforms like a Jacobi form of weight zero and index 
\be
m \= \frac{d}{2}+\sum_{\ell=1}^{d} \frac{1}{\wt k_{\ell}}\, .
\ee

\vspace{0.2cm}

\ndt {\bf Holomorphic anomaly:} \hfill \break
As is evident from Equation~\eqref{SquashElliptic2}, the elliptic genus $\chi_\text{ell}(\wt M_\text{tor}; \tau,z)$ 
is not holomorphic in $\t$. Indeed it satisfies the following holomorphic anomaly equation:
\bea \label{holanom}
\p_{\bar\t} \, \chi_\text{ell}(\wt M_\text{tor}; \tau,z)&& \=-\sum_{i,j=1}^{d}\bigintsss_{E_\tau^{(d-1)}} \, \prod_{\ell=1,\atop \ell\neq i}^d 
\Bigl(\frac{\dd^2 u'_\ell}{\t_{2}} \wt H_\ell (\tau, z, u'_\ell)\Bigr) \,  
\underset{v_{j}(u'_{i})=0}{\text{Res}} \Bigl(\chi_\text{ell}(M_\text{tor};\tau,z, u') \Bigr)    \cr
&& \qquad\qquad\qquad \qquad\qquad\qquad \times\frac{1}{k_{i}} \p_{\bar u^{i}}  \wt H_i (\tau, z, u'_i )\mid_{v_{j}(u'_{i})=0}\,.
\eea
Here $v_{i}(u')$ are certain linear combinations of $\{u'_\ell\}$ involving the flavor charges~$F_{i}^{\ell}$ (see~\cite{Gupta:2017bcp} for more details).
We will see later that the non-holomorphic behaviour captured by Equation~\eqref{holanom} is precisely that of a completed 
higher depth mock modular form. 

\vspace{0.2cm}

\ndt {\bf Some examples:}  \hfill \break
In~\eqref{SquashElliptic2} and~\eqref{unfoldedEq2}, we have given two equivalent expressions for the elliptic genus
of the squashed toric models. 
Below, we will illustrate these two formulas using two simple examples. We begin with the example of the squashed toric manifold~$\wt{\IC/\IZ}_2$.
This is described by a $U(1)$ gauge theory with two chiral multiplets of charges $+1$, $-1$, respectively. 
The theory has a~$U(1)$ flavor symmetry under which the two chiral multiplets carry charges~$F_{1}$, $F_{2}$, respectively.
The elliptic genus of this model is (with $b=F_{1}+F_{2}$) 
\be
\begin{split}
\chi_\text{ell}(\wt{\IC/\IZ}_{2}; \tau,z)  
&\= k \bigintsss_{\IC} \,    \frac{\dd^2 u'}{\t_{2}} \, 
e^{ - \frac{\pi k}{\tau_2} \bigl( u' +\frac{b\,z}{k} \bigr)
\bigl(\overline{u'}+\frac{b\,z}{k} \bigr)} \, 
\frac{\vth_1(\tau,-z+b\,u')}{\vth_1(\tau, b\,u')} \,.
\end{split}
\ee
We can absorb~$b$ in the definition of~$k$ by changing the integration variable from $u'\rightarrow \frac{u'}{b}$:
\begin{equation}\label{eq:C2Z2_structure}
\chi_\text{ell}(\wt{\IC/\IZ}_{2}; \tau,z)  
\= \wt k \bigintsss_{\IC} \,    \frac{\dd^2 u'}{\t_{2}} \, 
e^{ - \frac{\pi \wt k}{\tau_2} \bigl( u' +\frac{z}{\wt k} \bigr)
\bigl(\overline{u'}+\frac{z}{\wt k} \bigr)} \, 
\frac{\vth_1(\tau,-z+u')}{\vth_1(\tau, u')} \,,
\end{equation}
where $\wt k=\frac{k}{b^{2}}$\,.

The next example we consider is the case of the squashed toric model~$\wt A_{1}$. This model is described by 
a $U(1)$ gauge theory with three chiral multiplets of charges $1,-2,1$, respectively. It has $U(1)^{2}$ flavor symmetry under  
which the chiral multiplets have flavor charges~$F_{j}^{\ell},\, (\ell=1,2; \,j=1,2,3)$. 
The elliptic genus is 
\bea\label{Sq.A1Ex}
\chi_\text{ell}(\wt A_{1};\tau,z)&=&\frac{k_1k_2}{\tau^2_2}\int_{E_\tau} \dd^2u_1'\int_{E_\tau}\,
\dd^2u_2'\,\chi_\text{ell}(A_{1};\tau,z,u')\,\nn\\
&& \qquad \times \sum_{m_{1,2},w_{1,2}\in\mathbb{Z}}e^{2\pi i (b_1w_1+b_2w_2)z}
e^{-\frac{\pi k_1}{\tau_2}(w_1\tau+m_1+u'_1+\frac{b_1z}{k_1})(w_1\bar\tau+m_1+\bar{u'_1}+\frac{b_1z}{k_1})} \cr
&& \qquad \qquad \qquad \qquad 
\times \; e^{-\frac{\pi k_2}{\tau_2}(w_2\tau+m_2+u'_2+\frac{b_2z}{k_2})(w_2\bar\tau+m_2+\bar{u'_2} +\frac{b_2z}{k_2})}\,,
\eea
where $b_{\ell}=\sum_{j=1}^3F_{j}^{\ell}$ and $\chi_\text{ell}(A_{1};\tau,z,u')$ is the elliptic genus of the unsquashed~$A_{1}$ model, 
whose explicit expression is 
\be 
\begin{split}
& \chi_\text{ell}(A_{1};\tau,z,u') \= \\
& \qquad \frac{\vth_1(\tau,-z+v_{2}-v_{1})}{\vth_1(\tau,v_{2}-v_{1})}\frac{\vth_1(\tau,-z+2v_{1})}{\vth_1(\tau,2v_{1})}
+\frac{\vth_1(\tau,-z+v_{1}-v_{2})}{\vth_1(\tau,v_{1}-v_{2})}\frac{\vth_1(\tau,-z+2v_{2})}{\vth_1(\tau,2v_{2})}\,.
\end{split}
\ee
Here $2v_{1}=(2F_{1}+F_{2})\cdot u'$ and $2v_{2}=(2F_{3}+F_{2})\cdot u'$.
\vspace{0.2cm}

\subsection{Flavored Elliptic Genera of Squashed Toric Models}

As mentioned above, a squashed toric manifold of complex dimension $d$ is itself a toric 
manifold with~$U(1)^{d}$ toric symmetries that commute with the supersymmetry. 
We can therefore define a refined elliptic genus by introducing chemical potentials conjugate to~$d$ 
toric symmetry charges. This can be done by coupling the toric symmetry currents~$j^{\ell}_{\mu}$ 
to external gauge fields~$B^{\ell}_{\mu}$, $\ell=1,\cdots,d$. The functional integral will now depend on the holonomies
\be
\beta_{\ell} \= \oint_A B^{\ell}-\tau\oint_B B^{\ell} \,, \qquad \ell=1,\cdots,d \,,
\ee
in addition to the chemical potential for R-symmetry~$z$ which was defined in~\eqref{defz}.

We define the flavored elliptic genus, with~$q=e^{2\pi i \t}$, $\zeta_z = e^{2\pi i z}$, 
\be \label{defflavell}
\chi^\text{flav}_\text{ell}(\wt M_\text{tor};\t,z,\{\b\}) \=
\text{Tr}_{\CH_\text{RR}} \bigg[
(-1)^{F}q^{L_{0}}\, \bar q^{\bar L_{0}} \, \zeta_z^{J_{0}^{R}} \, \prod_{\ell=1}^{d}\,\exp \bigl(2 \pi i \b_\ell \int j_\ell \bigr)  \bigg] \,,
\ee
where~$\CH_{RR}$ is the Ramond-Ramond Hilbert space of the theory,~$L_{0}$ and~${\overline L_{0}}$ 
are the left- and right-moving Hamiltonians of the~$(2,2)$ algebra,~$J_{0}$ is the left-moving R-charge, and~$F$ 
is the fermion number operator. The toric charge~$\int j_\ell$ is an integral over a spatial slice in this Hamiltonian 
description. 
Note that the total charge $\int j_\ell$ is conserved, 
but the left and right moving pieces of the charge are not conserved individually. The chemical potential 
associated to corresponding charge $\b_{\ell}$, therefore, is a priori real. 
For the purpose of our calculation, we will keep it complex, but keep in mind that it is one real degree of freedom 
i.e.~$\overline{\b}$ cannot be varied independently of~$\b$.

We now compute this flavored elliptic genus using the technique of supersymmetric localization. 
In the GLSM picture, under the toric transformations all fields are 
neutral except $P_{\ell}$, on which they act as a shift of the imaginary part of $P_{\ell}$,
i.e.~$P_{\ell}\rightarrow P_{\ell}+i\alpha_{\ell}$ and $\bar P_{\ell}\rightarrow \bar P_{\ell}-i\alpha_{\ell}$, 
with $\alpha_{\ell}$ a real parameter. The toric symmetry currents of the squashed model are given by $j^{\ell}_{\m}=D_{\m}\text{Im}P_{\ell}$, 
where $D_{\m}$ is the gauge covariant derivative.

To calculate the flavoured elliptic genus we follow the steps described in~\cite{Gupta:2017bcp}. 
All the calculations go through as in~\cite{Gupta:2017bcp} but now there will be extra contributions to the part involving zero 
mode contributions of $\text{Im}P_{\ell}$. This is due to the fact that the covariant derivative 
of $P_{\ell}$ contains the background field $B_{\ell}$ in addition to the flavor symmetry gauge field $V^{'\ell}_{\m}$, i.e.,
\be
D_{\m}P_{\ell} \= \p_{\m}P_{\ell}+i(V^{'\ell}_{\m}+B^{\ell}_{\m})\,.
\ee
Thus including the contribution of the holonomy of $B^{\ell}_{\m}$, we find that the flavored elliptic genus is given by
\bea\label{Flav.SquashElliptic2}
\chi^\text{flav}_\text{ell}(\wt M_\text{tor}; \tau,z,\{\beta\})  \= 
\int_{E_\tau^{d}} \, \prod_{\ell=1}^d   \frac{\dd^2 u'_\ell}{\t_{2}} \,  
 H_\ell (\tau, z, u'_\ell,\beta_{\ell})  \; \chi_\text{ell}(M_\text{tor};\tau,z, u') \, ,
\eea
where we have defined the function 
\be 
 H_\ell (\tau, z, u,\beta) \= k_{\ell} \sum_{m,w \, \in \, \mathbb{Z}} \,
e^{2\pi i b_\ell w z - \frac{\pi k_\ell}{\tau_2} \bigl( w\tau+m+u+\beta+\frac{b_\ell z}{k_\ell} \bigr)
\bigl(w \bar{\tau}+m+\overline{u}+\overline\beta+\frac{b_\ell z}{k_\ell } \bigr)} \,.
\ee 

From the above expression one sees that the elliptic genus explicitly depends on $\overline{\b}_{\ell}$. 
We can easily compute its dependence on~$\overline{\b}_{\ell}$ as follows. 
Firstly one can rewrite the above integral~\eqref{Flav.SquashElliptic2} as follows. 
Using the elliptic transformations properties of~$H$ and~$\chi_\text{ell}(M_\text{tor})$, 
\be
 H_\ell (\tau, z, u+\l\t+\m,\beta+\eta\t+\n) \=e^{-2\pi ib_{\ell}(\l+\eta)z} H_\ell (\tau, z, u,\beta)\,,
\ee
\be\label{elliptictransf2}
\chi_\text{ell}(M_{\text{tor}};\tau,z,u'^{\ell}+\lambda^{\ell}\tau+\mu^{\ell})
\=e^{2\pi iz\sum_{\ell=1}^{d}b_{\ell}\lambda^{\ell}}\chi_\text{ell}(M_{\text{tor}};\tau,z,u')\,,
\ee
one can write \eqref{Flav.SquashElliptic2} as
\bea\label{unfoldedEq1}
\chi^\text{flav}_\text{ell}(\wt M_\text{tor}; \tau,z,\{\beta\})  &\=& 
\bigintsss_{\IC^{d}} \, \prod_{\ell=1}^d   \frac{\dd^2 u'^{\ell}}{\t_{2}} \, k_{\ell} \,
e^{ - \frac{\pi k_\ell}{\tau_2} \bigl( u'^{\ell}+\beta_{\ell}+\frac{b_\ell z}{k_\ell} \bigr)
\bigl(\overline{u'^{\ell}}+\overline{\beta_{\ell}}+\frac{b_\ell z}{k_\ell } \bigr)} \, 
\chi_\text{ell}(M_\text{tor};\tau,z, u') \,,\nn\\
&\=&
\bigintsss_{\IC^{d}} \, \prod_{\ell=1}^d   \frac{\dd^2 u'^{\ell}}{\t_{2}} \, k_{\ell} \,
e^{ - \frac{\pi k_\ell}{\tau_2} \bigl( u'^{\ell}+\frac{b_\ell z}{k_\ell} \bigr)
\bigl(\overline{u'^{\ell}}+\frac{b_\ell z}{k_\ell } \bigr)} \, 
\chi_\text{ell}(M_\text{tor};\tau,z, u'-\beta)\,,\nn\\
&\=&\int_{E_\tau^{d}} \, \prod_{\ell=1}^d   \frac{\dd^2 u'_\ell}{\t_{2}} \,  
 \wt H_\ell (\tau, z, u'_\ell)  \; \chi_\text{ell}(M_\text{tor};\tau,z, u'-\beta) \, ,
\eea
where $\wt H_\ell (\tau, z, u)$ is defined in~\eqref{def.Hell}.

Now in the expression~\eqref{unfoldedEq1} the chemical potentials $\beta_\ell$ appear only 
in $\chi_\text{ell}(M_\text{tor})$, and this dependence is meromorphic. 
Therefore if we hit it with a~$\overline{\b}_{\ell}$-derivative, integral receives contributions only
from the poles of~$\chi_\text{ell}(M_\text{tor})$ and in fact reduces to a residue calculation~\cite{Harvey:2013mda}.

\vspace{0.4cm}

\ndt {\bf Modular and elliptic properties:}  \hfill \break
Using the same technique as in the unflavored case, we can show that 
under the modular transformation, the flavored elliptic genus transforms as 
\be
\begin{split}
\chi^\text{flav}_\text{ell}(\wt M_\text{tor}; \tau+1,\bm z) & \=\chi^\text{flav}_\text{ell}(\wt M_\text{tor}; \tau,\bm z) \,,\\
\chi^\text{flav}_\text{ell}(\wt M_\text{tor}; -\frac{1}{\tau},\frac{\bm z}{\t})
&\=e^{\frac{2\pi i}{\t}{\bm z}^{T}\,\widehat M\,\bm z}\,\chi^\text{flav}_\text{ell}(\wt M_\text{tor}; \tau,\bm z)  \,.
\end{split}
\ee
Here  ${\bm z} =(z,\b_{1},....,\b_{d})$ and $\widehat M$ is a matrix with the following entries:
\be
\widehat M_{00}\=\frac{d}{2}+\sum_{\ell=1}^{d}\frac{b^{2}_{\ell}}{k_{\ell}}\,,\qquad 
\widehat M_{0i}\=\widehat M_{i0}\=\frac{b_{i}}{2},\quad \widehat M_{ij}\=0,\qquad i\=1,...,d \,.
\ee
To derive the elliptic property of the flavored elliptic genus we use the first line of \eqref{unfoldedEq1}. 
Using the elliptic properties of the $\chi_\text{ell}(M_\text{tor};\tau,z, u') $ and assuming that $\{\frac{b_{\ell}}{k_{\ell}}\}$,
$\ell = 1, \cdots, d$ are integers, one finds that
\be
\chi^\text{flav}_\text{ell}\bigl(\wt M_\text{tor}; \tau,{\bm z}  + {\bm \l} \bigr) 
\=\exp\bigl(-2\pi i({\bm \l}^T \,\widehat M \,{\bm \l} \,\t+2{\bm \l}^{T} \, \widehat M\,{\bm z}) \bigr) \,
\chi^\text{flav}_\text{ell}\bigl(\wt M_\text{tor}; \tau,{\bm  z} \bigr) \,,
\ee
where~$\bm \l\=(\l,\eta_{1},....,\eta_{d}) \in \IZ^{d+1}$\,.
These are precisely the transformation properties of a Jacobi form of~$d+1$ elliptic variables with weight zero and index~$\widehat M$.

\subsection{The General Structure for Flavored Elliptic Genera}\label{sec:general_structure}

Now, we would like to present the general structure for the integral form of the elliptic genus~\eqref{Flav.SquashElliptic2} 
which will be used in the Section~\ref{sec:generalcase} to evaluate it explicitly. We find that quite generically the 
flavored elliptic genus of a squashed toric model can be expressed as certain linear combinations of an integral of the following form 
\begin{equation}\label{GeneralintgForm.1}
\intt_{\IC^N} \frac{\ddd^{2N} \zzz'}{\tim^N} \,
\prod_{j=1}^N  \lb  \wt k_j \, 
\frac{\thone(\t,-z+\mmm^{(j)T} \zzz')}{\thone(\t,\mmm^{(j)T} \zzz')} \, 
e^{-\frac{\pi \wt k_j}{\tim} \bigl( \zzz'_j +\wt \b_{j} + \frac{z}{\wt k_j} \bigr) 
\bigl(\bar {\zzz'_j }+\bar{\wt \b}_{j} + \frac{z}{\wt k_j} \bigr) }
\rb ,
\end{equation}
where $\mmm^{(j)}$ for $j=1,...,N$ is an $N$-component column vector whose entries are functions of the charges $Q^{a}_{i}$ 
and $F_{i}^{\ell}$ only, and if we construct the~$N\times N$ matrix~$\CM$ whose columns are $\mmm^{(j)}$, then it satisfies
\begin{equation}\label{GeneralintgForm.2}
\CM \, \qr \= \qr,
\where \qr \coloneqq \lp 1, \ldots,  1\rp^T \in \IR^{N \times 1}\,.
\end{equation}
As we will explain below, the above equation is a consequence of the fact that~$\sum_{i=1}^{n}Q^{a}_{i}=0$.

Let us begin with examples. 
The elliptic genus of the flavored $\wt{\IC/\IZ}_2$ manifestly has the above structure, as can be seen from Equation \eqref{eq:C2Z2_structure}.
To see that this structure also holds in the case of flavored squashed~$A_{1}$ model, we unfold the integral in~\eqref{Sq.A1Ex} to $\IC^{2}$. 
Considering only the first term in~$\chi_\text{ell}(A_{1};\tau,z,u')$ (one can draw the same conclusion for the second term), we get
\begin{align}
&\chi^\text{flav}_\text{ell}(\wt A_{1};\tau,z,\{\b\})^{(1)}\=\frac{k_1k_2}{\tau^2_2}\int_{\IC^{2}}\dd^2u_1'\,\dd^2u_2'\,e^{-\frac{\pi k_1}{\tau_2}
\bigl(u'_1+\b_{1}+\frac{b_1z}{k_1}\bigr)  \bigl(\bar{u'_1} +\bar{\b}_{1}+\frac{b_1z}{k_1}\bigr) }\,
\nn\\
&\qquad \quad \times
e^{-\frac{\pi k_2}{\tau_2}\bigl(u'_2+\b_{2}+\frac{b_2z}{k_2}\bigr) \bigl(\bar{u'_2} +\bar{\b}_{2}+\frac{b_2z}{k_2}\bigr) }\,
\frac{\vth_1(\tau,-z+v_{2}-v_{1})}{\vth_1(\tau,v_{2}-v_{1})}\frac{\vth_1(\tau,-z+2v_{1})}{\vth_1(\tau,2v_{1})} \,.
\end{align}
Changing the integration variables as $u_{1}'=\frac{1}{b_{1}}\zzz_{1}'$ and $u_{2}'=\frac{1}{b_{2}}\zzz'_{2}$, we can rewrite the above integral as
\be
\chi^\text{flav}_\text{ell}(\wt A_{1};\tau,z,\{\b\})^{(1)}\=\int\limits_{\IC^{2}}  \frac{\dd^2\zzz_1'}{\tau_2} \,\frac{\dd^2\wt \zzz_2'}{\tau_2}
\prod_{j=1}^{2}\frac{\wt k_j \, \vth_1\bigl(\tau,-z+\m^{(j)T}\zzz'\bigr)}{\vth_1(\tau,\m^{(j)T}\zzz')}\,
e^{-\frac{\pi \wt k_j}{\tau_2}\bigl(\zzz'_j+\wt \b_{j}+\frac{z}{\wt k_j}\bigr) (\bar{\zzz'_j} +\bar {\wt \b}_{j}+\frac{z}{\wt k_j}\bigr) }\,.
\ee
Here
\be
\zzz'\= \biggl( \, \begin{matrix}\zzz_{1}' \\ \zzz_{2}'\end{matrix} \,\biggr)\,,\quad 
\wt\b_{j}\=b_{j}\b_{j}\,,\quad 
\m^{(1)}\=\begin{pmatrix}\frac{F^{1}_{3}-F^{1}_{1}}{b_{1}} \\ \frac{F^{2}_{3}-F^{2}_{1}}{b_{2}}\end{pmatrix},\quad 
\m^{(2)}\=\begin{pmatrix}\frac{2F^{1}_{1}+F^{1}_{2}}{b_{1}} \\ \frac{2F^{2}_{1}+F^{2}_{2}}{b_{2}}\end{pmatrix},
\ee
and $\wt k_{j}=\frac{k_{j}}{b^{2}_{j}}$\,.
Furthermore, if we construct a matrix~$\mathcal M$ whose column vectors are~$\m^{(j)}$, then the matrix~$\mathcal M$ satisfies
\be
\mathcal M \, \biggl(\, \begin{matrix}1\\1\end{matrix} \, \biggr) \= \biggl(\, \begin{matrix}1\\1\end{matrix} \, \biggr)\,.
\ee
Thus we see that the flavored elliptic genus of $\wt A_{1}$ model can be brought to the form of  \eqref{GeneralintgForm.1} and \eqref{GeneralintgForm.2}.

To see that these equations are valid more generically, let us begin with the theory of $U(1)^{n-d}$ gauge theory coupled to $n$ chiral multiplets. 
Then, for generic values of flavor charges $\{F_{i}^{\ell}\}$ and complex potentials $\{u'_{\ell}\}$, and for the non-degenerate 
situation\footnote{In the paper we will focus only on the case of non-degenerate poles.} 
(i.e.~the number of chiral multiplets becoming massless at a given point $u_*$ in $\IC^{n-d}$ 
is $(n-d)$), the Jeffrey-Kirwan residue of $Z_{\text{1-loop}}(\tau,z,u,u')$ at $u=u_*$ in \eqref{JKresidue1} is a linear combination of terms of the following form:\footnote{Typically, the residues are evaluated at the the zeros of $Q_{i}\cdot u+F_{ i}\cdot u'=0 \ \mathrm{mod}\,\mathbb{Z}\tau+\mathbb{Z}$ 
which are of the form $u=u_{*}(u') + a\tau+b \ \mathrm{mod}\,\mathbb{Z}\tau+\mathbb{Z}$ where $a,b\in \mathbb{Q}$. In the present case we take both $a$ and $b$ to be zero.}
\bea
\prod_{{\hat i} \in [n] \setminus S} \frac{\vth_1(\tau,-z+Q_{\hat i}\cdot u+F_{\hat i}\cdot u')}{\vth_1(\tau,Q_{\hat i}\cdot u+F_{\hat i}\cdot u')}
\Big|_{u=u_*}&\=&
\prod_{{\hat i} \in [n] \setminus S}\frac{\vth_1(\tau,-z+ Q^{a}_{\hat i} G_{a}^{\ell}u^{'\ell}
+F_{\hat i}\cdot u')}{\vth_1(\tau, Q^{a}_{\hat i} G_{a}^{\ell}u^{'\ell}+F_{\hat i}\cdot u')}\,\nn\\
&\= \;&\prod_{j=1}^{d}\frac{\vth_1(\tau,-z+\m^{(j) T}\zzz')}{\vth_1(\tau, \m^{(j) T}\zzz')} \,.
\eea
Here~$u^{'\ell}=\frac{1}{b_{\ell}}\zzz'_{\ell}$, $[n] = \{ 1, \ldots, n \}$, and the $n-d$ element subset $S \subset [n]$ and the $(n-d)\times d$ matrix $G$ determines the poles in~$\IC^{n-d}$ i.e.
\be
u_*=G\, u'\,,
\ee
and (for generic values of $u'$)
\be
\sum_{a=1}^{n-d}Q^{a}_{i} \, G^{\ell}_{a}+F^{\ell}_{i}=0\,, \qquad \forall\,\, \ell=1,..,d \,,
\ee
where $Q_{i}$ and $F_{i}$, $i \in S$ are charges for chiral multiplets which are massless at $u=u_*$. Thus, we see that 
\bea
\sum_{j=1}^{d}\m^{(j)}_{\ell}&\=&\frac{1}{b_{\ell}}\sum_{\hat i \in [n] \setminus S}(Q^{a}_{\hat i} \, G_{a}^{\ell}+F^{\ell}_{\hat i})\,\nn\\
&\=&\frac{1}{b_{\ell}}\sum_{\hat i \in [n] \setminus S} (Q^{a}_{\hat i} \, G_{a}^{\ell}+F^{\ell}_{\hat i})+
\frac{1}{b_{\ell}}\sum_{i \in S}(Q^{a}_{i} \, G^{\ell}_{a}+F^{\ell}_{i})
\=\frac{1}{b_{\ell}}\sum^{n}_{ i=1}F^{\ell}_{i}
\=1\,.
\eea
In the last line, we have used the fact that~$\sum_{i=1}^{n}Q^{a}_{i}=0$. 
The above equation is nothing but the observation~\eqref{GeneralintgForm.2}, i.e.
\begin{equation}
\CM \, \qr \= \qr, \quad
\text{with} \quad \qr \= \lp 1, \ldots,  1\rp^T \in \IR^{N \times 1}\,.
\end{equation}

We also see from the above discussion that the~$b_{\ell}$ dependence completely disappears from the integrand 
and its effect is to redefine the variable~$k_j$ to~$\wt k_j=k_j/b_j^2$ and $\beta_j$ to $\wt{\beta}_j = b_j \beta_j$. 
In the rest of the paper, we shall rename~$\wt k_j \to k_j$, as the properties of the elliptic genus, which we will describe 
in the following sections, do not depend on the fact that it arose as a ratio.

\subsection{A Simple Example and the First Appearance of Indefinite Theta Series \label{CigAL}}

In the simplest case of~$d=1$, the vacuum manifold~\eqref{sqVac} is the squashed version of the~$\IZ_2$ quotient of the complex plane, 
denoted by~$\wt{\IC/\IZ}_2$. We find according to Equation~\eqref{unfoldedEq1} that the flavored elliptic genus of this model is 
\be\label{chiflavcig}
\begin{split}
\chi^\text{flav}_\text{ell}(\wt{\IC/\IZ}_{2}; \tau,z,\beta)  
&\= k \bigintsss_{\IC} \,    \frac{\dd^2 u'}{\t_{2}} \, 
e^{ - \frac{\pi k}{\tau_2} \bigl( u'+b \beta +\frac{z}{k} \bigr)
\bigl(\overline{u'}+b\overline{\beta}+\frac{z}{k} \bigr)} \, 
\frac{\vth_1(\tau,-z+u')}{\vth_1(\tau, u')} \,.
\end{split}
\ee
The~$\b=0$ elliptic genus of this model was computed in~\cite{Gupta:2017bcp} and it was noticed that it coincides with 
the elliptic genus of the~$SL(2,\IR)_{k}/U(1)$ coset theory, based on which it was conjectured that~$\wt{\IC/\IZ}_{2}$ 
flows to the cigar. This can be now further corroborated by the above flavored computation---indeed it agrees with the 
corresponding expression for the cigar computed in~\cite{Ashok:2014nua} (with the replacement~$\beta \to -b \beta$).

The integral~\eqref{chiflavcig} was explicitly evaluated in~\cite{Ashok:2014nua}
and it was shown that the answer is related to the three-variable Appell-Lerch sum 
\be\label{AppellFunc}
A_{1,k}(\t,u,v)\; := \;\zeta_u^{k} \, \sum_{n\in\mathbb Z}\frac{q^{kn(n+1)}\, \zeta_v^{n}}{1-\zeta_u \, q^{n}}\,,\qquad \zeta_u\=e^{2\pi iu} \,, \qquad \zeta_v\=e^{2\pi iv} \,,
\ee
as we now explain. Firstly, the function~$A_{1,k}(\t,u,v)$ does not have good modular transformation properties, but it can be 
completed\footnote{The precise relation of the Appell-Lerch sum to mock modular forms and mock Jacobi forms 
has been spelled out in~\cite{Zwegers:2008zna, Dabholkar:2012nd,Zwegers:2011}.} 
to~$\wh A_{1,k}(\tau,u,v)$ which has the modular and elliptic transformation properties of a Jacobi form of weight 
one and index~$\smat{-k}{1/2}{1/2}{0}$. 
It was shown in~\cite{Ashok:2014nua} that 
the flavored elliptic genus of the~$\IZ_k$ orbifold of the cigar is related to this completion as follows, 
\be \label{cigorbAL}
\chi^\text{flav}_\text{ell} \Bigl(\frac{SL(2,\IR)_k}{U(1)} \Big/ \IZ_k;\tau,z,\b \bigr) \= \frac{i\theta_{1}(\t,z)}{\eta(\t)^{3}} \, 
\wh A_{1,k}\bigl( \tau,\frac{z}{k}, 2z+kb\beta \bigr) \,.
\ee
Equivalently by mirror symmetry~\cite{Hori:2001ax} the left-hand side of this equation can be read as the 
elliptic genus of the~$\CN=2$ Liouville theory with coupling constant~$1/k$.
The statement~\eqref{cigorbAL} can be inverted: the flavored elliptic genus of the cigar theory is given by the Atkin-Lehner 
operator~$W_k$~\cite{Skoruppa1988} acting on~$A_{1,k}(\tau,u,v)$. 

Now we note that the Appell-Lerch sum has the following Fourier expansion,
\be
A_{1,k}(\t,u,v) \= \half \sum_{n,m\in\IZ} \Bigl( \sgn(m+\varepsilon) - \sgn\bigl(-n-\tfrac{\Im(u)}{\Im(\t)} \bigr) \Bigr) \, 
q^{k n (n+1) + nm} \, \zeta_u^{m+k} \, \zeta_v^n \,,
\ee
where~$0<\varepsilon <1$ is arbitrary\footnote{The answer of course is independent of the choice of~$\varepsilon$.} 
and $\sgn(x) = +1$ for~$x>0$ and~$-1$ for~$x<0$.
This can be recognized as an indefinite theta function of a~$(1,1)$ lattice, as we will elaborate on in the following section. 
As Zwegers has explained~\cite{Zwegers:2008zna}, one can add a non-holomorphic correction term to it in order to 
obtain a completed function that transforms 
as a true modular object, thus giving an explicit construction of mock modular forms.  
As we will see, the elliptic genus of the squashed models in all the cases are a generalization of this observation, i.e., they 
are modular completions of some indefinite theta functions associated with an~$(n,n)$ lattice, which are constructions of 
higher-depth mock modular forms. 
Towards this end we turn to a review of the notion of mock modular forms at higher depth and indefinite theta functions for 
arbitrary lattices.


\section{Mock Modular Forms of Higher Depth \label{sec:MMF}}
\subsection{A Quick Review of Modular Forms and Theta Functions}\label{sec:modular_review}
We start the mathematical part of our discussion with a quick review of modular forms. The modular group $\mathrm{SL} (2, \IZ)$ is the group of integral $2\times2$ matrices with unit determinant. It is generated by the two elements
\begin{equation}
T \= \mat{1 & 1 \\ 0 & 1} 
\andd
S \= \mat{0 & 1 \\ -1 & 0}.
\end{equation} 
Its role as the group of large diffeomorphisms for two dimensional tori explains its appearance in many applications of string theory and two dimensional quantum field theory. 

Modular forms are complex valued functions defined on the upper half-plane, $\mathbb{H} = \{ z \in \IC : \ \mathrm{Im}(z) >0 \}$ that are symmetric under the modular group. More specifically, a (holomorphic) modular form of weight $k$ is a holomorphic function $f : \mathbb{H} \to \mathbb{C}$ that satisfies the following two conditions.
\begin{itemize}
\item Covariance under modular transformations:
\begin{equation}\label{eq:modularformtransf}
f \lp \frac{a \t + b}{c \t +d} \rp \= (c \t  +d)^k f(\t),
\where \mat{a & b \\ c& d} \in \mathrm{SL} (2, \IZ).
\end{equation}
\item The Fourier expansion\footnote{The equation \eqref{eq:modularformtransf} in particular requires $f(\t+1) = f(\t)$ and hence $f$ has a Fourier series.} of $f$ satisfies a growth condition, namely,
\begin{equation}
f (\t) \= \sum_{n \geq 0} a_n  q^n,
\where q \;\coloneqq\; e^{2 \pi i \t}.
\end{equation}
\end{itemize}
Very importantly, the set of weight $k$ modular forms, $M_k$ form a finite dimensional vector space over $\mathbb{C}$. At this point there are various avenues for possible generalizations. One possibility is to relax the growth condition as $f (\t) = \sum_{n \geq N} a_n q^n$ for some constant $N<0$, giving the notion of weakly holomorphic modular forms.
Another possibility is to require the modular transformation property \eqref{eq:modularformtransf} not for the whole $\SL(2, \IZ)$ but for one of its subgroups. Yet another possibility is to generalize the transformation as
\begin{equation}
f_\mu \lp \frac{a \t + b}{c \t +d} \rp \= (c \t  +d)^k \chi(\gamma)_\mu^{\ \nu} f_\nu(\t),
\quad \mbox{for } \mu,\nu = 1, \ldots, N,
\end{equation}
where $\gamma= \pmat{a & b \\ c& d} \in \mathrm{SL} (2, \IZ)$
and $\chi(\gamma)_\mu^{\ \nu}$ is a (projective) representation of the modular group compatible with weight $k$ and called a multiplier system. Such objects are called vector-valued modular forms.

A natural source of modular forms is theta series attached to integral, positive-definite lattices. One way to characterize an integral lattice is to view it as the set $\Lambda \equiv \IZ^N$ where $N$ is the rank of the lattice and associate an inner product\footnote{From here on, we will treat any element of $\mathbb{Q}^N$,  $\mathbb{Z}^N$,  $\mathbb{R}^N$, or  $\mathbb{C}^N$ as a column vector. } $(n,m) \mapsto n \cdot m \coloneqq n^T Q \, m$ for $n,m \in \IZ^N$ and where $Q$ is an $N \times N$ symmetric integral matrix. If we also have $n^2 \coloneqq n^T Q \, n \in 2 \IZ$ for any $n \in  \IZ^N$ the lattice is said to be even.  We also define the dual lattice $\Lambda^* \subset \mathbb{Q}^N$ by
\begin{equation}
\Lambda^* \; \coloneqq \; \{ r \in \mathbb{Q}^N : r^T Q \, m \in \IZ \mbox{ for any } m \in \Lambda \}.
\end{equation}
We should note that $\Lambda \subset \Lambda^*$ for integral lattices .

If the inner product is positive definite we can define the theta series $\Theta^{Q,p}_\mu : \mathbb{H} \to \mathbb{C}$ where $\mu \in \Lambda^* / \Lambda$ and $p \in \Lambda$ is a characteristic vector\footnote{In an integral lattice, a vector $p \in \Lambda$ is called a characteristic vector if $n^2 + n \cdot p \in 2 \IZ$ for every $n \in \Lambda$. Note that if $p'$ is another characteristic vector then $\frac{p-p'}{2}\in \Lambda^*$.}
\begin{equation}
\Theta^{Q,p}_\mu ( \t) \;\coloneqq \; \sum_{n \in \Lambda + \mu + \frac{p}{2}} (-1)^{p \cdot n}\, q^{\frac{1}{2} n^2}.
\end{equation}
The positive definiteness of $Q$ is vital in ensuring the convergence of the series. We will drop $p$ as a superscript if the lattice is even and take $p=0$. 

It is also natural to introduce elliptic variables to a theta function, i.e., we extend the theta series to a holomorphic function $\Theta^Q_\mu : \mathbb{H} \times \mathbb{C}^N \to \mathbb{C}$ defined as
\begin{equation}\label{eq:pos_theta_def}
\Theta^{Q,p}_\mu ( \t, \zzz) \;\coloneqq \; 
\sum_{n \in \Lambda + \mu + \frac{p}{2}}
 q^{\frac{1}{2} n^2} 
e^{2 \pi i \lp \zzz + \frac{p}{2} \rp \cdot n}.
\end{equation}
This function is covariant under elliptic and modular transformations.
\begin{itemize}
\item Elliptic transformations:
\begin{equation}\label{eq:pos_theta_ell1}
\Theta^{Q,p}_\mu ( \t, \zzz + m) \= (-1)^{m \cdot p} \, \Theta^{Q,p}_\mu ( \t, \zzz)
\quad \mbox{for } m \in \Lambda,
\end{equation}
\begin{equation}\label{eq:pos_theta_ell2}
\Theta^{Q,p}_\mu ( \t, \zzz + m \t) \= (-1)^{m \cdot p} q^{-\frac{1}{2}m^2}
e^{- 2\pi i \zzz \cdot m} \, \Theta^{Q,p}_\mu ( \t, \zzz)
\quad \mbox{for } m \in \Lambda.
\end{equation}
\item Modular Transformations:
\begin{equation}\label{eq:pos_theta_mod1}
\Theta^{Q,p}_\mu ( \t+1, \zzz) \= e^{\pi i \lp \mu + \frac{p}{2} \rp^2} 
\, \Theta^{Q,p}_\mu ( \t, \zzz),
\end{equation}
\begin{equation}\label{eq:pos_theta_mod2}
\Theta^{Q,p}_\mu \lp -\frac{1}{\t}, \frac{\zzz}{\t} \rp \= 
\frac{(- i \t)^{N/2}}{\sqrt{| \Lambda^* / \Lambda |}}
e^{\pi i \zzz^2/\t} e^{-\pi i p^2/2}
\sum_{\nu \in \Lambda^* / \Lambda } 
e^{- 2 \pi i \mu \cdot \nu} \Theta^{Q,p}_\nu \lp \t, \zzz \rp.
\end{equation}
\end{itemize}
The equations \eqref{eq:pos_theta_ell1}, \eqref{eq:pos_theta_ell2}, and \eqref{eq:pos_theta_mod1} quickly follow from the definition \eqref{eq:pos_theta_def} and one can prove Equation \eqref{eq:pos_theta_mod2} using Poisson resummation. These transformation properties are the defining properties of (vector valued) Jacobi forms (with a lattice index). The theory of scalar valued Jacobi forms with scalar index is developed in \cite{EichlerZagier}, where a holomorphic function $\phi: \mathbb{H} \times \mathbb{C} \to \mathbb{C}$ (with suitable growth conditions) is called a Jacobi form of weight $k$ and index $m$ if it satisfies
\begin{equation}\label{eq:jacobiformtransf}
\phi \lp \frac{a \t + b}{c \t +d}, \frac{z}{c \t + d} \rp \= (c \t  +d)^k \,
e^{2\pi i m \frac{cz^2}{c \t + d}} \, \phi(\t,z),
\where \mat{a & b \\ c& d} \in \mathrm{SL} (2; \IZ),
\end{equation}
and
\begin{equation}\label{eq:jacobiformtransf2}
\phi(\t,z+ \nu \t + \mu) \= e^{- 2 \pi i m (\nu^2 \t + 2 \nu z) } \, \phi(\t,z),
\where \nu,\mu \in \IZ.
\end{equation}
In particular, the theta function of a rank $2k$, even, unimodular lattice $\Lambda$ whose elliptic variable is restricted as $\zzz = \l z$ for some $\l \in \Lambda$ with $\l^2 = 2m$ gives a scalar Jacobi form of weight $k$ and index $m$. Moreover, thanks to the elliptic transformation property \eqref{eq:jacobiformtransf2}, Jacobi forms can be decomposed as 
\begin{equation}\label{eq:theta_decomposition}
\phi(\t,z) \= \sum_{\ell \, (\mathrm{mod}\, 2m )} h_\ell (\t) \, \vartheta_{m,\ell} (\t,z),
\end{equation}
where level $m$ theta functions, $\vartheta_{m,\ell} (\t,z)$, are defined in equation \eqref{eq:levelmtheta_def} and $h_\ell (\t)$ form a vector valued modular form of weight $k-\frac{1}{2}$ and with multiplier system dual to that of $\vartheta_{m,\ell} (\t,z)$. The decomposition in \eqref{eq:theta_decomposition} is called a theta expansion and gives an isomorphism between Jacobi forms of weight $k$ and vector valued modular forms of weight $k-\frac{1}{2}$ (whose multiplier system is fixed by the index $m$ of the Jacobi form).

\subsection{Mock Modular Forms (of Higher Depth)}\label{sec:MockModFormDepth}
In many contexts in physics, holomorphic modular forms appear as supersymmetric partition functions. This fact puts quite powerful restrictions on supersymmetric partition functions because holomorphic modular forms of fixed weight form finite dimensional vector spaces over $\mathbb{C}$.  In the absence of a restriction such as supersymmetry, even if modular covariance is physically expected, that may only restrict the relevant physical quantities to be real analytic modular forms. The space of real analytic modular forms is however very large compared to holomorphic modular forms and there is not as much mathematical control over such functions. 

A situation between these two extremes happen when the physical theory is supersymmetric and yet has a continuum in its spectrum which contributes to supersymmetric partition functions. In certain special examples of such theories, the relevant functions are in a class of real analytic modular forms called mock modular forms \cite{Zwegers:2008zna, BruinierFunke} (see \cite{Zagier2009, Bringmann:2017book} for further details) or a slight generalization called mixed mock modular forms \cite{Dabholkar:2012nd}. Examples include elliptic genus of supersymmetric $\mathrm{SL}(2;\IR) / \mathrm{U}(1)$ model \cite{Troost:2010ud, Eguchi:2010cb, Ashok:2011cy}, certain partition functions of topologically twisted $\mathcal{N} = 2$ and $\mathcal{N} = 4$ super Yang-Mills theories \cite{Vafa:1994tf, Moore:1997pc, Korpas:2017qdo}, and counting function of quarter-BPS states in $\mathcal{N}=4$ string theory \cite{Dabholkar:2012nd}.
Mathematically, a mixed mock modular form, $h(\t)$, of weight $k$ is the holomorphic part of a real analytic modular form, $\wh{h} (\t, \tb)$, of weight $k$ whose $\tb$-derivative satisfies
\begin{equation}\label{eq:mock_defn}
\frac{\del}{\del \tb} \wh{h} (\t, \tb) \;\in\; \bigoplus_j \lp \tim^{r_j} M_{k+r_j} \otimes \bar{M_{2+r_j}} \rp,
\end{equation}
where $M_{k_j}$ is the space of weight $k_j$ holomorphic modular forms (possibly with a multiplier system and/or on a subgroup of $\mathrm{SL}(2;\IZ)$). When $\tim^k \frac{\del}{\del \tb} \wh{h} (\t, \tb) \in \bar{M_{2-k}}$, the function $h(\t)$ is called a pure mock modular form. 

As we will find out, the flavored supersymmetric partition functions of squashed toric sigma models lie in a more general space of mock modular forms with depth. These spaces are defined recursively as follows \cite{ZagierZwegers}. Denoting the space of mock modular forms of depth $d$ and weight $k$ by $M_k^d$ and the space of their completions by $\wh{M}_k^d$, we define $M_k^0$ and $\wh{M}_k^0$ to be $M_k$, so depth zero modular forms are holomorphic modular forms. Then a holomorphic function $h:\mathbb{H} \to \mathbb{C}$ is a mock modular form of depth $d$ and weight $k$ if it has modular completion $\wh{h}$ whose $\tb$-derivative satisfies
\begin{equation}\label{eq:mock_depth_defn}
\frac{\del}{\del \tb} \wh{h} (\t, \tb) \;\in\; \bigoplus_j \lp \tim^{r_j} \wh{M}^{d-1}_{k+r_j} \otimes \bar{M_{2+r_j}} \rp
\end{equation}
and if $d$ is the smallest number consistent with this property.
This definition identifies depth one mock modular forms with mixed mock modular forms.
We can also define mock Jacobi forms (of any depth) as functions mapped to vector-valued mock modular forms of appropriate depth under the isomorphism induced by a theta expansion \eqref{eq:theta_decomposition}.
Besides their relevance for supersymmetric partition functions of two dimensional field theories discussed here, these objects have appeared in various other contexts in physics and mathematics \cite{Alexandrov:2016tnf, Alexandrov:2017qhn, bringmann2016higher, Manschot:2017xcr, Bringmann:2018cov}.

\subsection{Indefinite Theta Series}\label{sec:IndefThetaSeries}
In Section~\ref{sec:modular_review}, we introduced theta series as a rich source of modular forms. However, the series as defined in equation \eqref{eq:pos_theta_def} does not immediately extend to lattices with indefinite signature. That is because of the exponential growth of the summand for lattice points along negative directions (or non-decaying behavior along null directions) which renders the series divergent. A convergent series can be obtained by restricting the sum (asymptotically) to lattice points along positive directions. One way to accomplish this is to restrict the series to lattice points in a positive rectangular cone, that is we construct the series
\begin{equation}\label{eq:hol_theta_def}
\Theta^{Q,p}_\mu \lp C,C';  \t \rp \;\coloneqq \; 
\sum_{n \in \Lambda + \mu + \frac{p}{2}}
\lb 
\frac{1}{2^{N_-}} \prod_{j=1}^{N_-} \lp
\sign{ c_j \cdot n} - \sign{c'_j \cdot n}
\rp
\rb 
(-1)^{p \cdot n}
 q^{\frac{1}{2} n^2} ,
\end{equation}
where the signature of the rank $N$ lattice $\Lambda$ is $(N_+, N_-)$ with $N_+$ and $N_-$ denoting the number of positive and negative eigenvalues of the matrix $Q$, respectively and 
\begin{equation}\label{eq:sgndef}
\sgn (x) \= 
\begin{cases}
1 \quad &\mbox{if } x > 0,  \\
0 \quad &\mbox{if } x = 0, \\
-1 \quad &\mbox{if } x < 0. 
\end{cases}
\end{equation}
Here, $c_j$ and $c'_j$ are vectors (forming the columns of $N \times N_-$ matrices $C$ and $C'$) that ensure convergence by projecting out negative directions.\footnote{Both for this function and for the modular completion we will discuss next, conditions for convergence when $N_- = 1$ are developed in \cite{Zwegers:2008zna}. A set of sufficient conditions for convergence with rectangular cones is given in \cite{Alexandrov:2016enp} for $N_- = 2$ case, which is then generalized in \cite{Nazaroglu:2016lmr} for arbitrary $N_-$. More general conditions for convergence (also allowing different types of positive cones) can be understood and studied through the geometric picture of \cite{kudla2016theta, funke2017theta}. Further discussions and generalization to tetrahedral cones for generic $N_-$ can be found in \cite{westerholt2016indefinite, ZagierZwegers}. }

Generically, however, the series in \eqref{eq:hol_theta_def} fails to be modular invariant. As found out by Zwegers \cite{Zwegers:2008zna}, 
in the case $N_- = 1$ (Lorentzian lattices) such theta functions have a non-holomorphic modular completion and give concrete examples 
of mock modular forms as defined in Section~\ref{sec:MockModFormDepth}.\footnote{For specific lattices and specific choices of cones, 
these indefinite theta series are modular and do not require a non-holomorphic completion. The construction in \cite{Zwegers:2008zna} 
also makes clear when this happens.} Non-holomorphic completions in the case $N_- >1$ has been  
developed \cite{Alexandrov:2016enp, kudla2016theta, westerholt2016indefinite, Nazaroglu:2016lmr, funke2017theta, ZagierZwegers} 
and yield explicit examples of mock modular forms of higher depth. 

Before discussing technical details, let us give some intuition for both the failure of modular invariance and for the associated non-holomorphic completions of \eqref{eq:hol_theta_def}. As alluded earlier, the main tool to prove the S-invariance \eqref{eq:pos_theta_mod2} of theta series is the Poisson resummation formula. In the case of definite signature lattices, the S-transformation law follows from the well known self-duality of Gaussian function. However, in forming a convergent indefinite theta series in \eqref{eq:hol_theta_def}, we restricted to a proper subset of lattice points and essentially imposed a hard wall in our setup. When we Fourier transform, this hard wall is no longer strictly localized. This is why the self-duality property is lost and the indefinite theta series generically fail to be modular invariant. If we smoothen the hard walls with Gaussian factors we can recover self-duality and hence S-invariance, provided the penetration of the lattice sum to the dangerous negative-definite regions is not strong enough to ruin convergence.

To be more concrete let us get back to technical details and rewrite equation \eqref{eq:hol_theta_def} as
\begin{equation}\label{eq:hol_theta_def2}
\Theta^{Q,p}_\mu \lp C,C';  \t \rp \;\coloneqq \; 
\sum_{n \in \Lambda + \mu + \frac{p}{2}}
\lb 
\frac{1}{2^{N_-}} \sum_{P \subseteq [ N_- ] } (-1)^{|P|} \, \sign{-C^P \cdot n}
\rb 
(-1)^{p \cdot n}
 q^{\frac{1}{2} n^2} ,
\end{equation}
where $[n] \coloneqq \{1, \ldots, n \}$,  $C^P$ is an $N \times N_-$ matrix whose columns are the elements of 
$\{ c_j  :  j \in P \} \,  \cup \, \{ c'_j  : j \in [N_-] \setminus P \}$. For matrices $\mathcal{F} = \lp f^{(1)} \cdots f^{(r)} \rp$ and $\mathcal{G} = \lp g^{(1)} \cdots g^{(s)} \rp$ whose columns are vectors in $\Lambda \otimes \IR$ we use the notation $\mathcal{F} \cdot \mathcal{G} \coloneqq \mathcal{F}^T Q \, \mathcal{G}$, i.e.~the $r\times s$ matrix whose $(i,j)^{\mathrm{th}}$ entry is $f^{(i)} \cdot g^{(j)}$. Moreover, 
for a column matrix $f = (f_1, \ldots, f_r)^T \in \IR^{r\times 1}$ we define
\begin{equation}
\sign{f} \;\coloneqq \; \prod_{j=1}^r \sign{f_j}.
\end{equation}
The modular completion is then accomplished by replacing
\begin{equation}\label{eq:gen_error_replacement}
\sign{-C^P \cdot n} \;\to\; E^Q \lp C^P; \sqrt{2 \tim} n \rp,
\end{equation}
where we introduce the boosted generalized error function $E^Q$ as follows (we will follow the conventions of \cite{Nazaroglu:2016lmr} except for inverting the signature of the bilinear form). For a set of $r$ vectors $f^{(1)}, \ldots, f^{(r)}$ that span a negative definite subspace with respect to the inner product defined by $Q$ we define the boosted generalized error function $E^Q \lp F; x \rp$ for $F = \lp f^{(1)} \cdots f^{(r)} \rp$
as:
\begin{equation}\label{eq:boosted_gen_err}
E^Q \lp F; x \rp \;\coloneqq \; E_r \lp \mathcal{B} \cdot F; \mathcal{B} \cdot x \rp
\end{equation}
where $\mathcal{B}$ is an orthonormal basis for the subspace spanned by $F$, i.e. we have $\mathcal{B}^T Q \mathcal{B} = - I_r$ and $F = - \mathcal{B} \mathcal{B}^T Q F$. The generalized error functions $E_r$ are defined as
\begin{equation}
E_r (\CMM; \uu) \;\coloneqq \; \intt_{\IR^r} \ddd^r \uu' \, e^{- \pi (\uu - \uu')^T (\uu - \uu')}
\sign{\CMM^T \uu'},
\end{equation}
where $\CMM \in \IR^{r \times r}$ is a nondegenerate matrix and $\uu \in \IR^{r}$. Although we chose a basis $\mathcal{B}$ in equation~\eqref{eq:boosted_gen_err}, the symmetries of $r$-tuple error functions ensure that the right hand side gives the same result for any possible basis choice. Properties of generalized error functions are reviewed in Appendix~\ref{sec:gen_err}.

At this point let us reintroduce the elliptic variable $\zzz$ and summarize the modular completion to indefinite theta series with rectangular cones. We define
\begin{align}\label{eq:indef_theta_def} 
&\wh{\Theta}^{Q,p}_\mu \lp C,C';  \t,\zzz \rp \; \coloneqq \;   \notag \\ &\qquad
\sum_{n \in \Lambda + \mu + \frac{p}{2}}
\lb 
\frac{1}{2^{N_-}} \sum_{P \subseteq [ N_- ] } (-1)^{|P|} \, 
E^Q \lp C^P; \sqrt{2 \tim} \lp n  + \frac{\im{\zzz}}{\tim} \rp \rp
\rb 
 q^{\frac{1}{2} n^2} 
e^{2 \pi i \lp \zzz + \frac{p}{2} \rp \cdot n}.
\end{align}
Provided the vectors in $C$ and $C'$ are chosen in a way that ensures convergence, this function is covariant under elliptic and modular transformations.
\begin{itemize}
\item Elliptic transformations:
\begin{equation}\label{eq:indef_theta_ell1}
\wh{\Theta}^{Q,p}_\mu (C,C';  \t, \zzz + m) \= (-1)^{m \cdot p} \, \wh{\Theta}^{Q,p}_\mu (C,C';  \t, \zzz)
\quad \mbox{for } m \in \Lambda,
\end{equation}
\begin{equation}\label{eq:indef_theta_ell2}
\wh{\Theta}^{Q,p}_\mu (C,C';  \t, \zzz + m \t) \= (-1)^{m \cdot p} q^{-\frac{1}{2}m^2}
e^{- 2\pi i \zzz \cdot m} \, \wh{\Theta}^{Q,p}_\mu (C,C';  \t, \zzz)
\quad \mbox{for } m \in \Lambda.
\end{equation}
\item Modular Transformations:
\begin{equation}\label{eq:indef_theta_mod1}
\wh{\Theta}^{Q,p}_\mu (C,C';  \t+1, \zzz) \= e^{\pi i \lp \mu + \frac{p}{2} \rp^2} 
\, \wh{\Theta}^{Q,p}_\mu (C,C';  \t, \zzz),
\end{equation}
\begin{equation}\label{eq:indef_theta_mod2}
\wh{\Theta}^{Q,p}_\mu \lp C,C';  -\frac{1}{\t}, \frac{\zzz}{\t} \rp \= 
\frac{i^{N_-} (- i \t)^{N/2}}{\sqrt{| \Lambda^* / \Lambda |}}
e^{\pi i \zzz^2/\t} e^{-\pi i p^2/2}
\sum_{\nu \in \Lambda^* / \Lambda } 
e^{- 2 \pi i \mu \cdot \nu}
 \wh{\Theta}^{Q,p}_\nu \lp C,C'; \t, \zzz \rp.
\end{equation}
\end{itemize}

In this work we will also need a slight generalization where some of the vectors in $C$ and $C'$ defining the cone are null (viewing them as limits of appropriate negative vectors). In Appendix \ref{sec:gen_err}, the definition of generalized error functions is extended to the case where some of the vectors in its first argument are null. It is still required that these vectors define a negative semi-definite subspace which in turn implies any null vector in this set is orthogonal to all the others in that generalized error function. We will assume that this is the case for each matrix $C^P$ in the definition \eqref{eq:indef_theta_def}. In particular, if 
$N$ is a matrix of null vectors which together with negative vectors in a matrix $F$ spans a negative semidefinite subspace, then we view the corresponding generalized error function as
\begin{equation}\label{eq:gen_error_null_neg}
E^Q \lp (N, F); x \rp \;\coloneqq\; E^Q \lp F; x \rp \, \sign{- N \cdot x}.
\end{equation}
We define indefinite theta functions in this case using \eqref{eq:gen_error_null_neg} with $x = \sqrt{2 \tim} \lp n  + \frac{\im{\zzz}}{\tim} \rp$. Since the definition of generalized error functions in this case follows from null limits of negative vectors, the associated indefinite theta functions satisfy all the elliptic and modular properties given in Equations \eqref{eq:indef_theta_ell1}--\eqref{eq:indef_theta_mod2}\footnote{The convergence of theta series is a delicate issue when null vectors are involved in the construction of its cone. The $N_- = 1$ case is treated in \cite{Zwegers:2008zna} and $N_- = 2$ case with rectangular cones is treated in \cite{Alexandrov:2016enp}. Although more generic results are not available in the literature, one can proceed case by case. In our work we will not be dealing with generic choices of cones, indeed all indefinite theta functions involving null vectors appearing in the following arise from generalized Appell sums \cite{Manschot:2014cca} (see Equation \eqref{eq:genAppell}) for which convergence is immediate.}.

\vspace{0.4cm}

We next get back to the definition of the holomorphic part of indefinite theta functions given in \eqref{eq:hol_theta_def} and extend it to the case in which elliptic variables are turned on.
If the elliptic variable is of the form $\zzz = \zzz' + a \tau + b$ with $a, b \in \mathbb{Q}^N$ fixed but $\zzz' \in \IC^N$ allowed to vary then by the holomorphic part $\Theta^{Q,p}_\mu \lp C,C';  \t , \zzz' + a \tau +b\rp$ we mean Equation \eqref{eq:indef_theta_def} with its factors as in \eqref{eq:gen_error_null_neg} replaced by 
\begin{align}
 E^Q &\lp F; \sqrt{2 \tim} \lp n  + a + \frac{\im{\zzz'}}{\tim} \rp \rp \,
  \sign{- N \cdot \lp n  + a + \frac{\im{\zzz'}}{\tim} \rp}   \notag \\
  &\qquad \qquad \qquad \to \;
  \sign{-F \cdot (n+a)} \,
  \sign{- N \cdot \lp n  + a+ \frac{\im{\zzz'}}{\tim} \rp} .
  \label{eq:hol_replacement}
\end{align}

As an example let us now quickly review the fact that the $\mu$-function defined in \cite{Zwegers:2008zna},
\begin{equation}\label{eq:Zwegers_mu_defn}
\mu ( u, v; \tau) \;\coloneqq\; \frac{\yy_u^{1/2}}{\thone (\tau, v)} \sum_{n \in \IZ}
\frac{(-1)^n \, q^{\frac{1}{2} (n^2+n)} \, \yy_v^n  }{1 - q^n \, \yy_u},
\end{equation}
as well as its modular completion
can be understood in terms of indefinite theta functions we have been discussing. For this purpose let us expand the denominator of the summand in \eqref{eq:Zwegers_mu_defn} and rewrite it as
\begin{equation}\label{eq:mu_indef1}
\thone (\tau, v) \, \mu ( u, v; \tau) \= 
\sum_{n, k \in \IZ} (-1)^n \, q^{\frac{1}{2} (n^2+2nk+n)} \, \yy_v^n \, \yy_u^{k+\frac{1}{2}} \, 
\frac{1}{2}
\lb 
\sign{k+\epsilon} - \sign{-n - \frac{\im{u}}{\tim}}
\rb,
\end{equation}
where we are free to choose $0<\epsilon<1$. In particular, choosing $\epsilon = \frac{1}{2}$ lets us write the right hand side of equation \eqref{eq:mu_indef1} as an indefinite theta function,
\begin{equation}
\thone (\tau, v) \, \mu ( u, v; \tau) \= \Theta^{Q,p} \lp c,c';  \t, \zzz \rp,
\end{equation}
where
\begin{equation}\label{eq:mu_indef_vectors}
Q \= \mat{1 & 1 \\ 1 & 0}, \quad
p \= \mat{0 \\ 1}, \quad
c \= \mat{1 \\ -1}, \quad
c' \= \mat{0 \\ -1}, \quad
\zzz \= \mat{u \\ v - u}.
\end{equation}
Here we remember the definition of the modular completion in \eqref{eq:indef_theta_def}, the replacement in \eqref{eq:hol_replacement} used to get its holomorphic part, and also note that the lattice is unimodular having only one conjugacy class in $\Lambda^* / \Lambda$. This form also immediately yields the modular completion, $\wh{\Theta}^{Q,p} \lp c,c';  \t, \zzz \rp$, to be
\begin{equation}\label{eq:mu_indef_completion}
\sum_{n, k \in \IZ} (-1)^n q^{\frac{1}{2} (n^2+2nk+n)} \yy_v^n \yy_u^{k+\frac{1}{2}}
\frac{1}{2}
\lb 
\erf\lp \sqrt{2 \pi \tim} \lp k+\frac{1}{2} + \frac{\im(v-u)}{\tim} \rp \rp + \sign{n+ \frac{\im{u}}{\tim}}
\rb.
\end{equation}
The difference, $\wh{\Theta}^{Q,p} \lp c,c';  \t, \zzz \rp - \Theta^{Q,p} \lp c,c';  \t, \zzz \rp$,  can be brought in to the form found in \cite{Zwegers:2008zna} by explicitly performing the sum over $n$. Before moving on, let us note a few points that will generalize to other setups relevant for this work.
\begin{itemize}
\item Let us first reemphasize the qualitative difference between negative vectors and null vectors forming the cone. As can be seen in Equation \eqref{eq:gen_error_null_neg}, to obtain a modular completion, sign functions associated with negative vectors (denoted by $F$ in \eqref{eq:gen_error_null_neg}) should be replaced by generalized error functions whereas sign functions associated with null vectors (denoted by $N$ in \eqref{eq:gen_error_null_neg}) remain unchanged in the completion. This is exemplified in $\mu$-function, for which the cone of the associated indefinite theta function is formed by one negative vector, $c$, and one null vector, $c'$ given in \eqref{eq:mu_indef_vectors}. We see that in the modular completion \eqref{eq:mu_indef_completion}, the sign function for $c$ is replaced with an error function whereas the sign function for $c'$ remains unchanged. The functions $\mathcal{A}_{1,m} (\tau,z)$ defined in \cite{Dabholkar:2012nd} are similarly holomorphic parts of signature $(1,1)$ indefinite theta functions whose cones are formed by one negative vector and one null vector.

\item Expanding on this difference for the case of negative vectors, the holomorphic part obtained by the replacement \eqref{eq:hol_replacement} does not have good elliptic transformations because of the dropped $\frac{\im ( v-u)}{\tim}$ factor. The elliptic transformation equations \eqref{eq:indef_theta_ell1} and \eqref{eq:indef_theta_ell2} are obeyed by $\Theta^{Q,p} \lp c,c';  \t, \zzz \rp$ only for those transformations that leave $v-u$ constant.
Instead of defining the holomorphic part by~\eqref{eq:hol_replacement}, one could make a different choice of splitting the completed function
into two pieces, in which the analog of the holomorphic part retains the $F \cdot \im{\zzz'}$ factor in the sign function on the right hand 
side of \eqref{eq:hol_replacement}. 
In the case of $\mu$-function this would replace the $\sign{k+\epsilon}$ term with $\mathrm{sgn}( k+\frac{1}{2}+\frac{\im(v-u)}{\tim})$.
This restores the elliptic transformation property at the expense of holomorphicity in~$u-v$. 
Moreover, it produces an apparent wall-crossing behavior in the $u-v$ variable---which is canceled by an equal contribution 
from the remaining terms in the modular completion.\footnote{This kind of restoration of smoothness in $u-v$, by canceling contributions in the holomorphic part and the remaining part, is also observed in physical setups of \cite{Alexandrov:2014wca, Pioline:2015wza, Murthy:2018bzs}.}

\item In the case of null vectors, on the other hand, the factor $N \cdot \im{\zzz'}$ that remains in the holomorphic part after the replacement in \eqref{eq:hol_replacement} (which is $-\im{u}$ in the example) plays an important role in performing the sum over $k$ in \eqref{eq:mu_indef1} to get the denominator in the summand of \eqref{eq:Zwegers_mu_defn}. Note that the absence of a quadratic term in $k$ in the exponents of \eqref{eq:mu_indef1} was the key point in this resummation which in turn was ensured by the nullity of $c'$. The appearance of $N \cdot \im{\zzz'}$ factors is also related to the fact that in such cases the `holomorphic part' is in fact meromorphic in the elliptic variable $\zzz'$.
\item When the elliptic variable $v-u$ is restricted to an element of $\mathbb{Q} \t + \mathbb{Q}$, the $\bar{\tau}$-derivative of  $\wh{\Theta}^{Q,p} \lp c,c';  \t, \zzz \rp$ is of the form \eqref{eq:mock_defn} with $k=1$ and only one term that has $r_1 = - \frac{1}{2}$. Here the relevant element of $\bar{M_{\frac{3}{2}}}$ is a weight $\frac{3}{2}$ unary theta function (stemming from lattice points in the span of negative vector $c$) and the relevant element of $M_{\frac{1}{2}}$ is a weight $\frac{1}{2}$ Jacobi form (a holomorphic theta function in the orthogonal complement of~$c$).
\end{itemize}

This construction works also for generalized Appell functions defined by \cite{Manschot:2014cca} and appearing in partition functions of topological $\mathcal{N}=4$ super Yang-Mills theory. Let $\Lambda$ be a rank-$m$ lattice with a positive definite quadratic form $Q$ ($k \cdot n \coloneqq k^T Q \, n$ for $k,n \in \Lambda$), and let $m_j \in \Lambda^*$, $j= 1, \ldots,n$. Then for $v \in \IC^m$ and $u \in \IC^n$, these generalized Appell functions are schematically defined as
\begin{equation}\label{eq:genAppell}
A_{Q,m_j} (\tau, u, v) \; \coloneqq \; e^{2 \pi  i \ell(u)} \sum_{k \in \Lambda} 
\frac{q^{\frac{1}{2} k^2 + R} e^{2 \pi i v \cdot k}}{\prod_{j=1}^n \lp  
1 - q^{m_j \cdot k} e^{2 \pi i u_j}
\rp}
\end{equation}
for an appropriate constant $R$ and linear function $\ell(u)$. As in our discussion of the $\mu$-function, the denominator of the summand can be expanded in a way similar to \eqref{eq:mu_indef1}. This identifies the generalized Appell function $A_{Q,m_j}$ as the holomorphic part of an indefinite theta series on a lattice with $N_+ = m$ and $N_- = n$. This fact in turn yields a modular completion as in Equation \eqref{eq:indef_theta_def}. This special class of indefinite theta functions will appear in Section \ref{sec:EllGenMMF} as a building block for the flavored elliptic genera of squashed toric sigma models discussed in this work.

Finally, let us discuss why indefinite theta functions introduced in this section give examples of higher depth mock modular forms as introduced in Section \ref{sec:MockModFormDepth}.
The $\bar{\tau}$-derivative of the modular completion \eqref{eq:indef_theta_def} can be computed using equation \eqref{eq:gen_error_radial_derivative}.\footnote{The $\bar{\tau}$-derivative acting on the sign functions in \eqref{eq:indef_theta_def} due to the null vectors in $C$ and $C'$ give vanishing contributions as long as the elliptic variable is so that the arguments of sign functions are not zero. } We will assume that the projection of the elliptic variable $\zzz$ to the subspace spanned by the timelike vectors in $C$ and $C'$ is zero (or is of the form $a \tau + b$ for fixed $a, b \in \IR^N$). In this case we use \eqref{eq:gen_error_radial_derivative} to find
\begin{equation}
\frac{\del}{\del \bar{\tau}} E^Q \lp F; \sqrt{2 \tim} x \rp = \frac{-i}{\sqrt{2 \tim}} \sum_{j=1}^r 
\frac{f^{(j)} \cdot x}{\sqrt{-f^{(j)} \cdot f^{(j)}}} 
e^{2 \pi \tim \lp f^{(j)} \cdot x \rp^2/f^{(j)} \cdot f^{(j)}}
E^Q \lp F_{[r]/\{j\} \perp \{j\} } ; \sqrt{2 \tim} x \rp.
\end{equation}
As $f^{(j)}$ runs over the negative vectors in $C$ and $C'$, the contribution from the \linebreak ${(f^{(j)} \cdot x) \, e^{2 \pi \tim \lp f^{(j)} \cdot x \rp^2/f^{(j)} \cdot f^{(j)}}  }$ term can be summed to the complex conjugate of a weight $\frac{3}{2}$ unary theta function whereas the contribution from the $E^Q \lp F_{[r]/\{j\} \perp \{j\} } ; \sqrt{2 \tim} x \rp$ term yields an indefinite theta function in the orthogonal complement of $f^{(j)}$ with signature $(N_+, N_- - 1)$. Therefore, comparing with \eqref{eq:mock_depth_defn}, indefinite theta series for signature $(N_+, N_-)$ lattices generically\footnote{Special symmetries of the lattice and the positive cone may lead to mock modular forms with lower depth.} yield depth $N_-$ mock modular forms\footnote{
Note that even without the restrictions we imposed on the elliptic variable, the modular completion of the theta function will transform like a Jacobi form. However, it will not be strictly speaking a mock Jacobi form (with some depth) because it will depend non-holomorphically on components of the elliptic variable that have nonzero inner product with positive vectors in $C$ and $C'$. On components that have nonzero inner product with null vectors in $C$ and $C'$, on the other hand, the theta function has a meromorphic dependence. This meromorphic dependence may then be cancelled if multiplied with a holomorphic Jacobi form that vanishes on the location of these poles. 
In our physical setup we will in fact encounter both behaviors. The elliptic genus is holomorphic in the elliptic variable $z$ for $R$-symmetry whereas it will depend non-holomorphically on the chemical potentials $\beta_j$ for toric flavor symmetries as can be deduced from the building block integral given in \eqref{GeneralintgForm.1}.}.


\section{Elliptic Genera of Squashed Toric Models as Mock Modular Forms of Higher Depth  \label{sec:EllGenMMF}}

Having reviewed the relevant mathematical background, in this section we show that the (flavored) elliptic genera of the squashed toric 
models~\cite{Gupta:2017bcp} discussed in Section~\ref{sec:STM} are built out of indefinite theta series of generic signature and yield
higher depth mock modular forms.  It is worth emphasizing that our physical expression derives the completions of these mock modular 
forms and hence include both the holomorphic part from discrete states and the non-holomorphic part due to continuum.

\subsection{A Warm-up Example: Squashed $\IC / \IZ_2$ Model}
Before studying the general case, let us review the squashed $\CZtwo$ model whose elliptic genus is known to be a mixed mock modular form (i.e. a depth one mock modular form). This will illustrate our computation in an easier setting, which in its early stages follows \cite{Ashok:2013zka}.
We start with the following expression for the  elliptic genus:
\begin{equation}
\chl \lp  \CZtwot ; \t, z \rp \= 
k \intt_{E_\t} \frac{\ddd^2 \zzz}{\tim} 
\frac{\thone (\t, -z + \zzz)}{\thone (\t, \zzz)}
\sum_{m ,\ww \in \IZ} e^{2 \pi i \ww z} \,
e^{-\frac{\pi k }{\tim} \lp \zzz + \ww \t + m + \frac{z}{k} \rp
\lp \zzb + \ww \tb + m + \frac{z}{k} \rp}.
\end{equation}
\textbullet \  Let us first perform Poisson resummation over $m$ in the above sum. Writing $\zzz = \zs \t + \zt$ where $\zs, \zt \in \IR$, we have 
\begin{align}
\sum_{m ,\ww \in \IZ} e^{2 \pi i \ww z} \,
&e^{-\frac{\pi k }{\tim} \lp (\ww +\zs) \t + m + \zt + \frac{z}{k} \rp
\lp  (\ww +\zs) \tb + m + \zt + \frac{z}{k}\rp} 
\notag \\
&\quad = \sqrt{\frac{\tim}{k}}
\sum_{n, \ww \in \IZ} 
\yy_z^{\ww + \frac{n}{k}} \, e^{2 \pi i \zt n} \,
q^{\frac{k}{4} \lp  \ww +\zs  + \frac{n}{k} \rp^2} \,
\qb^{\frac{k}{4} \lp  \ww +\zs - \frac{n}{k} \rp^2}.
\end{align}
So at this stage we find $\chl \lp  \CZtwot ; \t, z \rp$ to be
\begin{equation}  
\sqrt{k \tim} \intt_0^1 \ddd \zs \intt_0^1 \ddd \zt \;
\frac{\thone (\t, -z + \zs \t + \zt)}{\thone (\t, \zs \t + \zt)}
\sum_{n, \ww \in \IZ} 
\yy_z^{\ww + \frac{n}{k}} \, e^{2 \pi i \zt n} \,
q^{\frac{k}{4} \lp  \ww +\zs  + \frac{n}{k} \rp^2} \,
\qb^{\frac{k}{4} \lp  \ww +\zs - \frac{n}{k} \rp^2}.
\end{equation}
\textbullet \  Now let us expand both theta functions. For the theta function in the numerator we are simply going to use
\begin{equation}
\thone ( \t, z) \= i \sum_{m \in \IZ} (-1)^m q^{\frac{1}{2} \lp m + \frac{1}{2} \rp^2}
\yy_z^{m+\frac{1}{2}},
\end{equation}
and for the theta function in the denominator we are going to employ the identity (see for example \cite{Ashok:2013zka} for its proof)
\begin{equation}
\frac{1}{\thone (\t, z) } \= \frac{i}{\eta (\t)^3} \sum_{r \in \IZ} \yy_z^{r+\frac{1}{2}} \, 
S_r (\t)
\quad \mbox{ for } |q| < | \yy_z | < 1,
\end{equation}
where
\begin{equation}
S_r ( \t) \; \coloneqq \; \sum_{m=0}^\infty (-1)^m \, q^{m (m + 2r + 1)/2}.
\end{equation}
Note that for $\zzz = \zs \t + \zt$ we have $| \yy_\zzz |  =| q |^\zs$ and therefore the identity holds for $0< \zs < 1$.
Thus we now have $\chl \lp  \CZtwot ; \t, z \rp$ as
\begin{align}
\chl \lp  \CZtwot ; \t, z \rp \= &\sqrt{k \tim} \  \frac{(-1)}{\eta (\t)^3}  \intt_0^1 \ddd \zs \intt_0^1 \ddd \zt
\ \  \ \mathclap{\sum_{r,m,n, \ww \in \IZ}} \  \  (-1)^m \, q^{\frac{1}{2} \lp m + \frac{1}{2} \rp^2} \,
\yy_z^{-m-\frac{1}{2}} \, q^{\zs \lp m+\frac{1}{2} \rp}  
\, e^{2 \pi i \zt \lp m + \frac{1}{2} \rp}
\notag \\
& \times
S_r (\t) \, q^{\zs \lp r+\frac{1}{2} \rp}  \, e^{2 \pi i \zt \lp r + \frac{1}{2} \rp} \,
\yy_z^{\ww + \frac{n}{k}} \, e^{2 \pi i \zt n} \, q^{(\ww+\zs) n} \,
\lp q \qb \rp^{\frac{k}{4} \lp  \ww +\zs - \frac{n}{k} \rp^2}.
\end{align}
\textbullet \  Now we are perform the integral over $t$. This imposes $m+r+n+1 = 0$ using which we can perform the sum over $r$ and write $\chl \lp  \CZtwot ; \t, z \rp$ as
\begin{equation}
 - \frac{\sqrt{k \tim}}{\eta (\t)^3}  \intt_0^1 \ddd \zs 
\sum_{m,n, \ww \in \IZ} (-1)^m \, q^{\frac{1}{2} \lp m + \frac{1}{2} \rp^2} \,
\yy_z^{-m-\frac{1}{2}} \, S_{-m-n-1} (\t) \,  
\yy_z^{\ww + \frac{n}{k}} \, q^{n \ww} \,
\lp q \qb \rp^{\frac{k}{4} \lp  \ww +\zs - \frac{n}{k} \rp^2}.
\end{equation}
\textbullet \  Our next step is to perform the sum over $m$. We will do this using the following identity (again see \cite{Ashok:2013zka} for its proof):
\begin{equation}
\frac{i \, \thone (\t, z)}{1 - \yy_z^{-1} \, q^{-p}} \=
\sum_{m \in \IZ} (-1)^m \, q^{\frac{1}{2} \lp m + \frac{1}{2} \rp^2} \,
\yy_z^{-m-\frac{1}{2}} \, S_{-m-p-1} (\t),
\end{equation}
which holds for any $p \in \IZ$. This then gives us
\begin{equation}
\chl \lp  \CZtwot ; \t, z \rp \= 
\sqrt{k \tim} \  \lp \frac{-i \, \thone (\t, z)}{\eta (\t)^3} \rp 
\sum_{n, \ww \in \IZ} 
\frac{q^{n \ww} \, \yy_z^{\ww + \frac{n}{k}}}{1 - \yy_z^{-1} \, q^{-n}}
\intt_0^1 \ddd \zs  \lp q \qb \rp^{\frac{k}{4} \lp  \ww +\zs - \frac{n}{k} \rp^2}.
\end{equation}
\textbullet \  Finally, the integral over $s$ can be taken using error functions:
\begin{align}
\sqrt{k \tim} \intt_0^1 \ddd \zs  
&\lp q \qb \rp^{\frac{k}{4} \lp  \ww +\zs - \frac{n}{k} \rp^2}
\= 
\frac{\sqrt{k \tim}}{2} \intt_{-\infty}^\infty \ddd \zs  
\lb \sgn (s) - \sgn (s-1) \rb
e^{-\pi k \tim \lp  \ww +\zs - \frac{n}{k} \rp^2} \notag \\
&\quad = \frac{1}{2} 
\lb 
\erf \lp \sqrt{k \pi \tim} \lp  \frac{n}{k} - \ww  \rp \rp
- \erf \lp \sqrt{k \pi \tim} \lp  \frac{n}{k} - \ww -1 \rp \rp
\rb .
\end{align}
This yields the final result:
\begin{align}
\chl \lp  \CZtwot ; \t, z \rp \= 
&\frac{-i\thone (\t, z)}{\eta (\t)^3} 
\sum_{n, \ww \in \IZ} 
\frac{q^{n \ww} \, \yy_z^{\ww + \frac{n}{k}}}{1 - \yy_z^{-1} \, q^{-n}} \notag \\
& \quad \times \frac{1}{2} 
\lb 
\erf \lp \sqrt{k \pi \tim} \lp  \frac{n}{k} - \ww  \rp \rp
- \erf \lp \sqrt{k \pi \tim} \lp  \frac{n}{k} - \ww -1 \rp \rp
\rb.
\end{align}

\subsubsection*{The Holomorphic (Discrete) Part of the Squashed $\IC / \IZ_2$ Elliptic Genus}
As $\tim \to \infty$ we have $\erf \lp \sqrt{k \pi \tim} x \rp \to \sgn (x)$,
so the holomorphic part of $\chl \lp  \CZtwot ; \t, z \rp$ is given by
\begin{equation}
\chl^\hol \lp  \CZtwot ; \t, z \rp =
\frac{-i \, \thone (\t, z)}{\eta (\t)^3} 
\sum_{n, \ww \in \IZ} 
\frac{q^{n \ww} \, \yy_z^{\ww + \frac{n}{k}}}{1 - \yy_z^{-1} \, q^{-n}} \ 
\frac{1}{2} 
\lb 
\sgn \lp  \frac{n}{k} - \ww  \rp 
- \sgn \lp  \frac{n}{k} - \ww -1 \rp
\rb.
\end{equation}
Multiplying the numerator and the denominator of the summand by $\yy_z \, q^n$, shifting $w \to w -1$, plugging in $n = k \ww - \g$ where $\g \in \IZ$, and defining 
\begin{equation}
a_{\g, k} \;\coloneqq\;
\begin{cases}
1/2 \quad &\mbox{if } \g = 0,k, \\
1 \quad &\mbox{if } 0<\g<k, \\
0 \quad &\mbox{otherwise}, 
\end{cases}
\end{equation}
we can get this to a more familiar form\footnote{This is slightly different from the Equation 3.9 
of \cite{Ashok:2011cy} due to the ambiguity in defining the contribution of discrete states that 
are touching the continuum (equivalently, defining $\sgn(0)$ to be $-1$ would produce their result). 
Note that this ambiguity also appears in \cite{Ashok:2011cy} in the choice of integration contours 
due to the poles that are that are touching those contours. }
\begin{equation}
\chl^\hol \lp  \CZtwot ; \t, z \rp \=
\frac{i \, \thone (\t, z)}{\eta (\t)^3} 
\sum_{\g=0}^{k} a_{\g, k}
\sum_{\ww \in \IZ} 
\frac{q^{k \ww^2 - \g \ww}\,\yy_z^{2\ww  - \frac{\g}{k}}}{1 - \yy_z \, q^{k \ww - \g}}\,,
\end{equation}
which is nothing but the Atkin-Lehner operator~$W_k$~\cite{Skoruppa1988} acting on~$A_{1,k}(\tau,u,v)$,
as mentioned below Equation~\eqref{AppellFunc}.

\subsection{The General Case}\label{sec:generalcase}
Our expression for the flavored elliptic genera of squashed toric sigma models leads us to the following integral (see Section \ref{sec:general_structure} for details):
\begin{equation}\label{eq:fk_definition}
f_{\kkk} ( \CM; \t, z, \uu) \= 
\intt_{\IC^N} \frac{\ddd^{2N} \zzz'}{\tim^N} \,
\prod_{j=1}^N  \lb  k_j \, 
\frac{\thone(\t,-z+\mmm^{(j)T} \zzz')}{\thone(\t,\mmm^{(j)T} \zzz')} \, 
e^{-\frac{\pi k_j}{\tim} \lp \zzz'_j  + \wt{\beta}_j(\uu) + \frac{z}{k_j} \rp 
\lp \zzb'_j + \bar{\wt{\beta}_j(\uu)} + \frac{z}{k_j} \rp }
\rb ,
\end{equation}
where $\t \in \IH$, $z \in \IC$ and $\uu \in \IC^N$, $\kkk \coloneqq (k_1, \ldots, k_N)$ and $\zzz'$ is understood to be a column vector whose entries are $\zzz'_1, \ldots, \zzz'_N \in \IC$.
Here, $\CM$ is a real $N \times N$ non-degenerate matrix whose columns are $\mmm^{(j)}$ and which we assume to satisfy
\begin{equation}\label{eq:CMmatrixproperty}
\CM \qr \= \qr,
\where \qr \coloneqq \lp 1, \ldots,  1\rp^T \in \IR^{N \times 1}\,.
\end{equation}
Finally, $\wt{\beta}(\uu) \coloneqq \CM^{-T} \uu$ corresponds to the chemical potentials conjugate to the toric symmetries\footnote{ 
In this section, the variable $u$ is used to denote certain convenient linear combinations of chemical potentials $\beta$ for toric flavor symmetries and  should not be confused with chemical potentials for gauge symmetries defined in Equation \eqref{defuupr}.}.

Before attempting the integral, it will be convenient to introduce some notation.
\begin{equation}\label{eq:integralnotation1}
\CK \coloneqq \mathrm{diag}\lp \frac{1}{\sqrt{2k_1}} , \ldots, \frac{1}{\sqrt{2k_N}}\rp,
\quad
\CE \coloneqq \CK^{-1} \CM^{-T}, \quad 
g \coloneqq \CE^T \CE = \CM^{-1} \CK^{-2} \CM^{-T}.
\end{equation}
Using this notation and introducing a new variable $\zzz \coloneqq \CM^T \zzz'$ we can write
\begin{align}
\prod_{j=1}^N 
&e^{-\frac{\pi k_j}{\tim} \lp \zzz'_j + \beta_j(\uu) + \frac{z}{k_j} \rp \lp \zzb'_j +  \bar{\beta_j(\uu)}+  \frac{z}{k_j} \rp }  \notag \\
&\=
\exp \lp  -\frac{\pi}{2 \tim} 
\lp \zzb'^T \CK^{-1} + \bar{\beta(\uu)}{}^T \CK^{-1} + 2 z \qr^T \CK \rp
 \lp  \CK^{-1} \zzz' + \CK^{-1} \beta(\uu) + 2\CK \qr z \rp \rp \notag \\
&\=
\exp \lp  -\frac{\pi}{2 \tim} 
\lp \zzb^T \CE^T + \uub^T \CE^T + 2z \qr^T \CK \rp 
\lp  \CE \zzz  + \CE \uu + 2 \CK \qr z \rp \rp.
\end{align}
Noting that $\displaystyle\prod_{j=1}^N  k_j = \frac{1}{2^N (\det \CK)^2}$ and changing variables $\zzz' \mapsto \zzz = \CM^T \zzz'$ in the integral, we obtain
\begin{align}
&f_{\kkk} ( \CM; \t, z, \uu) \= \notag \\ &\quad
\frac{(\det \CE)^2}{2^N}
\intt_{\IC^N} \frac{\ddd^{2N} \zzz}{\tim^N} \,
\prod_{j=1}^N  \lb 
\frac{\thone(\t,-z+ \zzz_j)}{\thone(\t,\zzz_j)} \rb 
e^{-\frac{\pi}{2 \tim} 
\lp \zzb^T \CE^T + \uub^T \CE^T+  2z \qr^T \CK \rp
\lp  \CE \zzz + \CE \uu + 2 \CK \qr z  \rp} .
\end{align}
Using equation \eqref{eq:thetaell} we have
\begin{equation}
\frac{\thone(\t,-z+ \zzz_j + w_j \t + m_j)}{\thone(\t,\zzz_j + w_j \t + m_j)}
\= \yy_z^{w_j} \frac{\thone(\t,-z+ \zzz_j)}{\thone(\t,\zzz_j)},
\end{equation}
and hence 
\begin{align}
&f_{\kkk} ( \CM; \t, z, \uu) \= 
\frac{(\det \CE)^2}{2^N}
\intt_{E_\t^N} \frac{\ddd^{2N} \zzz}{\tim^N} \,
\prod_{j=1}^N  \lb 
\frac{\thone(\t,-z+ \zzz_j )}{\thone(\t,\zzz_j)} \rb 
\sum_{\mm, \ww \in \IZ^N} \yy_z^{\qr^T \ww}    \\
&\qquad \times
\exp\lp -\frac{\pi}{2 \tim}  
\lb (\zzb + \uub+ \tb \ww + \mm )^T \CE^T + 2z \qr^T \CK \rb
 \lb  \CE (\zzz + \uu + \ww \t + \mm)+ 2 \CK \qr z \rb \rp. \notag
\end{align}
Writing $\zzz = \zs \t + \zt$ where $\zs, \zt \in \lb 0, 1 \rb^N$ we have
\begin{align}
&f_{\kkk} ( \CM; \t, z, \uu) \= 
\frac{(\det \CE)^2}{2^N}\ 
\mathclap{\intt_{\lb 0, 1 \rb^N}} \ \ddd^{N} \zs 
\  \mathclap{\intt_{\lb 0, 1 \rb^N}} \  \ddd^{N}  \zt \,
\prod_{j=1}^N  \lb 
\frac{\thone(\t,-z + \zs_j \t + \zt_j)}{\thone(\t,  \zs_j \t + \zt_j)} \rb 
\sum_{\mm, \ww \in \IZ^N} \yy_z^{\qr^T \ww} \\
&\times
\exp\lp -\frac{\pi}{2 \tim}  
\lb ( \uub+ \tb (\ww + \zs) + \mm + \zt)^T \CE^T + 2z \qr^T \CK \rb
 \lb  \CE (\uu + (\ww + \zs) \t + \mm + \zt)+ 2 \CK \qr z \rb \rp. \notag
\end{align}
Using Poisson summation \eqref{eq:PoissonSum} for the sum over $\mm$ with 
\begin{equation}
A \= \frac{1}{2 \tim} \CE^T \CE 
\andd
B \= \frac{1}{2 i \tim} \CE^T \lb \CE ( \mathrm{Re}(u) + (\ww + \zs) \tre + \zt) + 2 \CK \qr z \rb,
\end{equation}
we obtain
\begin{align}
&f_{\kkk} ( \CM; \t, z, \uu) \= 
\lp \frac{\tim}{2} \rp^{\frac{N}{2}}
|\det \CE | \ 
\mathclap{\intt_{\lb 0, 1 \rb^N}} \ \ddd^{N} \zs 
\  \mathclap{\intt_{\lb 0, 1 \rb^N}} \  \ddd^{N}  \zt \,
\prod_{j=1}^N  \lb 
\frac{\thone(\t,-z + \zs_j \t + \zt_j)}{\thone(\t, \zs_j \t + \zt_j)} \rb 
\sum_{\nnn, \ww \in \IZ^N} \yy_z^{2 \qr^T g^{-1} p_L } 
 \notag \\
&\label{eq:integralMomentum}
\qquad 
\times
 e^{2 \pi i \nnn^T ( \zt + \frac{\mathrm{Im}(\uub \t)}{\tim} )}  \, 
q^{\frac{1}{2} p_L ( \zs + \frac{\mathrm{Im}(\uu)}{\tim} )^T g^{-1} p_L (\zs + \frac{\mathrm{Im}(\uu)}{\tim})} \, 
\qb^{\frac{1}{2} p_R(\zs + \frac{\mathrm{Im}(\uu)}{\tim})^T g^{-1} p_R (\zs + \frac{\mathrm{Im}(\uu)}{\tim})},
\end{align}
where we defined `twisted' left and right moving momenta that depends on $\nnn$ and $\ww$ implicitly as in toroidal compactifications of string theory (on a torus with metric $g = \CE^T \CE$):
\begin{equation}\label{eq:twistedmomenta}
p_L (\zs) \;\coloneqq\; \nnn + \frac{1}{2} g (\ww+\zs), \quad
p_R (\zs) \;\coloneqq\; \nnn - \frac{1}{2} g (\ww+\zs), \andd
p_L \;\coloneqq\; p_L (0).
\end{equation}
At this point we note that 
\begin{equation}
\frac{1}{2} p_L(\zs)^T g^{-1} p_L (\zs) - \frac{1}{2} p_R(\zs)^T g^{-1} p_R (\zs) \= 
\nnn^T (\ww + \zs).
\end{equation}
Therefore, the sum over $\nnn, \ww \in \IZ^N$ in \eqref{eq:integralMomentum} basically defines a Siegel-Narain theta function over the even, unimodular, signature $(N,N)$ lattice $U^{\oplus N}$, for which the projections to the positive and negative definite subspaces are defined by the matrix $\CE$. The lattice $U$ is defined through its Gram matrix $\mat{0&1\\1&0}$.

Next we are going to expand the theta functions in the denominator using the identity
\begin{equation}
\frac{1}{\thone (\t, z) } \= \frac{i}{\eta (\t)^3} \sum_{\rr \in \IZ} \yy_z^{\rr+\frac{1}{2}} \, 
S_\rr (\t)
\  \mbox{ for } |q| < | \yy_z | < 1,
\end{equation}
where $S_\rr ( \t) \coloneqq \sum_{m=0}^\infty (-1)^m \, q^{m (m + 2\rr + 1)/2}$.
Note that for $z = \zs \t + \zt$ with $0< \zs, \zt <1$ we have $| \yy_z | = |q|^\zs$ and the expansion is valid. We will also expand the theta functions in the numerators as
\begin{equation}
\thone (\t, z) 
\= i \sum_{\mm \in \IZ} (-1)^\mm \, q^{\frac{1}{2} \lp \mm + \frac{1}{2} \rp^2}
\yy_z^{\mm+\frac{1}{2}}. 
\end{equation}
This produces the following expression for $f_{\kkk} ( \CM; \t, z, \uu)$:
\begin{align}
&
\lp \frac{\tim}{2} \rp^{\frac{N}{2}}
|\det \CE | \lp \frac{-1}{\eta (\t)^3}  \rp^N
\ 
\mathclap{\intt_{\lb 0, 1 \rb^N}} \ \ddd^{N} \zs 
\  \mathclap{\intt_{\lb 0, 1 \rb^N}} \  \ddd^{N}  \zt \qquad 
\mathclap{\sum_{\rr,\mm,\nnn,\ww \in \IZ^N}} \ \ \ 
(-1)^{\sum_{j=1}^N \mm_j} q^{\frac{1}{2} \sum_{j=1}^N (\mm_j+\frac{1}{2})^2}
\yy_z^{- \sum_{j=1}^N  (\mm_j+\frac{1}{2})}
\notag
\\
&\quad \times 
q^{\sum_{j=1}^N \zs_j (\mm_j+\frac{1}{2})} e^{2 \pi i \sum_{j=1}^N \zt_j (\mm_j+\frac{1}{2})}
\lp \prod_{j=1}^N  
S_{\rr_j} (\t) \rp 
q^{\sum_{j=1}^N \zs_j (\rr_j+\frac{1}{2})} e^{2 \pi i \sum_{j=1}^N \zt_j (\rr_j+\frac{1}{2})}
 \notag \\
&\quad 
\times
\yy_z^{2 \qr^T g^{-1} p_L } 
 e^{2 \pi i \nnn^T ( \zt + \frac{\mathrm{Im}(\uub \t)}{\tim} )}  \, 
q^{\nnn^T (\ww + \zs+\frac{\mathrm{Im}(\uu)}{\tim}) } \, 
(q \qb)^{\frac{1}{2} p_R(\zs + \frac{\mathrm{Im}(\uu)}{\tim})^T g^{-1} p_R (\zs + \frac{\mathrm{Im}(\uu)}{\tim})}.
\end{align}

Performing the integrals over the variables $\zt_j$ imposes $\mm_j + \rr_j + \nnn_j +1 = 0$ using which we can take the sum over $\rr$ and get
\begin{align}
&f_{\kkk} ( \CM; \t, z, \uu) \= \notag \\
& \quad
\lp \frac{\tim}{2} \rp^{\frac{N}{2}}
|\det \CE | \lp \frac{-1}{\eta (\t)^3}  \rp^N
\sum_{\nnn,\ww \in \IZ^N}
\prod_{j=1}^N \lb \sum_{\mm_j \in \IZ} (-1)^{\mm_j} 
q^{\frac{1}{2} \lp \mm_j + \frac{1}{2} \rp^2}  
\yy_z^{- \lp \mm_j+\frac{1}{2}\rp} S_{-\mm_j - \nnn_j -1}(\t)\rb 
 \notag \\
&\qquad 
\times
\yy_z^{2 \qr^T g^{-1} p_L } 
 e^{2 \pi i \nnn^T \frac{\mathrm{Im}(\uub \t)}{\tim}}  \, 
q^{\nnn^T (\ww + \frac{\mathrm{Im}(\uu)}{\tim}) } \ \ 
\mathclap{\intt_{\lb 0, 1 \rb^N}} \  \ddd^{N} \zs  \, 
(q \qb)^{\frac{1}{2} p_R(\zs + \frac{\mathrm{Im}(\uu)}{\tim})^T g^{-1} p_R (\zs + \frac{\mathrm{Im}(\uu)}{\tim})}.
\end{align}
Next we are going to use the identity 
\begin{equation}
\frac{i \thone(\t,z)}{1 - \yy_z^{-1} q^{-p}} \= \sum_{\mm \in \IZ}
(-1)^{\mm} 
q^{\frac{1}{2} \lp \mm + \frac{1}{2} \rp^2}  
\yy_z^{- \lp \mm+\frac{1}{2}\rp} S_{-\mm - p -1}(\t),
\quad p \in \IZ,
\end{equation}
and also note that $\t \nnn^T  \frac{\mathrm{Im}(\uu)}{\tim} + \nnn^T \frac{\mathrm{Im}(\uub \t)}{\tim} = n^T u$. Moreover, we are going to insert 
\begin{equation}
\frac{1}{2^N} \sum_{c \in \{ 0,1\}^N} (-1)^{\sum_{j=1}^N c_j} \, \sign{\zs-c}
\end{equation}
in the integral,
this factor gives unity inside $\lb 0,1 \rb^N$ and is zero on the rest of $\IR^N$. So we can extend the $\zs_j$ integrals to the whole $\IR$ using this factor and get
\begin{align}
f_{\kkk} ( \CM; \t, z, \uu) \=&
\lp \frac{\tim}{2} \rp^{\frac{N}{2}}
|\det \CE | \lp \frac{-i \thone (\t, z)}{\eta (\t)^3}  \rp^N \sum_{\nnn,\ww \in \IZ^N}
\frac{q^{\nnn^T \ww}\, \yy_z^{2 \qr^T g^{-1} p_L }  \, e^{2\pi i \nnn^T \uu}}{\prod_{j=1}^N \lp 1 - \yy_z^{-1} q^{-\nnn_j} \rp}  \\
& \times
\frac{1}{2^N} \sum_{c \in \{ 0,1\}^N} (-1)^{\sum_{j=1}^N c_j} {\intt_{\IR^N}} \ddd^{N} \zs  \, \sign{\zs-c}
e^{-2  \pi \tim  p_R(\zs + \frac{\mathrm{Im}(\uu)}{\tim})^T g^{-1} p_R (\zs + \frac{\mathrm{Im}(\uu)}{\tim})}. \notag
\end{align}

Lastly, we will perform the integral over $\zs$:
\begin{equation}
\intt_{\IR^N} \ddd^{N} \zs  \, \sign{\zs}
e^{-2  \pi \tim  p_R(\zs +c+ \frac{\mathrm{Im}(\uu)}{\tim})^T g^{-1} p_R (\zs +c+ \frac{\mathrm{Im}(\uu)}{\tim})}.
\end{equation}
Looking at the definitions \eqref{eq:integralnotation1}, \eqref{eq:twistedmomenta} and 
defining $\l \coloneqq \sqrt{2 \tim} \lb \CE^{-T} \nnn - \frac{1}{2} \CE \lp \ww + c + \frac{\mathrm{Im}(\uu)}{\tim} \rp \rb$ and $\l' \coloneqq \sqrt{\frac{\tim}{2}} \CE \zs$, the integral becomes 
\begin{equation}
\lp \frac{2}{\tim} \rp^{\frac{N}{2}}
\frac{1}{|\det \CE |} \intt_{\IR^N} \ddd^{N} \l' \, \sign{\CE^{-1} \l'} 
e^{- \pi (\l - \l')^T (\l - \l')}.
\end{equation}
This integral then yields a generalized error function \cite{Alexandrov:2016enp, Nazaroglu:2016lmr} (see the definition in \eqref{eq:gen_err_defn}) and we get
\begin{equation}
\lp \frac{2}{\tim} \rp^{\frac{N}{2}}
\frac{1}{|\det \CE |} E_N ( \CE^{-T}; \l).
\end{equation}

So our final answer is 
\begin{align}
&f_{\kkk} ( \CM; \t, z, \uu) \=
\lp \frac{-i \thone (\t, z)}{\eta (\t)^3}  \rp^N \sum_{\nnn,\ww \in \IZ^N}
\frac{q^{\nnn^T \ww}\, \yy_z^{2 \qr^T g^{-1} p_L }  \, e^{2\pi i \nnn^T \uu}}{\prod_{j=1}^N \lp 1 - \yy_z^{-1} q^{-\nnn_j} \rp}  \notag \\
& \qquad \times
\frac{1}{2^N} \sum_{c \in \{ 0,1\}^N} (-1)^{\sum_{j=1}^N c_j} 
E_N \lp \CE^{-T};  \sqrt{2 \tim}  \lb \CE^{-T} \nnn - \frac{1}{2}\CE \lp \ww + c + \frac{\mathrm{Im}(\uu)}{\tim} \rp \rb \rp.
\label{eq:fk_integral}
\end{align} 
In fact, it is easy to see that the holomorphic limit of the series, obtained by taking the $\tim \to \infty$ limit in the generalized error functions, yields holomorphic parts of indefinite theta series. That is because in this limit, the error functions reduce to sign functions as in \eqref{eq:gen_err_asymp} and the second line of equation \eqref{eq:fk_integral} is nonzero only for finitely many $\ww$ for fixed $\nnn$ (or vice versa). In the example of squashed $A_1$ model we will also check that the completions of these indefinite theta series agree with the completion given in equation \eqref{eq:fk_integral}.

\subsection{Details for the Squashed $A_1$ Model}
In the preceding sections we showed that the computation of (flavored) elliptic genera of squashed toric sigma models reduces to the evaluation of integrals $f_{\kkk} ( \CM; \t, z, \uu)$ defined in \eqref{eq:fk_definition}. Our final result in \eqref{eq:fk_integral} then yields the answer for the class of theories considered in this paper and shows that they are built out of indefinite theta series (of generic signature). In this section, we will elaborate on the example of squashed $A_1$ model to illustrate our result in a concrete setting. 

To simplify matters we will focus on a particular choice of gauged flavor charges given in Table \ref{tab:FlavorCharges} and also considered 
in \cite{Gupta:2017bcp} as an example, and further restrict to the case $\kk_1 = \kk_2 = \kk$.
\begin{table}[h]
\centering
\begin{tabular}{c | c c}
\toprule
Field  &  $U(1)_1$ & $U(1)_2$ \\ 
\hline 
$\phi_1$ & 1 & 0 \\
$\phi_2$ & 1 & 1 \\
$\phi_3$ & 0 & 1 \\
\bottomrule
\end{tabular}
\caption{}
\label{tab:FlavorCharges}
\end{table}
With these choices and setting fugacities for global symmetries to zero we have
\begin{align} \label{chiA1}
\chl (\Ait, \kk ; \t, z)  \= 2\kk^2 
\intt_{\IC} \frac{\ddd^2 \zzz'_1}{\tim}  \intt_{\IC} \frac{\ddd^2 \zzz'_2}{\tim} 
&\frac{\thone \lp \t, -z  - \frac{1}{2} \zzz'_1 +\frac{1}{2} \zzz'_2 \rp}{\thone \lp \t,  - \frac{1}{2} \zzz'_1 + \frac{1}{2} \zzz'_2\rp} \,
\frac{\thone \lp \t, -z + \frac{3}{2}  \zzz'_1 + \frac{1}{2} \zzz'_2 \rp}{\thone \lp \t, \frac{3}{2}  \zzz'_1 + \frac{1}{2}  \zzz'_2 \rp} 
\notag
 \\
& \times 
e^{-\frac{\pi \kk }{\tim} \lp \zzz'_1 + \frac{z}{\kk} \rp \lp \zzb'_1 + \frac{z}{\kk} \rp} \,
e^{-\frac{\pi \kk }{\tim} \lp \zzz'_2 + \frac{z}{\kk} \rp \lp \zzb'_2 + \frac{z}{\kk} \rp}.
\end{align}

Comparing to equation \eqref{eq:fk_definition} we find\footnote{Note that dummy integration variables are scaled so that equation \eqref{eq:CMmatrixproperty} is satisfied.}
$N=2$ and 
\begin{equation}
\chl (\Ait, \kk ; \t, z)  \= 2 f_{\kkk} \lp \CM; \t, z, 0 \rp,
\where \kkk \= (\kk,\kk) \ \mathrm{and} \ 
\CM \= \mat{-\frac{1}{2} & \frac{3}{2} \\ \frac{1}{2} & \frac{1}{2}}.
\end{equation}
For the values of $\kkk$ and $\CM$ here and with respect to the notation introduced in \eqref{eq:CMmatrixproperty}, \eqref{eq:integralnotation1}, and \eqref{eq:twistedmomenta} we have
\begin{equation}
\CK = \frac{1}{\sqrt{2k}} I_2, \quad
\CE = \sqrt{2k} \mat{-\frac{1}{2} & \frac{1}{2} \\ \frac{3}{2} & \frac{1}{2}}, \quad
g = k \mat{5 & 1 \\ 1 & 1},\quad
2 Q_2^T g^{-1} p_L = \ww_1 + \ww_2 + \frac{2 \nnn_2}{\kk}.
\end{equation}
Then, according to our result in \eqref{eq:fk_integral} we find
\begin{align}
\label{eq:chi_ell_A1_completion}
&\chl (\Ait, \kk ; \t, z) \= 2
\lp \frac{-i \thone (\t, z)}{\eta (\t)^3}  \rp^2 \sum_{\nnn,\ww \in \IZ^2}
\frac{q^{\nnn_1 \ww_1 + \nnn_2 \ww_2}\, 
\yy_z^{\ww_1 + \ww_2 + \frac{2 \nnn_2}{\kk} } }
{ \lp 1 - \yy_z^{-1} q^{-\nnn_1} \rp  \lp 1 - \yy_z^{-1} q^{-\nnn_2} \rp}  \\
& \qquad \times
\frac{1}{4} \sum_{c_1, c_2 \in \{ 0,1\}} (-1)^{c_1+c_2} 
E_2 \lp \CE^{-T};  \frac{1}{2k} \sqrt{\frac{\tim}{2}} \CE 
\mat{\nnn_1 - \nnn_2 - 2 \kk (\ww_1 + c_1) \\ 5\nnn_2 - \nnn_1 - 2 \kk(\ww_2 + c_2) }  \rp.
\notag
\end{align} 
 
To isolate the holomorphic (discrete) part of this elliptic genus we will use the fact that, according to \eqref{eq:gen_err_asymp} we have $E_2 ( \mathcal{F}; \sqrt{2 \tim} \, x ) \to \sgn (\mathcal{F}^T \, x)$ as 
$\tim \to \infty$ with sign functions defined as in equation \ref{eq:sgndef}. We will give the details for the special case of $\kk=2$.
\begin{align}
\chl^\hol (\Ait, 2 ; \t, z) \= 2
&\lp \frac{-i \thone (\t, z)}{\eta (\t)^3}  \rp^2 \sum_{\nnn,\ww \in \IZ^2}
\frac{q^{\nnn_1 \ww_1 + \nnn_2 \ww_2}\, 
\yy_z^{\ww_1 + \ww_2 + \nnn_2 } }
{ \lp 1 - \yy_z^{-1} q^{-\nnn_1} \rp  \lp 1 - \yy_z^{-1} q^{-\nnn_2} \rp}  \notag \\
& \qquad \times
\frac{1}{4} \sum_{c_1, c_2 \in \{ 0,1\}} (-1)^{c_1+c_2} \,
\sgn\mat{
\nnn_1 - \nnn_2 - 4 (\ww_1 + c_1) \\ 5\nnn_2 - \nnn_1 -4(\ww_2 + c_2) 
}.
\label{eq:k2_hol_genus}
\end{align}
The second line gives nonzero contributions only when
\begin{equation}
4 \ww_1 \leq \nnn_1 - \nnn_2 \leq 4 \ww_1 +4
\andd
4 \ww_2 \leq - \nnn_1 + 5 \nnn_2 \leq 4 \ww_2 + 4.
\end{equation}
Writing $\nnn_1 = 5 \ww_1 + \ww_2 + p_1$ and $\nnn_2 = \ww_1 + \ww_2 + p_2$ (where we decomposed $\nnn$ as $\frac{1}{2} g w + p$) the inequalities reduce to
\begin{equation}\label{eq:p_region}
0 \leq p_1 - p_2 \leq 4, 
\andd
0 \leq - p_1 + 5 p_2 \leq 4.
\end{equation}
There are seven points in $p \in \IZ^2$ that satisfy these conditions. Four of those seven points,
\begin{equation}\label{eq:pchoiceedge}
\mat{p_1 \\ p_2} \= \mat{0 \\ 0}, \   \mat{1 \\ 1}, \   \mat{5 \\ 1}, \   \mat{6 \\ 2},
\end{equation}
are on the vertices of the parallelogram shaped region in \eqref{eq:p_region} and yield $\frac{1}{4}$ for the second line of \eqref{eq:k2_hol_genus}. The remaining three points, on the other hand, yield unity:
\begin{equation}\label{eq:pchoicebulk}
\mat{p_1 \\ p_2} \= \mat{2 \\ 1}, \   \mat{3 \\ 1}, \   \mat{4 \\ 1}.
\end{equation}
Therefore, defining 
\begin{align}
g(p_1, p_2 ; \t, z) \;\coloneqq\;
2 \, &\lp \frac{- i \, \thone (\t, z)}{\eta (\t)^3} \rp^2
\sum_{\ww \in \IZ^2} 
q^{5 \ww_1^2 + 2 \ww_1 \ww_2 + \ww_2^2} \, \yy_z^{2 \ww_1 + 2 \ww_2}
\notag \\
& \qquad \times
\frac{q^{p_1 \ww_1 +p_2 \ww_2} \, \yy_z^{p_2}}{\lp 1 - \yy_z^{-1} \, q^{-5 \ww_1 - \ww_2 - p_1} \rp \, \lp 1 - \yy_z^{-1} \, q^{-\ww_1 - \ww_2 - p_2} \rp} \,,
\label{eq:gfunc_hol}
\end{align}
we find the discrete part of $\chl (\Ait, 2 ; \t, z)$ to be
\begin{align}
\chl^\hol (\Ait, 2 ; \t, z)&\= \frac{1}{4} \lb g(0,0; \t,z) +  g(1,1; \t,z) +  g(5,1; \t,z) +  g(6,2; \t,z)  \rb 
\notag \\ &\quad 
+ g(2,1; \t,z) + g(3,1; \t,z) + g(4,1; \t,z) .
\label{eq:chiell_hol_g}
\end{align}

The sum in \eqref{eq:gfunc_hol} can be brought into the form of a signature $(2,2)$ indefinite theta function by expanding the factors in the denominator as
\begin{align}
g(p_1, p_2 ; \t, z) =
2 \, &\lp \frac{- i \, \thone (\t, z)}{\eta (\t)^3} \rp^2
\sum_{\ww, r \in \IZ^2} \frac{1}{4}
\lb \sign{-r_1+\e_1} - \sign{5 \ww_1 + \ww_2+p_1+\frac{\mathrm{Im}(z)}{\tim}} \rb \
\notag \\
& \quad \times
\lb \sign{-r_2+\e_2} - \sign{\ww_1 + \ww_2+p_2+\frac{\mathrm{Im}(z)}{\tim}} \rb
 \yy_z^{2 \ww_1 + 2 \ww_2 + r_1+r_2+p_2}
\notag \\
& \quad \times
q^{5 \ww_1^2 + 2 \ww_1 \ww_2 + \ww_2^2 + r_1 (5 \ww_1 + \ww_2) + r_2 (\ww_1+\ww_2)} \, 
q^{p_1 (\ww_1+r_1) +p_2 (\ww_2+r_2)},
\label{eq:gfunc_hol2}
\end{align}
where $\e_1, \e_2 \in (0,1)$ are arbitrary. In fact, it will be convenient to define
\begin{align}
&h(p_1, p_2 ; \t, z) \;\coloneqq\; \notag \\
 &\quad 2 \, \lp \frac{- i \, \thone (\t, z)}{\eta (\t)^3} \rp^2
\sum_{\ww, r \in \IZ^2} \frac{1}{4}
\lb \sign{-4r_1+p_1-p_2} - \sign{5 \ww_1 + \ww_2+p_1+\frac{\mathrm{Im}(z)}{\tim}} \rb \
\notag \\
& \quad \times
\lb \sign{-4r_2-p_1+5p_2} - \sign{\ww_1 + \ww_2+p_2+\frac{\mathrm{Im}(z)}{\tim}} \rb
 \yy_z^{2 \ww_1 + 2 \ww_2 + r_1+r_2+p_2}
\notag \\
& \quad \times
q^{5 \ww_1^2 + 2 \ww_1 \ww_2 + \ww_2^2 + r_1 (5 \ww_1 + \ww_2) + r_2 (\ww_1+\ww_2)} \, 
q^{p_1 (\ww_1+r_1) +p_2 (\ww_2+r_2)},
\label{eq:hfunc_hol2}
\end{align}
which is essentially \eqref{eq:gfunc_hol2} with $\e_1 = \frac{p_1-p_2}{4}$ and $\e_2 = \frac{5p_2-p_1}{4}$.  
The upshot is that $h(p_1, p_2 ; \t, z) = g(p_1, p_2 ; \t, z)$ for $p_1$ and $p_2$ chosen as in \eqref{eq:pchoicebulk}, 
while the points on the boundary combine to\footnote{For example, the expression for $g(5, 1 ; \t, z)$ can be made similar to that of $g(0,0 ; \t, z)$ 
by shifting $\ww_1 \mapsto \ww_1-1$ and $r_1 \mapsto r_1+1$ after which which we essentially have $g(0,0;\t,z)$ but 
with $\e_1 \mapsto \e_1 - 1$. Repeating this for $(p_1, p_2) \in \{ (1,1), \, (6,2) \}$ and noting 
that $\sign{x+\e} + \sign{x+\e-1} = 2 \sign{x}$ for $\e \in (0,1)$ yields equation \eqref{eq:h_g_relation}.}
\begin{equation}\label{eq:h_g_relation}
h(0, 0 ; \t, z) \=  \frac{1}{4} \lb g(0,0; \t,z) +  g(1,1; \t,z) +  g(5,1; \t,z) + g(6,2; \t,z)  \rb \, ,
\end{equation}
so that 
\begin{equation}\label{eq:chiell_hol_h}
\chl^\hol (\Ait, 2 ; \t, z)\= 
h(0,0; \t,z) + h(2,1; \t,z) + h(3,1; \t,z) + h(4,1; \t,z) \, .
\end{equation}

More importantly, the series in $h(p_1, p_2 ; \t, z)$ can be realized as the holomorphic part of an indefinite theta function. For this purpose, let us introduce the lattice $\wt{\Lambda} \simeq \IZ^4$ and let us define
\begin{equation}\label{eq:lattice_info1}
\wt{n} \coloneqq \mat{\ww_1 \\ \ww_2 \\ r_1 \\ r_2}, \quad
\wt{Q} \coloneqq \mat{10 & 2 & 5 &1 \\ 2 & 2 & 1 &1 \\ 5 &1 &0 & 0 \\ 1 & 1 & 0 & 0}, \quad
\zzz \coloneqq \mat{0 \\ z \\ 0 \\ 0},  \andd
\mu (p) \coloneqq \frac{1}{4}
 \mat{p_1 - p_2 \\ -p_1+5p_2 \\ -p_1 +p_2 \\ p_1 - 5p_2}.
\end{equation}
Then taking $\wt{Q}$ to define the quadratic form on $\wt{\Lambda}$, we have
\begin{equation}
\frac{1}{2} \wt{n}^2 \coloneqq \frac{1}{2} \wt{n}^T \wt{Q} \, \wt{n}  = 5 \ww_1^2 + 2 \ww_1 \ww_2 + \ww_2^2 + r_1 (5 \ww_1 + \ww_2) + r_2 (\ww_1+\ww_2),  \quad
\mu(p)^2 \= 0,
\end{equation}
\begin{equation}
\mu (p) \cdot \zzz = p_2 z, \quad
\wt{n} \cdot \lb \zzz + \mu (p) \t \rb = (2 \ww_1 + 2 \ww_2 +r_1 +r_2) z + \lb p_1 (\ww_1+r_1) +p_2 (\ww_2+r_2) \rb \t.
\end{equation}
In particular, note that $\mu(p) \in \wt{\Lambda}^*$.
We will define an indefinite theta function on $\wt{\Lambda}$ via a rectangular cone defined by vectors
\begin{equation}\label{eq:cone_info1}
\wt{c}_1 \;\coloneqq\; \mat{-1 \\ 1 \\ 2 \\ -2}, \quad
\wt{c}_2 \;\coloneqq\; \mat{1 \\ -5 \\ -2 \\ 10}, \quad
\wt{c}'_1 \;\coloneqq\; \mat{0 \\ 0 \\ 1 \\ 0}, \quad
\wt{c}'_2 \;\coloneqq\; \mat{0 \\ 0 \\ 0 \\ 1}.
\end{equation}
These vectors satisfy
\begin{align}
&\qquad \qquad 
\wt{c}_1^2 \= -8, \quad \wt{c}_2^2 \= -40, 
\quad \wt{c}_1'^2 \=0, \quad \wt{c}_2'^2 \= 0, \\
&\wt{c}_1 \cdot \wt{c}_2 \= 8, \quad 
\wt{c}_1' \cdot \wt{c}_2' \= \wt{c}_1 \cdot \wt{c}_2' \= \wt{c}_2 \cdot \wt{c}_1' \= 0,
\quad
\wt{c}_1 \cdot \wt{c}_1' \= \wt{c}_2 \cdot \wt{c}_2' \= -4. 
\end{align}
Note that null vectors $\wt{c}_1'$ and $\wt{c}_2'$ are orthogonal to every vector in this set except for $\wt{c}_1$ and $\wt{c}_2$, respectively, as was required in Section \ref{sec:IndefThetaSeries}.\footnote{These vectors also obey the convergence conditions stated in Section 4.3 of \cite{Alexandrov:2016enp}.} Finally noting that
\begin{align}
&\wt{c}_1 \cdot \lp \wt{n}+ \mu(p) + \frac{\mathrm{Im}(\zzz)}{\tim} \rp\= -4 r_1 + p_1 - p_2, \notag \\
&\wt{c}_1' \cdot \lp \wt{n}+ \mu(p) + \frac{\mathrm{Im}(\zzz)}{\tim} \rp\= 5 \ww_1 + \ww_2 + p_1+\frac{\mathrm{Im}(z)}{\tim}, \notag \\
&\wt{c}_2 \cdot \lp \wt{n}+ \mu(p) + \frac{\mathrm{Im}(\zzz)}{\tim} \rp\= -4 r_2 - p_1 +5 p_2, \notag \\ 
&\wt{c}_2' \cdot \lp \wt{n}+ \mu(p) + \frac{\mathrm{Im}(\zzz)}{\tim} \rp\= \ww_1 + \ww_2+p_2+\frac{\mathrm{Im}(z)}{\tim},
\end{align}
we write $h(p_1, p_2 ; \t, z)$ as
\begin{align}
&
 2 \, \lp \frac{- i \, \thone (\t, z)}{\eta (\t)^3} \rp^2
\sum_{\wt{n} \in \wt{\Lambda} } \frac{1}{4}
\lb \sign{\wt{c}_1 \cdot  \lb \wt{n} + \mu(p) \rb } 
- \sign{\wt{c}_1' \cdot \lp \wt{n} + \mu(p) + \frac{\mathrm{Im}(\zzz)}{\tim} \rp} \rb \
\notag \\
& \quad \times
\lb \sign{\wt{c}_2 \cdot  \lb \wt{n} + \mu(p) \rb} 
- \sign{\wt{c}_2' \cdot \lp \wt{n} + \mu(p)+ \frac{\mathrm{Im}(\zzz)}{\tim} \rp} \rb
e^{2 \pi i \zzz \cdot  (\wt{n} + \mu(p))} \,
 q^{\frac{1}{2} (\wt{n} + \mu(p))^2 }.
\label{eq:hfunc_hol3}
\end{align}

We can use this representation to obtain the modular completion $\wh{h}(p_1, p_2 ; \t, z) $ for $h(p_1, p_2 ; \t, z)$, 
using the replacement given in \eqref{eq:gen_error_replacement}:
\begin{align}
 &\wh{h}(p_1, p_2 ; \t, z)  \= 
 2 \, \lp \frac{- i \, \thone (\t, z)}{\eta (\t)^3} \rp^2
\sum_{\ww, r \in \IZ^2} \frac{1}{4}
\Big[ E_2 \lp  \mat{-1 & 3 \\ 1 & 1}; \sqrt{\frac{\tim}{8}} 
\mat{2 r_1 - 2 r_2 -p_1 + 3p_2 \\ -6r_1 -2r_2 + p_1 + p_2} \rp \notag\\
&\qquad 
- \sign{\ww_1 + \ww_2+p_2+\frac{\mathrm{Im}(z)}{\tim}} 
\erf \lp \sqrt{4\pi \tim} \lp -r_1 +\frac{p_1 - p_2}{4} \rp \rp \notag \\
&\qquad 
- \sign{5 \ww_1 + \ww_2+p_1+\frac{\mathrm{Im}(z)}{\tim}} 
\erf \lp \sqrt{\frac{4\pi \tim}{5}} \lp -r_2 +\frac{5p_2 - p_1}{4} \rp \rp \notag \\
&\qquad
+\sign{\ww_1 + \ww_2+p_2+\frac{\mathrm{Im}(z)}{\tim}} 
\sign{5 \ww_1 + \ww_2+p_1+\frac{\mathrm{Im}(z)}{\tim}}  \Big]
\notag \\
& \qquad \times
\yy_z^{2 \ww_1 + 2 \ww_2 + r_1+r_2+p_2}
q^{5 \ww_1^2 + 2 \ww_1 \ww_2 + \ww_2^2 + r_1 (5 \ww_1 + \ww_2) + r_2 (\ww_1+\ww_2)} \, 
q^{p_1 (\ww_1+r_1) +p_2 (\ww_2+r_2)} \,.
\label{eq:hcompletion}
\end{align}
Since in Equation \eqref{eq:chiell_hol_h} we have written $\chl^\hol (\Ait, 2 ; \t, z)$ in terms of $h(p_1, p_2 ; \t, z)$ 
the modular completions should also satisfy the same relation, i.e. we should have
\begin{equation}\label{eq:chiell_comp_h}
\chl (\Ait, 2 ; \t, z)\= 
\wh{h}(0,0; \t,z) + \wh{h}(2,1; \t,z) + \wh{h}(3,1; \t,z) + \wh{h}(4,1; \t,z) \,.
\end{equation}
A direct proof of this equality is also possible starting with the expression of $\chl (\Ait, 2 ; \t, z)$ 
given in \eqref{eq:chi_ell_A1_completion}.\footnote{A quick sketch of this proof is as follows: 
One starts by rewriting the sum over $\nnn,\ww \in \IZ^2$ in \eqref{eq:chi_ell_A1_completion} 
as a sum over  $r,\ww' \in \IZ^2$ where $\nnn = \frac{1}{2} g \ww' + p$, $\ww = \ww' + r$ and 
with $p$ running over the set $\pmat{0 \\ 0},\pmat{2 \\ 1}, \   \pmat{3 \\ 1}, \   \pmat{4 \\ 1}$. 
Then let us use $\chl (\Ait, 2, p_1,p_2 ; \t, z)$ to denote the contribution to $\chl (\Ait, 2 ; \t, z)$ 
from a certain $p$. Now we note that $\chl^\hol (\Ait, 2, p_1,p_2 ; \t, z) = h(p_1, p_2 ; \t, z)$ and hence try to prove
\begin{equation*}
\chl (\Ait, 2, p_1,p_2 ; \t, z)  - \chl^\hol (\Ait, 2, p_1,p_2 ; \t, z) 
=  \wh{h}(p_1, p_2 ; \t, z)  - h(p_1, p_2 ; \t, z) .
\end{equation*}
Now note that in equation \eqref{eq:chi_ell_A1_completion}, the shift $\ww \mapsto \ww - c$ would allow us to 
cancel the factor in the denominator. This replacement however is not legal because the sum over $c$ can not 
switch places with the sum over $\ww$  due to divergence issues. In the new equation we are trying to prove 
we can however generate factors free from this divergence issue by using equation \eqref{eq:EQ_decomposition}.
}

Before concluding this section, let us give an alternative expression for~$\chl (\Ait, 2 ; \t, z)$
by noticing that we can combine the sum over $\wt{\Lambda}$ and the sum over $p$ as a sum over a 
new lattice $\Lambda$ generated by $\wt{\Lambda}$ and $\mu(2,1)$.\footnote{We have $2 \mu(2,1) \equiv \mu(3,1) \pmod{\wt{\Lambda}}$, 
$3 \mu(2,1) \equiv \mu(4,1) \pmod{\wt{\Lambda}}$, and $4 \mu(2,1) \equiv 0 \pmod{\wt{\Lambda}}$.} 
In particular, we pick the following basis for $\Lambda$:
\begin{equation}
\l_1 \= \mat{-1/4 \\ 1/4 \\ -3/4 \\ -1/4}, \quad
\l_2 \= \mat{0 \\ 1 \\ 0 \\ 0}, \quad
\l_3 \= \mat{-1/4 \\ 1/4 \\ 1/4 \\ -1/4}, \quad 
\l_4 \= \mat{0 \\ 0 \\ 0 \\ 1} \in \wt{\Lambda}^*,
\end{equation}
so that an element of $\Lambda$ is written in the form $\sum_{j=1}^4 n_j \l_j$ where $n_j \in \IZ$ for $j=1,2,3,4$ 
specifying an identification of $\Lambda$ with $\IZ^4$. Mapping the information in \eqref{eq:lattice_info1} 
and \eqref{eq:cone_info1} (and removing tildes to denote the component vectors in this basis) we find
\begin{equation}\label{eq:lattice_info2}
Q \coloneqq \mat{2 & -1 & 1 &0 \\ -1 & 2 & 0 &1 \\ 1 &0 &0 & 0 \\ 0 & 1 & 0 & 0}, \ 
\zzz(z) \coloneqq \mat{0 \\ z \\ 0 \\ 0}, \ 
n \coloneqq \mat{n_1 \\ n_2 \\ n_3 \\ n_4} \  \mathrm{where} \ 
\begin{cases}
& \ww_1 = - \frac{n_1+n_3}{4}, \\
& \ww_2 =  \frac{n_1+n_3}{4} + n_2, \\
& r_1 = \frac{-3n_1+n_3}{4}, \\
& \ww_1 = - \frac{n_1+n_3}{4} + n_4,
\end{cases}
\end{equation}
\begin{equation}\label{eq:cone_info2}
c_1 \;\coloneqq\; \mat{-1 \\ 0 \\ 5 \\ -1}, \quad
c_2 \;\coloneqq\; \mat{1 \\ -4 \\ -5 \\ 9}, \quad
c'_1 \;\coloneqq\; \mat{-1 \\ 0 \\ 1 \\ 0}, \quad
c'_2 \;\coloneqq\; \mat{0 \\ 0 \\ 0 \\ 1}.
\end{equation}
With these definitions we can finally state the discrete part of the elliptic genus as
\begin{equation} \label{eq:chi_hol_theta}
\chl^\hol (\Ait, 2 ; \t, z) \= 
2 \, \lp \frac{- i \, \thone (\t, z)}{\eta (\t)^3} \rp^2
\Theta^{Q} \lp C,C';  \t ,\zzz (z) \rp .
\end{equation}
Note that the lattice $\Lambda$ with $Q$ as its quadratic form is even and unimodular. The completed elliptic genus is similarly
\begin{equation}\label{eq:chi_completion}
\chl (\Ait, 2 ; \t, z) \= 
2 \, \lp \frac{- i \, \thone (\t, z)}{\eta (\t)^3} \rp^2
\wh{\Theta}^{Q} \lp C,C';  \t ,\zzz (z) \rp .
\end{equation}
The fact that $\Lambda$ is rank four and $\zzz(z)^2 = 2 z^2$ implies the completed theta function transforms 
like a weight $2$, index $1$ Jacobi form. Combined with the prefactor $\lp \frac{- i \, \thone (\t, z)}{\eta (\t)^3} \rp^2$ 
which is a weight $-2$, index $1$ Jacobi form,\footnote{Note that the zeros of $\thone (\t,z)^2$ cancel the double 
poles of $\Theta^{Q} \lp C,C';  \t ,\zzz (z) \rp$ on $\IZ \t + \IZ$. These double poles stem from the two null 
vectors $c_1'$ and $c_2'$ defining the theta function. } the representation in equation \eqref{eq:chi_completion} 
tells us that the elliptic genus $\chl (\Ait, 2 ; \t, z)$ is the completion of a depth two mock Jacobi form of weight $0$ 
and index $2$, consistent with our general discussion.

In conclusion, in this section we have given several equivalent expressions for the elliptic genus of squashed $A_1$ 
model (at $\kk_1 = \kk_2 =2$ and for the charge configuration given in Table \ref{tab:FlavorCharges}) 
in \eqref{eq:chi_ell_A1_completion}, \eqref{eq:chiell_comp_h}, and \eqref{eq:chi_completion}. We have also given expressions 
for its holomorphic (discrete) part in \eqref{eq:chiell_hol_g}, \eqref{eq:chiell_hol_h}, and \eqref{eq:chi_hol_theta}. 
In particular, we can compute its leading Fourier coefficients using these expressions:
\begin{equation}
\chl^\hol (\Ait, 2 ; \t, z) \= 
\lp 1 + \frac{1}{2} \yy_z^{\pm 1} \rp +
\lp 6 -\frac{3}{2} \yy_z^{\pm 1} -2  \yy_z^{\pm 2} + \frac{1}{2} \yy_z^{\pm 3}\rp q + 
 O (q^2).
\end{equation}
The function~$\chl^\hol (\Ait ,2 ; \t, z)$ has a theta decomposition 
\be
\chl^\hol (\Ait, 2 ; \t, z) \= \sum_{\mu = -1}^2 h_\mu (\t) \, \vth_{2,\mu} (\t,z) \,,
\ee
with
\begin{align}
h_{- 1} (\t) &\= h_1 (\t)   \\
h_0 (\t) &\= 1 + 6 q + 34 q^2 + 144 q^3 + 534 q^4 + 1776 q^5  + \ldots \\
h_1 (\t) &\= q^{-1/8} \lp \frac{1}{2} - \frac{3}{2} q- \frac{25}{2} q^2 -78 q^3 - \frac{683}{2} q^4 - 1270 q^5 + \ldots \rp \\
h_2 (\t) &\= q^{1/2} \lp -2 - 4 q + 12 q^2 +120 q^3 +594 q^4  + \ldots \rp .
\end{align}
Finally, at~$z=0$ we expect to recover the Witten index, which should be~$q$-independent. 
Indeed we have checked that $\chl^\hol (\Ait ,2 ; \t, z=0) = 2$ to $O(q^{14})$ precision.
(In fact one can deduce that this is the correct value of the Witten index directly from 
the expression~\eqref{chiA1}, which reduces to a Gaussian integral at~$z=0$.)
We note that this is also in agreement with the Witten index of the unsquashed~$A_1$. 
The squashing deformation thus does not change the Witten index even for this non-compact 
model, in spite of quite drastically changing the boundary conditions, similar to the observation 
in~\cite{Gupta:2017bcp} for the compact case.


\section{A Possible Relation with Vafa-Witten Partition Functions on~$\IC\IP^2$ \label{sec:VWrel}}

In this paper we discussed a class of two-dimensional GLSMs whose target space is a squashed toric CY of complex dimension~$n$. 
We found that the elliptic genus of these models are built out of the modular completions of indefinite theta functions associated to lattices 
of signature~$(n,n)$. In this manner every squashed toric CY is associated to a particular mock modular form of depth~$n$, thus
generalizing the classic relation between CY manifolds and modular forms discussed in the introduction. 

In fact, mock modular forms of higher depth also arise in another interesting physical context, namely in the 
discussion~\cite{Manschot:2014cca, Manschot:2017xcr} of partition functions 
of twisted~$\CN=4$ super Yang-Mills theory (Vafa-Witten (VW) theory~\cite{Vafa:1994tf}) on~$\IC\IP^2$. 
The twisted super Yang-Mills theory has different topological sectors which are labelled by the value of magnetic
't Hooft flux. For gauge group~$U(N)$, the 't Hooft flux on~$\IC \IP^2$ takes values in~$\IZ_N$. 
The partition function contains, as a factor, the generating function of~$\bar{\chi}(\mathcal M_{j,m})$ which is
the Euler characteristic (or rather a related rational invariant) of the moduli space of instantons with instanton 
number $m$ and 't Hooft flux~$j$.

In this section we comment on a potential relation between these two apparently different physical phenomena. 
For~$N=2$ we have a precise mathematical relation between the two functions described above. 
For generic~$N$, the main new point is an observation about the similarity of the modular structure of the elliptic genera computed 
in this paper and the~$U(N)$ VW partition functions on~$\IC\IP^2$.

We begin with the Vafa-Witten partition function on $\IC \IP^{2}$ for the $U(2)$ gauge group. 
Based on the work of Yoshioka and Klyachko \cite{Yoshioka1994,Klyachko1991}, the relevant partition functions~$Z_j(\t)$, $j=0,1$, 
are~\cite{Vafa:1994tf} 
\begin{equation}\label{Z01}
Z_{0}(\t) \=\frac{3}{\eta(\t)^{6}}\,\sum_{n=0}^{\infty} \, H(4n) \, q^{n}
\andd
Z_{1}(\t) \=\frac{3}{\eta(\t)^{6}}\,\sum_{n=1}^{\infty}\, H(4n-1) \, q^{n-\frac{1}{4}} \,,
\end{equation}
where~$H(n)$ for~$n>0$ are Hurwitz-Kronecker class numbers, i.e.~the number of~$SL(2,\IZ)$-equivalence classes of  
positive integral binary quadratic forms of discriminant $-n$, weighted by the reciprocal of the number of their automorphisms, 
and~$H(0)=-1/12$. 
It was then shown in \cite{Bringmann:2010sd} that these generating functions\footnote{In~\cite{Bringmann:2010sd}, 
there is an extra term in the expression for~$Z_0(\t)$ compared to Equation~\eqref{Z01}, 
which reflects the ambiguity in defining the Euler characteristic of the moduli space of instantons when 
the 't Hooft flux and the rank are not relatively prime~\cite{Manschot:2017xcr}. This extra term breaks the covariance 
of the vector-valued partition function~$Z_j(\t)$  in~\eqref{Z01} under the full~$SL(2,\IZ)$.} 
are given in terms of derivatives of the Zwegers~$\mu$-function~\cite{Zwegers:2008zna} defined as 
\be
\mu(u,v;\t) \=\frac{e^{\pi iu}}{\vartheta_{1}(\tau,v)}\sum_{n\in\IZ} 
\frac{(-1)^{n}e^{\pi i(n^{2}+n)\t+2\pi inv}}{1-e^{2\pi in\t+2\pi iu}}\,,\qquad u,v\in \IC\,.
\ee
The relation is as follows
\bea \label{Zmurel}
&&Z_{0}(\t)\=-\frac{1}{\eta(\t)^{6}} \, \frac{1}{2\pi i} \, \frac{d}{dz}\biggl(q^{-1/4}\, \z_z^{3/2} \, \m \Bigl( 2z-\t,\frac{1}{2}-z;2\t \Bigr) \biggr)\bigg|_{z=0}\,,\nn\\
&&Z_{1}(\t)\=-\frac{1}{\eta(\t)^{6}} \, \frac{1}{2\pi i} \,\frac{d}{dz} \,\m \Bigl( 2z-\t,\frac{1}{2}-\t-z;2\t \Bigr) \bigg|_{z=0}\,.
\eea

Now we note that the $\mu$-function can be written in terms of the Appell-Lerch sum~\eqref{AppellFunc} 
for~$k=\frac{1}{2}$ as
\be \label{muALrel}
\mu(u,v;\t)\=\frac{1}{\vartheta_{1}(\tau,v)} \,A_{1,\frac{1}{2}} \bigl(\t,u,v+\frac{1}{2} \bigr)\,.
\ee
In Section~\ref{CigAL} we explained the relation between the Appell-Lerch sum~$A_{1,k}$
and the elliptic genus of the squashed $\wt{\IC/\IZ_2}$ manifold, or equivalently, that of 
the~$SL(2,\IR)_{\frac12}/U(1)$ cigar coset theory (or its mirror~$\CN=2$ Liouville theory). 
Putting these two relations together, we can recast the observation~\eqref{Zmurel} as a relation 
between the elliptic genus of the squashed $\wt{\IC/\IZ_2}$/cigar/$\CN=2$ Liouville 
and the~$U(2)$ VW partition function. 

From a physical point of view, the VW theory can be thought of as arising from a fivebrane  
in string theory wrapped on~$\IC\IP^2 \times T^2$ \cite{Minahan:1998vr}. 
The VW theory appears as the effective theory when the~$T^2$ is small, and the VW partition function is supposed to 
equal the modified elliptic genus of the $(0,4)$ supersymmetric effective 2d theory on the~$T^2$. The details of this 
effective 2d theory are not currently understood (beyond calculations protected by anomalies). 
Our above observation suggests that the 2d theory is closely related to the $\wt{\IC/\IZ_2}$/cigar/$\CN=2$ Liouville 
theory with~$k=\half$. (There are additional~$\vartheta_{1}$ and~$\eta$ factors in Equations
\eqref{cigorbAL} and~\eqref{muALrel} which point to additional degrees of freedom apart from 
those captured by the cigar.) 

From the geometric point of view, 
it is tempting to conjecture that the Euler characteristic of the moduli space of instantons of $U(2)$ theory 
are related to certain topological invariants of the $\wt{\IC/\IZ_2}$/cigar/$\CN=2$ Liouville theory for some deeper reason. 
A geometric relation may begin with the fact that instanton moduli spaces are toric manifolds, and perhaps 
a certain regularization of these manifolds required to define the Euler characteristic effectively squashes 
the manifold. 
Some support for this comes from the fact that the relation~\eqref{Zmurel} can be refined further.
The work of~\cite{Bringmann:2010sd} shows that the generating function of the Poincare polynomial~$P(t)$ of the 
instanton moduli space is given in terms of the~$\mu$-function. Equation~\eqref{Zmurel} is the limit of this 
relation as~$t=-1$ for which the Poincare polynomial reduces to the Euler characteristic.

Another, more speculative, idea is that of a holographic relation in the context of NS fivebranes in string 
theory~\cite{Aharony:1998ub, Giveon:1999zm, Murthy:2003es}. 
The cigar/$\CN=2$ Liouville theory is known to capture the near-horizon geometry of wrapped NS fivebranes. 
The VW theory would then be viewed as a boundary theory and some twisted version of the cigar theory as the spacetime hologram. 
In this regard, we note that at $k=\frac{1}{2}$, the central charge of 
the cigar/$\mathcal N=2$ Liouville theory is 
\be
c\=3 \bigl(1+\frac{2}{k} \bigr)\=15\,,
\ee
which is the central charge for the critical superstring theory with the two-dimensional target space being the cigar 
coset~\cite{Kutasov:1990ua}. 
It would be really interesting if all these field theory phenomena has some holographic 
interpretation in terms of a string theory observable on this pure cigar target space.

Finally, we note that for generic $N$, Vafa-Witten partition functions on $\IC \IP^{2}$ are expressed in terms 
of generalized Appell functions \cite{Manschot:2014cca} (see Equation \eqref{eq:genAppell}). These are specific 
indefinite theta series for which a subset of the vectors determining its rectangular cone are null vectors. 
In Section \ref{sec:generalcase} we found a similar structure for the elliptic genera of squashed toric models. 
Given that these generalized Appell functions are the key parts determining the (mock) modular behavior of both 
the VW partition function and the squashed toric model elliptic genus, it would be interesting to look for a 
GLSM description of the $2d$, $\CN=(0,4)$ effective theory of wrapped branes using ingredients similar to the 
ones used in squashed toric models.

\section*{Acknowledgements}
We thank Kathrin Bringmann for useful discussions and collaboration during the initial stages of this work. The work of R.G.~and 
S.M.~was supported by an ERC Consolidator Grant N.~681908, ``Quantum black holes: A microscopic window into the microstructure 
of gravity'', and by the STFC grant ST/P000258/1. The research of C.N.~is supported by the European Research Council under the European 
Union’s Seventh Framework Programme (FP/2007-2013) / ERC Grant agreement n.~335220 - AQSER.

\appendix
\section{Definitions and Conventions \label{sec:app_defn}}
For $\t \in \IH$ we will use $\tre \coloneqq \Re (\t)$, $\tim \coloneqq \im (\t)$ and $q \coloneqq e^{2 \pi i \t}$. For an elliptic variable $z$ we use the notation $\yy_z \coloneqq e^{2 \pi i z}$. We are also use the fundamental parallelogram $E_\t \coloneqq \{ \zs \t + \zt \, : \ 0 \leq \zs \leq 1, \ 0 \leq \zt \leq 1 \}$. For a column vector $\uu \in \IR^{N \times 1}$ we define
\begin{equation}
\sign{\uu} \;\coloneqq\; \prod_{j=1}^N \sign{\uu_j}
\andd
\prod \uu \;\coloneqq\;\prod_{j=1}^N \uu_j.
\end{equation}

We now list some of the basic modular objects we need in our exposition.
\begin{enumerate}[label=(\roman*)]
\item The Dedekind eta function is defined as
\begin{equation}
\eta (\t) \coloneqq q^{1/24} \prod_{n=1}^\infty (1 - q^n) 
\= q^{1/24} \sum_{n \in \IZ} (-1)^n \, q^{n(3n-1)/2}.
\end{equation}
It satisfies
\begin{equation}
\eta (\t + 1) \= e^{\pi i /12} \eta (\t)
\andd
\eta (-1/\t) \= e^{- \pi i / 4} \, \t^{1/2} \, \eta (\t).
\end{equation}
\item The level $m$ theta functions are defined as
\begin{equation}\label{eq:levelmtheta_def}
\vartheta_{m,r} (\t,z) \;\coloneqq\; \sum_{n \in \IZ + r/2m} q^{mn^2} \, \yy_z^{2mn}.
\end{equation}
\item The Jacobi theta function $\thone (\t, z)$ is defined as
\begin{align}
\thone (\t, z) &\;\coloneqq\; i \, q^{1/8} \yy_z^{1/2} \prod_{n=1}^\infty (1 -q^n)
(1 - \yy_z \, q^n) (1 -\yy_z^{-1} \, q^{n-1})  \\
\label{eq:thetadef}
&\= i \sum_{m \in \IZ} (-1)^m q^{\frac{1}{2} \lp m + \frac{1}{2} \rp^2}
\yy_z^{m+\frac{1}{2}}.
\end{align}
The Jacobi theta function is an odd function of its elliptic variable, $\thone (\t, -z) = - \thone (\t, z) $, and transforms under elliptic transformations as 
\begin{equation}\label{eq:thetaell}
\thone (\t, z + \a \t + \b) \= (-1)^{\a + \b} \, q^{-\a^2/2} \, \yy_z^{-\a} \,
\thone (\t, z) 
\mbox{ for any } \a,\b \in \IZ.
\end{equation}
Under modular transformations we have
\begin{equation}
\thone (\t + 1, z) \= e^{\pi i/ 4} \, \thone (\t, z)
\end{equation}
and
\begin{equation}
\thone \lp -1/\t,z/\t \rp \= e^{- 3 \pi i/4 } \, \t^{1/2} \, e^{\pi i z^2 / \t} \,
\thone (\t, z).
\end{equation}
One final property that will be useful for us is
\begin{equation}\label{eq:thetaderivative}
\frac{1}{2 \pi i} \thone' (\t, 0) \;\coloneqq\;
\frac{1}{2 \pi i} \left.  \frac{\del}{\del z} \thone (\t, z)  \right|_{z=0} \= i \, \eta (\t)^3.
\end{equation}
\end{enumerate}

We also use Poisson summation in our discussion. For convenience we give its statement for Gaussian functions:
\begin{equation}\label{eq:PoissonSum}
\sum_{\mm \in \IZ^N} \exp \lp - \pi \mm^T A \mm - 2 \pi i B^T \mm \rp
\= \frac{1}{\sqrt{\det A}} \sum_{\nnn \in \IZ^N} 
\exp \lp -\pi (B + \nnn)^T A^{-1} (B + \nnn) \rp,
\end{equation}
where $A$ is an $N \times N$ positive definite matrix and $B$ is an $N$ component column vector.

\section{Generalized Error Functions}\label{sec:gen_err}
Generalized error functions, first introduced in \cite{Alexandrov:2016enp}, are important ingredients in the construction of indefinite theta series. In this section we introduce their definitions and review some of their properties following the conventions of \cite{Nazaroglu:2016lmr} (except for changing the sign of the bilinear form for boosted error functions). 

Let $\CMM \in \IR^{r \times r}$ be a nondegenerate matrix and let $\vv \in \IR^{r \times 1}$. The `$r$-tuple error function' $E_r (\CMM; \vv)$ is then defined as
\begin{equation}\label{eq:gen_err_defn}
E_r (\CMM; \vv) \;\coloneqq\; \intt_{\IR^r} \ddd^r \vv' \, e^{- \pi (\vv - \vv')^T (\vv - \vv')}
\sign{\CMM^T \vv'}.
\end{equation}
For generic $\vv$, we have
\begin{equation}\label{eq:gen_err_asymp}
E_r (\CMM; \vv) \;\to\; \sign{\CMM^T \vv} \mbox{ as } | \vv | \;\to\; \infty.
\end{equation}
We also define the `complementary $r$-tuple error function' (which is piecewise smooth) as
\begin{equation}\label{eq:comp_gen_err_defn}
M_r (\CMM; \vv) \;\coloneqq\; \lp \frac{i}{\pi} \rp^r \left| \det \CMM \right|^{-1}
 \intt_{\IR^r - i \vv} \ddd^r \zzz \, 
 \frac{
 e^{-\pi \zzz^T \zzz - 2 \pi i\zzz^T \vv}
 }{\prod \lp \CMM^{-1} \zzz \rp}.
\end{equation}
When $r=1$ we have
$E_1 (1; \vv) = \erf (\vv \sqrt{\pi})$
 and 
$M_1 (1; \vv) = - \sign{\vv} \, \mathrm{erfc} ( |\vv| \sqrt{\pi} )$,
which in particular satisfy $E_1 (1; \vv)  = \sign{\vv} + M_1 (1; \vv)$. For $r=2$, the generalized error functions $E_2 \lp \pmat{1 & -\a \\ 0 & 1}^{-T}; \pmat{\vv_1 \\ \vv_2} \rp$ and $M_2 \lp \pmat{1 & -\a \\ 0 & 1}^{-T}; \pmat{\vv_1 \\ \vv_2} \rp$ reduce to $E_2 \lp \a; \vv_1, \vv_2 \rp$ and $M_2 \lp \a; \vv_1, \vv_2 \rp$ as defined in \cite{Alexandrov:2016enp}, respectively.

Orthogonal transformations leave these functions invariant, i.e. for $\Lambda \in O(r; \IR)$ we have
\begin{equation}\label{eq:gen_err_orth_inv}
E_r (\Lambda \CMM; \Lambda \vv) \= E_r (\CMM; \vv)
\andd
M_r (\Lambda \CMM; \Lambda \vv) \= M_r (\CMM; \vv).
\end{equation}
We also define boosted generalized error functions for a quadratic form $(x,y) \mapsto x \cdot y \coloneqq x^T Q y$, where $x,y \in \IR^n$, by
\begin{equation}\label{eq:boost_gen_err_app}
E^Q \lp F; x \rp \;\coloneqq\; E_r \lp \mathcal{B} \cdot F; \mathcal{B} \cdot x \rp
\andd
M^Q \lp F; x \rp \;\coloneqq\; M_r \lp \mathcal{B} \cdot F; \mathcal{B} \cdot x \rp,
\end{equation}
where the columns of $F= \lp f^{(1)} \cdots f^{(r)} \rp \in \IR^{n \times r}$ span a negative definite subspace with respect to the quadratic form defined by $Q$ and $\mathcal{B} \in \IR^{n \times r}$ is a matrix whose columns form an orthonormal basis for the subspace spanned by the columns of $F$, i.e. $\mathcal{B}^T Q \mathcal{B} = - I_r$ and $F = - \mathcal{B} \mathcal{B}^T Q F$. As in the main text, we use the notation $\mathcal{F} \cdot \mathcal{G} \coloneqq \mathcal{F}^T Q \, \mathcal{G} \in \IR^{r \times s}$ for any $\mathcal{F} \in \IR^{n \times r}$ and $\mathcal{G} \in \IR^{n \times s}$. Moreover, we call a vector $f \in \IR^n$ positive, negative, or null if $f^2 > 0$, $f^2<0$, or $f^2 = 0$, respectively. Note that thanks to equation \eqref{eq:gen_err_orth_inv}, the right hand sides of definitions in \eqref{eq:boost_gen_err_app} are independent of the choice of basis $\mathcal{B}$.\footnote{We will take $E^Q$ and $M^Q$ to be unity when the set $F$ is empty.}

We also introduce the following notation:
\begin{itemize}
\item $\wt{F} = \lp \wt{f}^{(1)} \cdots \wt{f}^{(r)} \rp \in \IR^{n \times r}$ is a matrix whose columns form a dual basis to those of $F$ in the subspace they span. In other words, $\wt{F} \cdot F = -I_r$ and  $\wt{F} = - \mathcal{B} \mathcal{B}^T Q \wt{F}$. We have $\wt{F} = - \mathcal{B} (\mathcal{B}^T Q F)^{-T}$.
\item For $S \subseteq [r] \coloneqq \{ 1, \ldots, r\}$ we define $F_S$ to be the $n \times |S|$ matrix whose columns are $f^{(j)}$ with $j \in S$ and ordered in increasing $j$ order. We similarly define $\wt{F}_S$ as the matrix whose columns are $\wt{f}^{(j)}$ with $j \in S$.
\item $F_{S \perp S'}$ is defined by orthogonally projecting the columns of $F_S$ to the subspace orthogonal to that spanned by $F_{S'}$. The columns of $F_{S \perp S'}$ are $f^{(j)} - F_{S'} (F_{S'} \cdot F_{S'})^{-1} F_{S'} \cdot f^{(j)}$ with $j \in S$.
\end{itemize}
 
 With this notation at hand, the generalized error functions satisfy the following properties (see \cite{Nazaroglu:2016lmr} for the proofs of these facts).

\noindent i) $E^Q \lp F; x \rp$ and $M^Q \lp F; x \rp$ are invariant under permutations and positive scalings of the columns of $F$. Also they are odd under sign flips of these columns.

\noindent ii) If $F$ is the disjoint union of two sets of columns, $F_1$ and $F_2$, which span orthogonal subspaces with respect to $Q$, we have
\begin{equation}\label{eq:gen_error_factorization}
E^Q \lp F; x \rp = E^Q \lp F_1; x \rp \, E^Q \lp F_2; x \rp
\andd
M^Q \lp F; x \rp = M^Q \lp F_1; x \rp \, M^Q \lp F_2; x \rp .
\end{equation}

\noindent iii) The complementary error function $M^Q \lp F; x \rp$ is exponentially suppressed (uniformly) along the directions spanned by $F$:
\begin{equation}\label{eq:MQ_suppression}
\left| M^Q \lp F; x \rp  \right| \;\leq\; r! \, e^{-\pi \lp \mathcal{B} \cdot x \rp^T \lp \mathcal{B} \cdot x \rp}.
\end{equation}

\noindent iv) We have the following decompositions (generalizing $E_1 (1; \vv)  = \sign{\vv} + M_1 (1; \vv)$):
\begin{equation}
M^Q \lp F; x \rp \= 
\sum_{S \subseteq [r]} \sign{\wt{F}_{[r]/S} \cdot x} \, E^Q \lp F_S; x \rp
\end{equation}
and
\begin{equation}\label{eq:EQ_decomposition}
E^Q \lp F; x \rp \= 
\sum_{S \subseteq [r]} \sign{- F_{[r]/S \perp S} \cdot x} \, M^Q \lp F_S; x \rp .
\end{equation}
The decomposition in \eqref{eq:EQ_decomposition} together with equation \eqref{eq:MQ_suppression} implies generically (that is except on some lower dimensional subspaces)
\begin{equation}
E^Q \lp F; \lambda x \rp \;\to\; \sign{- F \cdot x} \mbox{ as } \lambda \to \infty .
\end{equation}

\noindent v) The generalized error function $E^Q$ satisfies a second order differential equation called Vign\'eras equation:
\begin{equation}
\lb 2 \pi x^T \del_x - \del_x^T Q^{-1} \del_x \rb E^Q \lp F; x \rp \= 0,
\end{equation}
where $\del_x = \lp \del_{x_1}, \ldots, \del_{x_n} \rp^T$. This differential equation ensures a self-duality property for $E^Q \lp F; x \rp$ under Fourier transform \cite{Vigneras1977}. This is precisely the point that allows the replacement of sign functions by generalized error functions in \eqref{eq:gen_error_replacement} yield a modular invariant object. Another key property that ensures convergence after this replacement (subject to some conditions on the vectors forming the positive cone) is the decomposition in \eqref{eq:EQ_decomposition} together with the inequality \eqref{eq:MQ_suppression}, which implies that the difference between the sign function product and the generalized error function consists of exponentially suppressed pieces along negative directions.

\noindent vi) In computing the anti-holomorphic dependence of the completed indefinite theta function it will be useful to note
\begin{equation}\label{eq:gen_error_radial_derivative}
x^T \del_x E^Q \lp F; x \rp \= -2 \sum_{j=1}^r 
\frac{f^{(j)} \cdot x}{\sqrt{-f^{(j)} \cdot f^{(j)}}} 
e^{\pi \lp f^{(j)} \cdot x \rp^2/f^{(j)} \cdot f^{(j)}}
E^Q \lp F_{[r]/\{j\} \perp \{j\} } ; x \rp.
\end{equation}

Let us finally relax the definition of boosted generalized error functions by allowing null vectors in $F$ (as limits of negative vectors). We will still require that the subspace spanned by the vectors in $F$ to be negative semi-definite. That constrains each null vector to be orthogonal to any other vector in $F$. So using the factorization property \eqref{eq:gen_error_factorization} and remembering that for a negative vector $n$, $E^Q \lp n; x \rp = \erf \lp - \sqrt{\pi} \, n \cdot x / \sqrt{-n \cdot n} \rp$ becomes $\sign{- n \cdot x}$ in the limit $n\cdot n \to 0$ we define
\begin{equation}
E^Q \lp (N, F); x \rp \= E^Q \lp F; x \rp \, \sign{- N \cdot x},
\end{equation}
where $N$ is a set of null vectors which together with negative vectors in $F$ span a negative semidefinite subspace with respect to the inner product defined by $Q$.

\bigskip

%

\begin{thebibliography}{10}

\bibitem{Schellekens:1986yi}
A.~N. Schellekens and N.~P. Warner, \emph{{Anomalies and Modular Invariance in
  String Theory}},
  \href{https://doi.org/10.1016/0370-2693(86)90760-4}{\emph{Phys. Lett.}
  {\bfseries B177} (1986) 317--323}.

\bibitem{Schellekens:1986xh}
A.~N. Schellekens and N.~P. Warner, \emph{{Anomalies, Characters and Strings}},
  \href{https://doi.org/10.1016/0550-3213(87)90108-8}{\emph{Nucl. Phys.}
  {\bfseries B287} (1987) 317}.

\bibitem{Pilch:1986en}
K.~Pilch, A.~N. Schellekens and N.~P. Warner, \emph{{Path Integral Calculation
  of String Anomalies}},
  \href{https://doi.org/10.1016/0550-3213(87)90109-X}{\emph{Nucl. Phys.}
  {\bfseries B287} (1987) 362--380}.

\bibitem{Witten:1986bf}
E.~Witten, \emph{{Elliptic Genera and Quantum Field Theory}},
  \href{https://doi.org/10.1007/BF01208956}{\emph{Commun. Math. Phys.}
  {\bfseries 109} (1987) 525}.

\bibitem{Witten:1987cg}
E.~Witten, \emph{{The index of the Dirac operator in loop space}},
  {\emph{http://alice.cern.ch/format/showfull?sysnb=0088339} (1987) }.

\bibitem{Alvarez:1987wg}
O.~Alvarez, T.~P. Killingback, M.~L. Mangano and P.~Windey, \emph{{String
  Theory and Loop Space Index Theorems}},
  \href{https://doi.org/10.1007/BF01239011}{\emph{Commun. Math. Phys.}
  {\bfseries 111} (1987) 1}.

\bibitem{Alvarez:1987de}
O.~Alvarez, T.~P. Killingback, M.~L. Mangano and P.~Windey, \emph{{The
  Dirac-Ramond operator in string theory and loop space index theorems}},
  \href{https://doi.org/10.1016/0920-5632(87)90110-1}{\emph{Nucl. Phys. Proc.
  Suppl.} {\bfseries 1A} (1987) 189--215}.

\bibitem{Gupta:2017bcp}
R.~K. Gupta and S.~Murthy, \emph{{Squashed toric sigma models and mock modular
  forms}},  
\href{https://doi.org/10.1007/s00220-017-3069-5}{\emph{Commun. Math. Phys.}
  {\bfseries 360} (2018) 405},  
  [\href{https://arxiv.org/abs/1705.00649}{{\ttfamily 1705.00649}}].

\bibitem{Hori:2001ax}
K.~Hori and A.~Kapustin, \emph{{Duality of the fermionic 2-D black hole and N=2
  liouville theory as mirror symmetry}},
  \href{https://doi.org/10.1088/1126-6708/2001/08/045}{\emph{JHEP} {\bfseries
  08} (2001) 045}, [\href{https://arxiv.org/abs/hep-th/0104202}{{\ttfamily
  hep-th/0104202}}].

\bibitem{Zwegers:2008zna}
S.~Zwegers, \emph{{Mock Theta Functions}}, Ph.D. thesis, 2002.
\newblock [\href{https://arxiv.org/abs/0807.4834}{{\ttfamily 0807.4834}}].

\bibitem{BruinierFunke}
J.~H. Bruinier and J.~Funke, \emph{On two geometric theta lifts},
  \href{https://doi.org/10.1215/S0012-7094-04-12513-8}{\emph{Duke Math. J.}
  {\bfseries 125} (2004) 45--90}.

\bibitem{Zagier2009}
D.~Zagier, \emph{Ramanujan's mock theta functions and their applications (after
  {Z}wegers and {O}no-{B}ringmann)}, {\emph{Ast\'erisque} (2009) Exp. No. 986,
  vii--viii, 143--164 (2010)}.

\bibitem{Dabholkar:2012nd}
A.~Dabholkar, S.~Murthy and D.~Zagier, \emph{{Quantum Black Holes, Wall
  Crossing, and Mock Modular Forms}},
  [\href{https://arxiv.org/abs/1208.4074}{{\ttfamily 1208.4074}}].

\bibitem{Bringmann:2017book}
K.~Bringmann, A.~Folsom, K.~Ono and L.~Rolen, \emph{Harmonic {M}aass forms and
  mock modular forms: theory and applications}, vol.~64 of \emph{American
  Mathematical Society Colloquium Publications}.
\newblock American Mathematical Society, Providence, RI, 2017.

\bibitem{ZagierZwegers}
D.~Zagier and S.~Zwegers, unpublished.

\bibitem{Alexandrov:2016enp}
S.~Alexandrov, S.~Banerjee, J.~Manschot and B.~Pioline, \emph{{Indefinite theta
  series and generalized error functions}},
  [\href{https://arxiv.org/abs/1606.05495}{{\ttfamily 1606.05495}}].

\bibitem{kudla2016theta}
S.~Kudla, \emph{Theta integrals and generalized error functions}, 
[\href{https://arxiv.org/abs/1608.03534}{{\ttfamily  1608.03534}}].

\bibitem{westerholt2016indefinite}
M.~Westerholt-Raum, \emph{Indefinite theta series on tetrahedral cones},
[\href{https://arxiv.org/abs/1608.08874}{{\ttfamily  1608.08874}}].

\bibitem{Nazaroglu:2016lmr}
C.~Nazaroglu, \emph{{$r$-Tuple Error Functions and Indefinite Theta Series of
  Higher-Depth}}, \href{http://dx.doi.org/10.4310/CNTP.2018.v12.n3.a4}{\emph{Communications in Number Theory and Physics}
  {\bfseries 12} (2018) 581-608},
   [\href{https://arxiv.org/abs/1609.01224}{{\ttfamily 1609.01224}}].

\bibitem{funke2017theta}
J.~Funke and S.~Kudla, \emph{{Theta integrals and generalized error functions,
  II}},  [\href{https://arxiv.org/abs/1708.02969}{{\ttfamily 1708.02969}}].

\bibitem{Witten:1993yc}
E.~Witten, \emph{{Phases of N=2 theories in two-dimensions}},
  \href{https://doi.org/10.1016/0550-3213(93)90033-L}{\emph{Nucl. Phys.}
  {\bfseries B403} (1993) 159--222},
  [\href{https://arxiv.org/abs/hep-th/9301042}{{\ttfamily hep-th/9301042}}].

\bibitem{Murthy:2013mya}
S.~Murthy, \emph{{A holomorphic anomaly in the elliptic genus}},
  \href{https://doi.org/10.1007/JHEP06(2014)165}{\emph{JHEP} {\bfseries 06}
  (2014) 165}, [\href{https://arxiv.org/abs/1311.0918}{{\ttfamily 1311.0918}}].

\bibitem{Zwegers:2011}
S.~Zwegers, \emph{{Appell-Lerch Sums}},  2011, 
\href{http://indico.ictp.it/event/a10129/session/42/contribution/25/material/0/0.pdf}{{http://indico.ictp.it/event/a10129/session/42/contribution/25/material/0/0.pdf }}.

\bibitem{Ashok:2014nua}
S.~K. Ashok, E.~Dell'Aquila and J.~Troost, \emph{{Higher Poles and Crossing
  Phenomena from Twisted Genera}},
  \href{https://doi.org/10.1007/JHEP08(2014)087}{\emph{JHEP} {\bfseries 08}
  (2014) 087}, [\href{https://arxiv.org/abs/1404.7396}{{\ttfamily 1404.7396}}].

\bibitem{daSilva:2001}
A.~C. da~Silva, \emph{{Symplectic Toric Manifolds}}.
\newblock Birkh{\"a}user series Advanced Courses in Mathematics, CRM Barcelona,
  Birkhauser (Springer), 2003.

\bibitem{Benini:2013nda}
F.~Benini, R.~Eager, K.~Hori and Y.~Tachikawa, \emph{{Elliptic genera of
  two-dimensional N=2 gauge theories with rank-one gauge groups}},
  \href{https://doi.org/10.1007/s11005-013-0673-y}{\emph{Lett. Math. Phys.}
  {\bfseries 104} (2014) 465--493},
  [\href{https://arxiv.org/abs/1305.0533}{{\ttfamily 1305.0533}}].

\bibitem{Benini:2013xpa}
F.~Benini, R.~Eager, K.~Hori and Y.~Tachikawa, \emph{{Elliptic Genera of 2d
  ${\mathcal{N}}$ = 2 Gauge Theories}},
  \href{https://doi.org/10.1007/s00220-014-2210-y}{\emph{Commun. Math. Phys.}
  {\bfseries 333} (2015) 1241--1286},
  [\href{https://arxiv.org/abs/1308.4896}{{\ttfamily 1308.4896}}].

\bibitem{Harvey:2013mda}
J.~A. Harvey and S.~Murthy, \emph{{Moonshine in Fivebrane Spacetimes}},
  \href{https://doi.org/10.1007/JHEP01(2014)146}{\emph{JHEP} {\bfseries 01}
  (2014) 146}, [\href{https://arxiv.org/abs/1307.7717}{{\ttfamily 1307.7717}}].

\bibitem{Skoruppa1988}
N.-P. Skoruppa and D.~Zagier, \emph{Jacobi forms and a certain space of modular
  forms}, \href{https://doi.org/10.1007/BF01394347}{\emph{Inventiones
  mathematicae} {\bfseries 94} (Feb, 1988) 113--146}.

\bibitem{EichlerZagier}
M.~Eichler and D.~Zagier, \emph{The theory of {J}acobi forms}, vol.~55 of
  \emph{Progress in Mathematics}.
\newblock Birkh\"auser Boston, Inc., Boston, MA, 1985,
  \href{https://doi.org/10.1007/978-1-4684-9162-3}{10.1007/978-1-4684-9162-3}.

\bibitem{Troost:2010ud}
J.~Troost, \emph{{The non-compact elliptic genus: mock or modular}},
  \href{https://doi.org/10.1007/JHEP06(2010)104}{\emph{JHEP} {\bfseries 06}
  (2010) 104}, [\href{https://arxiv.org/abs/1004.3649}{{\ttfamily 1004.3649}}].

\bibitem{Eguchi:2010cb}
T.~Eguchi and Y.~Sugawara, \emph{{Non-holomorphic Modular Forms and
  SL(2,R)/U(1) Superconformal Field Theory}},
  \href{https://doi.org/10.1007/JHEP03(2011)107}{\emph{JHEP} {\bfseries 03}
  (2011) 107}, [\href{https://arxiv.org/abs/1012.5721}{{\ttfamily 1012.5721}}].

\bibitem{Ashok:2011cy}
S.~K. Ashok and J.~Troost, \emph{{A Twisted Non-compact Elliptic Genus}},
  \href{https://doi.org/10.1007/JHEP03(2011)067}{\emph{JHEP} {\bfseries 03}
  (2011) 067}, [\href{https://arxiv.org/abs/1101.1059}{{\ttfamily 1101.1059}}].

\bibitem{Vafa:1994tf}
C.~Vafa and E.~Witten, \emph{{A Strong coupling test of S duality}},
  \href{https://doi.org/10.1016/0550-3213(94)90097-3}{\emph{Nucl. Phys.}
  {\bfseries B431} (1994) 3--77},
  [\href{https://arxiv.org/abs/hep-th/9408074}{{\ttfamily hep-th/9408074}}].

\bibitem{Moore:1997pc}
G.~W. Moore and E.~Witten, \emph{{Integration over the u plane in Donaldson
  theory}}, \href{https://doi.org/10.4310/ATMP.1997.v1.n2.a7}{\emph{Adv. Theor.
  Math. Phys.} {\bfseries 1} (1997) 298--387},
  [\href{https://arxiv.org/abs/hep-th/9709193}{{\ttfamily hep-th/9709193}}].

\bibitem{Korpas:2017qdo}
G.~Korpas and J.~Manschot, \emph{{Donaldson-Witten theory and indefinite theta
  functions}}, \href{https://doi.org/10.1007/JHEP11(2017)083}{\emph{JHEP}
  {\bfseries 11} (2017) 083},
  [\href{https://arxiv.org/abs/1707.06235}{{\ttfamily 1707.06235}}].

\bibitem{Alexandrov:2016tnf}
S.~Alexandrov, S.~Banerjee, J.~Manschot and B.~Pioline, \emph{{Multiple
  D3-instantons and mock modular forms I}},
  \href{https://doi.org/10.1007/s00220-016-2799-0}{\emph{Commun. Math. Phys.}
  {\bfseries 353} (2017) 379--411},
  [\href{https://arxiv.org/abs/1605.05945}{{\ttfamily 1605.05945}}].

\bibitem{Alexandrov:2017qhn}
S.~Alexandrov, S.~Banerjee, J.~Manschot and B.~Pioline, \emph{{Multiple
  D3-instantons and mock modular forms II}},
  \href{https://doi.org/10.1007/s00220-018-3114-z}{\emph{Commun. Math. Phys.}
  {\bfseries 359} (2018) 297--346},
  [\href{https://arxiv.org/abs/1702.05497}{{\ttfamily 1702.05497}}].

\bibitem{bringmann2016higher}
K.~Bringmann, J.~Kaszian and L.~Rolen, \emph{Higher-depth mock modular forms
  arising in gromov-witten theory of elliptic orbifolds}, 
 [\href{https://arxiv.org/abs/1608.08588}{{\ttfamily 1608.08588}}].

\bibitem{Manschot:2017xcr}
J.~Manschot, \emph{{Vafa-Witten theory and iterated integrals of modular
  forms}},  [\href{https://arxiv.org/abs/1709.10098}{{\ttfamily 1709.10098}}].

\bibitem{Bringmann:2018cov}
K.~Bringmann and C.~Nazaroglu, \emph{{An exact formula for $\mathrm{U}(3)$
  Vafa-Witten invariants on $\mathbb{P}^2$}},
  [\href{https://arxiv.org/abs/1803.09270}{{\ttfamily 1803.09270}}].
  
\bibitem{Alexandrov:2014wca} 
  S.~Alexandrov, G.~W.~Moore, A.~Neitzke and B.~Pioline,
  \emph{$\mathbb R^3$ Index for Four-Dimensional $N=2$ Field Theories},
  Phys.\ Rev.\ Lett.\  {\bf 114}, 121601 (2015),
  [\href{https://arxiv.org/abs/1406.2360}{{\ttfamily 1406.2360}}].
  
\bibitem{Pioline:2015wza} 
  B.~Pioline,
  \emph{Wall-crossing made smooth},
  JHEP {\bf 1504}, 092 (2015),
  [\href{https://arxiv.org/abs/1501.01643}{{\ttfamily 1501.01643}}].
  
\bibitem{Murthy:2018bzs} 
  S.~Murthy and B.~Pioline,
  \emph{Mock modularity from black hole scattering states},
  [\href{https://arxiv.org/abs/1808.05606}{{\ttfamily 1808.05606}}].

\bibitem{Manschot:2014cca}
J.~Manschot, \emph{{Sheaves on $\mathbb{P}^2$ and generalized Appell
  functions}}, \href{https://doi.org/10.4310/ATMP.2017.v21.n3.a3}{\emph{Adv.
  Theor. Math. Phys.} {\bfseries 21} (2017) 655--681},
  [\href{https://arxiv.org/abs/1407.7785}{{\ttfamily 1407.7785}}].

\bibitem{Ashok:2013zka}
S.~K. Ashok and J.~Troost, \emph{{Elliptic genera and real Jacobi forms}},
  \href{https://doi.org/10.1007/JHEP01(2014)082}{\emph{JHEP} {\bfseries 01}
  (2014) 082}, [\href{https://arxiv.org/abs/1310.2124}{{\ttfamily 1310.2124}}].

\bibitem{Yoshioka1994}
K.~Yoshioka, \emph{The {B}etti numbers of the moduli space of stable sheaves of
  rank {$2$} on {$\bold P^2$}},
  \href{https://doi.org/10.1515/crll.1994.453.193}{\emph{J. Reine Angew. Math.}
  {\bfseries 453} (1994) 193--220}.

\bibitem{Klyachko1991}
A.~A. Klyachko, \emph{Moduli of vector bundles and numbers of classes},
  \href{https://doi.org/10.1007/BF01090685}{\emph{Functional Analysis and Its
  Applications} {\bfseries 25} (Jan, 1991) 67--69}.

\bibitem{Bringmann:2010sd}
K.~Bringmann and J.~Manschot, \emph{{From sheaves on $P^2$ to a generalization
  of the Rademacher expansion}},
  [\href{https://arxiv.org/abs/1006.0915}{{\ttfamily 1006.0915}}].

\bibitem{Minahan:1998vr}
J.~A. Minahan, D.~Nemeschansky, C.~Vafa and N.~P. Warner, \emph{{E strings and
  N=4 topological Yang-Mills theories}},
  \href{https://doi.org/10.1016/S0550-3213(98)00426-X}{\emph{Nucl. Phys.}
  {\bfseries B527} (1998) 581--623},
  [\href{https://arxiv.org/abs/hep-th/9802168}{{\ttfamily hep-th/9802168}}].

\bibitem{Aharony:1998ub}
O.~Aharony, M.~Berkooz, D.~Kutasov and N.~Seiberg, \emph{{Linear dilatons, NS
  five-branes and holography}},
  \href{https://doi.org/10.1088/1126-6708/1998/10/004}{\emph{JHEP} {\bfseries
  10} (1998) 004}, [\href{https://arxiv.org/abs/hep-th/9808149}{{\ttfamily
  hep-th/9808149}}].

\bibitem{Giveon:1999zm}
A.~Giveon, D.~Kutasov and O.~Pelc, \emph{{Holography for noncritical
  superstrings}},
  \href{https://doi.org/10.1088/1126-6708/1999/10/035}{\emph{JHEP} {\bfseries
  10} (1999) 035}, [\href{https://arxiv.org/abs/hep-th/9907178}{{\ttfamily
  hep-th/9907178}}].

\bibitem{Murthy:2003es}
S.~Murthy, \emph{{Notes on noncritical superstrings in various dimensions}},
  \href{https://doi.org/10.1088/1126-6708/2003/11/056}{\emph{JHEP} {\bfseries
  11} (2003) 056}, [\href{https://arxiv.org/abs/hep-th/0305197}{{\ttfamily
  hep-th/0305197}}].

\bibitem{Kutasov:1990ua}
D.~Kutasov and N.~Seiberg, \emph{{Noncritical superstrings}},
  \href{https://doi.org/10.1016/0370-2693(90)90233-V}{\emph{Phys. Lett.}
  {\bfseries B251} (1990) 67--72}.

\bibitem{Vigneras1977}
M.-F. Vign{\'e}ras, \emph{S{\'e}ries theta des formes quadratiques ind{\'e}finies}.
\emph{Springer Lecture Notes}, 627:227 – 239, 1977.

\end{thebibliography}

\providecommand{\href}[2]{#2}\begingroup\raggedright\endgroup

\end{document}